\documentclass[aps,eqsecnum,preprint,floats,epsf,epsfig,nofootinbib,letter]{revtex4}
\usepackage{epsfig}

\begin{document}
\def\B{{\cal B}}
\def\ov{\overline}
\def\pr{{\sl Phys. Rev.}~}
\def\prl{{\sl Phys. Rev. Lett.}~}
\def\pl{{\sl Phys. Lett.}~}
\def\np{{\sl Nucl. Phys.}~}
\def\zp{{\sl Z. Phys.}~}
\def\lsim{ {\ \lower-1.2pt\vbox{\hbox{\rlap{$<$}\lower5pt\vbox{\hbox{$\sim$}
}}}\ } }
\def\gsim{ {\ \lower-1.2pt\vbox{\hbox{\rlap{$>$}\lower5pt\vbox{\hbox{$\sim$}
}}}\ } }

\font\el=cmbx10 scaled \magstep2{\obeylines\hfill August, 2015}

\vskip 1.5 cm

\centerline{\large\bf Charmed Baryons Circa 2015}

\vskip 1.5 cm

\bigskip
\bigskip
\centerline{\bf Hai-Yang Cheng}
\medskip
\centerline{Institute of Physics, Academia Sinica}
\centerline{Taipei, Taiwan 115, Republic of China}
\medskip

\bigskip
\bigskip
\centerline{\bf Abstract}
\bigskip
\small

This is basically the update of \cite{Cheng:2009}, a review on charmed baryon physics around 2007. Topics of this review include the spectroscopy, strong decays, lifetimes, nonleptonic and semileptonic weak decays, and electromagnetic decays of charmed baryons.

\pagebreak

\tableofcontents

\newpage

\section{Introduction}
Charm baryon spectroscopy provides an excellent ground to study the dynamics of light quarks in the environment of a heavy quark.  In the past decade,
many new excited charmed baryon states have been
discovered by BaBar, Belle, CLEO and LHCb. A very rich source of charmed baryons comes both from $B$
decays and from the $e^+e^-\to c\bar c$ continuum. A lot of efforts have been devoted to identifying the quantum numbers of these new states and understand their properties.

Consider the strong decays $\Sigma_Q\to\Lambda_Q\pi$ and $\Xi_Q^{'*}\to\Xi_Q\pi$ with $Q=c,b$. It turns out that the mass differences between $\Sigma_c$ and $\Lambda_c$ and between $\Xi_c^{'*}$ and $\Xi_c$ in the charmed baryon sector are large enough to render the strong decays of $\Sigma_c$ and $\Xi_c^{'*}$ kinematically allowed. As a consequence, the charmed baryon system offers a unique and excellent laboratory for testing the ideas and predictions of heavy quark symmetry of the heavy quark and chiral symmetry of the light quarks. This will have interesting implications for the low-energy dynamics of heavy baryons interacting with the Goldstone bosons.

Theoretical interest in the charmed baryon hadronic weak decays was peaked around early nineties and then faded away. Until today, we still don't have a good phenomenological model, not mentioning the QCD-inspired approach as in $B$ meson decays, to describe the complicated physics of baryon decays. We do need cooperative efforts from both experimentalists and theorists to make progress in this arena.

This review is basically the update of \cite{Cheng:2009} around 2007. The outline of the content is the same as that of \cite{Cheng:2009} except we add discussions on
the spectroscopy and lifetimes of doubly charmed baryons.

Several excellent review articles on charmed baryons can be found in
\cite{Korner94,Bigireview,Roberts,Klempt,Crede,Bphysics}.

\section{Spectroscopy}
\subsection{Singly charmed baryons}
The singly charmed baryon is composed of a charmed quark and
two light quarks. Each
light quark is a triplet of the flavor SU(3). There are two different SU(3)
multiplets of charmed baryons: a symmetric sextet {\bf 6} and an
antisymmetric antitriplet ${\bf \bar 3}$. The $\Lambda_c^+$,
$\Xi_c^+$ and $\Xi_c^0$ form an ${\bf \bar 3}$ representation and
they all decay weakly. The $\Omega_c^0$, $\Xi'^+_c$, $\Xi'^0_c$
and $\Sigma_c^{++,+,0}$ form a {\bf 6} representation; among them,
only $\Omega_c^0$ decays weakly. We follow the Particle
Data Group's convention \cite{PDG} to use a prime to distinguish the
$\Xi_c$ in the {\bf 6} from the one in the ${\bf \bar 3}$.

\begin{figure}[t]
\centerline{\psfig{file=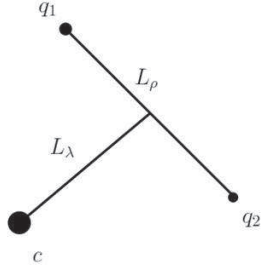,width=1.5in}}
\caption{Singly charmed baryon where $L_\rho$ describes relative orbital excitation of the two light quarks and $L_\lambda$ the orbital excitation of the center of the mass of the two light quarks relative to the charmed quark. } \label{fig:orbital}
\end{figure}

In the quark model, the orbital angular momentum of the light
diquark can be decomposed into ${\bf L}_\ell={\bf L}_\rho+{\bf
L}_\lambda$, where ${\bf L}_\rho$ is the orbital angular momentum
between the two light quarks and ${\bf L}_\lambda$ the orbital
angular momentum between the diquark and the charmed quark.
Although the separate spin angular momentum
$S_\ell$ and orbital angular momentum $L_\ell$ of the light
degrees of freedom are not well defined, they are included for
guidance from the quark model. In the heavy quark limit, the spin
of the charmed quark ${\bf S}_c$ and the total angular momentum of the
two light quarks ${\bf J}_\ell={\bf S}_\ell+{\bf L}_\ell$ are separately conserved. The total angular momentum is given by ${\bf J}={\bf S}_c+{\bf J}_\ell$.
It is convenient to use $S_\ell$, $L_\ell$ and $J_\ell$ to enumerate the spectrum of states.
Moreover, one can
define two independent relative momenta ${\bf p_\rho}={1\over \sqrt{2}}({\bf
p}_1-{\bf p}_2)$ and ${\bf p_\lambda}={1\over \sqrt{6}}({\bf p}_1+{\bf p}_2-2{\bf
p}_c)$ from the two light quark momenta ${\bf p}_1$, ${\bf p}_2$
and the heavy quark momentum ${\bf p}_c$.
Denoting the quantum
numbers $L_\rho$ and $L_\lambda$ as the eigenvalues of ${\bf L}_\rho^2$ and
${\bf L}_\lambda^2$, the $\rho$-orbital momentum $L_\rho$ describes relative
orbital excitations of the two light quarks, and the $\lambda$-orbital
momentum $L_\lambda$ describes orbital excitations of the center of the
mass of the two light quarks relative to the heavy quark (see Fig. \ref{fig:orbital}). The $p$-wave heavy baryon can be either in the
$(L_\rho=0,L_\lambda=1)$ $\lambda$-state or the $(L_\rho=1,L_\lambda=0)$ $\rho$-state. It is
obvious that the orbital $\lambda$-state ($\rho$-state) is symmetric
(antisymmetric) under the interchange of ${\bf p}_1$ and ${\bf
p}_2$. In the following, we shall use the notation
${\cal B}_{cJ_\ell}(J^P)$ ($\tilde{\cal B}_{cJ_\ell}(J^P)$) to denote the states symmetric (antisymmetric) in the orbital wave functions under the exchange of two light quarks. The lowest-lying orbitally excited baryon states are the $p$-wave charmed baryons with their quantum numbers listed in Table
\ref{tab:pwave}.

\begin{table}[t]
\caption{The $p$-wave charmed baryons denoted by ${\cal B}_{cJ_\ell}(J^P)$ and $\tilde {\cal B}_{cJ_\ell}(J^P)$
where $J_\ell$ is the total angular momentum of
the two light quarks.  The orbital $\rho$-states with $L_\rho=1$ and $L_\lambda=0$   have odd orbital wave functions under the
permutation of the two light quarks and are denoted by a tilde. } \label{tab:pwave}
\begin{center}
\begin{tabular}{|c|cccc||c|cccc|} \hline
~~~~~State~~~~~ & SU(3) & ~~$S_\ell$~~ & ~~$L_\ell(L_\rho,L_\lambda)$~~&
~~$J_\ell^{P_\ell}$~~ & ~~~~~State~~~~~ & SU(3) & ~~$S_\ell$~~ &
~~$L_\ell(L_\rho,L_\lambda)$~~&
~~$J_\ell^{P_\ell}$~ \\
 \hline
 $\Lambda_{c1}({1\over 2}^{-},{3\over 2}^{-})$ & ${\bf \bar 3}$ & 0 & 1\,(0,1) &
 $1^-$ &
 $\Sigma_{c0}({1\over 2}^{-})$ & ${\bf 6}$ & 1 & 1\,(0,1) & $0^-$
 \\
 $\tilde\Lambda_{c0}({1\over 2}^{-})$ & ${\bf \bar 3}$ & 1 & 1\,(1,0) & $0^-$ &
 $\Sigma_{c1}({1\over 2}^{-},{3\over 2}^{-})$ & ${\bf 6}$ & 1 & 1\,(0,1) & $1^-$
 \\
 $\tilde\Lambda_{c1}({1\over 2}^{-},{3\over 2}^{-})$ & ${\bf \bar 3}$ & 1 & 1\,(1,0) & $1^-$
 &
 $\Sigma_{c2}({3\over 2}^{-},{5\over 2}^{-})$ & ${\bf 6}$ & 1 & 1\,(0,1) & $2^-$
 \\
 $\tilde\Lambda_{c2}({3\over 2}^{-},{5\over 2}^{-})$ & ${\bf \bar 3}$ & 1 & 1\,(1,0) & $2^-$
 &
 $\tilde \Sigma_{c1}({1\over 2}^{-},{3\over 2}^{-})$ & ${\bf 6}$ & 0 & 1\,(1,0) & $1^-$
 \\
 \hline
 $\Xi_{c1}({1\over 2}^{-},{3\over 2}^{-})$ & ${\bf \bar 3}$ & 0 & 1\,(0,1) & $1^-$  &
 $\Xi'_{c0}({1\over 2}^{-})$ & ${\bf 6}$ & 1 & 1\,(0,1) & $0^-$ \\
 $\tilde\Xi_{c0}({1\over 2}^{-})$ & ${\bf \bar 3}$ & 1 & 1\,(1,0) & $0^-$
 &  $\Xi'_{c1}({1\over 2}^{-},{3\over 2}^{-})$ & ${\bf 6}$ & 1 & 1\,(0,1) &
 $1^-$\\
 $\tilde\Xi_{c1}({1\over 2}^{-},{3\over 2}^{-})$ & ${\bf \bar 3}$ & 1 & 1\,(1,0) & $1^-$
 &  $\Xi'_{c2}({3\over 2}^{-},{5\over 2}^{-})$ & ${\bf 6}$ & 1 & 1\,(0,1) &
 $2^-$\\
 $\tilde\Xi_{c2}({3\over 2}^{-},{5\over 2}^{-})$ & ${\bf \bar 3}$ & 1 & 1\,(1,0) & $2^-$
 &  $\tilde\Xi'_{c1}({1\over 2}^{-},{3\over 2}^{-})$ & ${\bf 6}$ & 0 & 1\,(1,0) & $1^-$ \\
 \hline
\end{tabular}
\end{center}
\end{table}

\begin{table}[t]
\caption{The first positive-parity excitations of charmed baryons denoted by
${\cal B}_{cJ_\ell}(J^P)$,  $\hat {\cal B}_{cJ_\ell}(J^P)$ and $\tilde {\cal B}^{L_\ell}_{cJ_\ell}(J^P)$. Orbital $L_\rho=L_\lambda=1$ states with antisymmetric orbital wave
functions  are denoted by a tilde. States with the symmetric orbital wave
functions $L_\rho=2$ and $L_\lambda=0$ are
denoted by a hat.  For convenience, we drop the superscript $L_\ell$ for tilde states in the sextet.
} \label{tab:pp}
\begin{center}
\begin{tabular}{|c|cccc||c|cccc|} \hline
~~~~~State~~~~~ & SU(3)$_F$ & ~~$S_\ell$~~ & ~~$L_\ell(L_\rho,L_\lambda)$~~&
~~$J_\ell^{P_\ell}$~~ & ~~~~~State~~~~~ & SU(3)$_F$ & ~~$S_\ell$~~
& ~~$L_\ell(L_\rho,L_\lambda)$~~&
~~$J_\ell^{P_\ell}$~ \\
 \hline
 $\Lambda_{c2}({3\over 2}^+,{5\over 2}^+)$ & ${\bf \bar 3}$ & 0 & 2\,(0,2) &
 $2^+$ & $\Sigma_{c1}({1\over 2}^+,{3\over 2}^+)$ & ${\bf 6}$ & 1 & 2\,(0,2) & $1^+$ \\
 $\hat\Lambda_{c2}({3\over 2}^+,{5\over 2}^+)$ & ${\bf \bar 3}$ & 0 & 2\,(2,0) &
 $2^+$ &  $\Sigma_{c2}({3\over 2}^+,{5\over 2}^+)$ & ${\bf 6}$ & 1 & 2\,(0,2) & $2^+$ \\
 $\tilde\Lambda_{c1}({1\over 2}^+,\frac32^+)$ & ${\bf \bar 3}$ & 1 & 0\,(1,1) & $1^+$ &
 $\Sigma_{c3}({5\over 2}^+,{7\over 2}^+)$ & ${\bf 6}$ & 1 & 2\,(0,2) & $3^+$ \\
 $\tilde\Lambda^1_{c0}({1\over 2}^+)$ & ${\bf \bar 3}$ & 1 & 1\,(1,1) & $0^+$ &
 $\hat\Sigma_{c1}({1\over 2}^+,{3\over 2}^+)$ & ${\bf 6}$ & 1 & 2\,(2,0) & $1^+$  \\
 $\tilde\Lambda^1_{c1}({1\over 2}^+,{3\over 2}^+)$ & ${\bf \bar 3}$ & 1 & 1\,(1,1) & $1^+$ &
 $\hat\Sigma_{c2}({3\over 2}^+,{5\over 2}^+)$ & ${\bf 6}$ & 1 & 2\,(2,0) & $2^+$  \\
 $\tilde\Lambda^1_{c2}({3\over 2}^+,{5\over 2}^+)$ & ${\bf \bar 3}$ & 1 & 1\,(1,1) & $2^+$ &
 $\hat\Sigma_{c3}({5\over 2}^+,{7\over 2}^+)$ & ${\bf 6}$ & 1 & 2\,(2,0) & $3^+$  \\
 $\tilde\Lambda^2_{c1}({1\over 2}^+,{3\over 2}^+)$ & ${\bf \bar 3}$ & 1 & 2\,(1,1) & $1^+$ &
 $\tilde\Sigma_{c0}({1\over 2}^+)$ & ${\bf 6}$ & 0 & 0\,(1,1) & $0^+$ \\
 $\tilde\Lambda^2_{c2}({3\over 2}^+,{5\over 2}^+)$ & ${\bf \bar 3}$ & 1 & 2\,(1,1) & $2^+$ &
  $\tilde\Sigma_{c1}({1\over 2}^+,{3\over 2}^+)$ & ${\bf 6}$ & 0 & 1\,(1,1) & $1^+$ \\
 $\tilde\Lambda^2_{c3}({5\over 2}^+,{7\over 2}^+)$ & ${\bf \bar 3}$ & 1 & 2\,(1,1) & $3^+$ &
 $\tilde\Sigma_{c2}({3\over 2}^+,{5\over 2}^+)$ & ${\bf 6}$ & 0 & 2\,(1,1) & $2^+$ \\
 \hline
 $\Xi_{c2}({3\over 2}^+,{5\over 2}^+)$ & ${\bf \bar 3}$ & 0 & 2\,(0,2) &
 $2^+$ & $\Xi'_{c1}({1\over 2}^+,{3\over 2}^+)$ & ${\bf 6}$ & 1 & 2\,(0,2) & $1^+$ \\
 $\hat\Xi_{c2}({3\over 2}^+,{5\over 2}^+)$ & ${\bf \bar 3}$ & 0 & 2\,(2,0) &
 $2^+$ & $\Xi'_{c2}({3\over 2}^+,{5\over 2}^+)$ & ${\bf 6}$ & 1 & 2\,(0,2) & $2^+$ \\
 $\tilde\Xi_{c1}({1\over 2}^+,\frac32^+)$ & ${\bf \bar 3}$ & 1 & 0\,(1,1) & $1^+$ &
 $\Xi'_{c3}({5\over 2}^+,{7\over 2}^+)$ & ${\bf 6}$ & 1 & 2\,(0,2) & $3^+$ \\
 $\tilde\Xi^1_{c0}({1\over 2}^+)$ & ${\bf \bar 3}$ & 1 & 1\,(1,1) & $0^+$ &
 $\hat\Xi'_{c1}({1\over 2}^+,{3\over 2}^+)$ & ${\bf 6}$ & 1 & 2\,(2,0) & $1^+$  \\
 $\tilde\Xi^1_{c1}({1\over 2}^+,{3\over 2}^+)$ & ${\bf \bar 3}$ & 1 & 1\,(1,1) & $1^+$ &
 $\hat\Xi'_{c2}({3\over 2}^+,{5\over 2}^+)$ & ${\bf 6}$ & 1 & 2\,(2,0) & $2^+$ \\
 $\tilde\Xi^1_{c2}({3\over 2}^+,{5\over 2}^+)$ & ${\bf \bar 3}$ & 1 & 1\,(1,1) & $2^+$ &
 $\hat\Xi'_{c3}({5\over 2}^+,{7\over 2}^+)$ & ${\bf 6}$ & 1 & 2\,(2,0) & $3^+$ \\
 $\tilde\Xi^2_{c1}({1\over 2}^+,{3\over 2}^+)$ & ${\bf \bar 3}$ & 1 & 2\,(1,1) & $1^+$ &
 $\tilde\Xi'_{c0}({1\over 2}^+)$ & ${\bf 6}$ & 0 & 0\,(1,1) & $0^+$\\
 $\tilde\Xi^2_{c2}({3\over 2}^+,{5\over 2}^+)$ & ${\bf \bar 3}$ & 1 & 2\,(1,1) & $2^+$ &
  $\tilde\Xi'_{c1}({1\over 2}^+,{3\over 2}^+)$ & ${\bf 6}$ & 0 & 1\,(1,1) & $1^+$ \\
 $\tilde\Xi^2_{c3}({5\over 2}^+,{7\over 2}^+)$ & ${\bf \bar 3}$ & 1 & 2\,(1,1) & $3^+$ &
 $\tilde\Xi'_{c2}({3\over 2}^+,{5\over 2}^+)$ & ${\bf 6}$ & 0 & 2\,(1,1) & $2^+$ \\
 \hline
\end{tabular}
\end{center}
\end{table}

The next orbitally excited states are the positive-parity
excitations with $L_\rho+L_\lambda=2$.  There exist multiplets
(e.g. $\Lambda_{c2}$ and $\hat\Lambda_{c2}$) with the
symmetric orbital wave function, corresponding to
$L_\lambda=2,L_\rho=0$ and $L_\lambda=0,L_\rho=2$ (see Table \ref{tab:pp}). We use a hat to distinguish them.
Since the orbital $L_\lambda=L_\rho=1$ states
are antisymmetric under the interchange of two light quarks,
we shall use a tilde to denote them.
Moreover, the notation $\tilde {\cal B}^{L_\ell}_{cJ_\ell}(J^P)$ is reserved for tilde states in the ${\bf \bar 3}$ as the quantum number $L_\ell$ is needed to distinguish different states.

\begin{figure}[h]
\centerline{\psfig{file=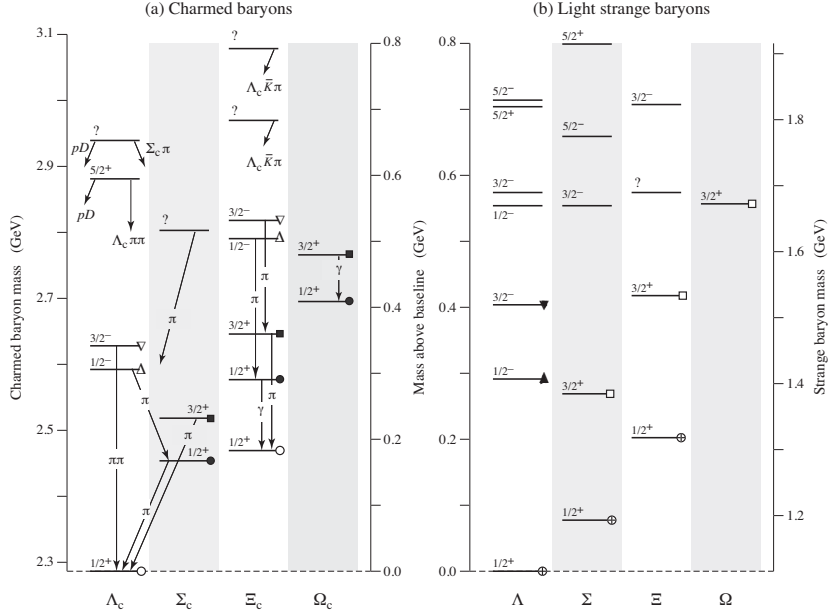,width=4.3in}}
\caption{Charmed baryons and their excitations
\cite{PDG}.} \label{fig:charmspect}
\end{figure}

The observed mass spectra and decay widths of charmed baryons are
summarized in Table \ref{tab:spectrum} (see also Fig.
\ref{fig:charmspect}).
Notice that except for the parity of the lightest
$\Lambda_c^+$ and the heavier one $\Lambda_c(2880)^+$, none of the other $J^P$ quantum numbers given in
Table \ref{tab:spectrum} has been measured. One has to rely on the
quark model to determine the spin-parity assignments.

\begin{table}[!]
\caption{Mass spectra and widths (in units of MeV) of
charmed baryons. Experimental values are taken from the Particle
Data Group \cite{PDG}. For the widths of the $\Sigma_c(2455)^{0/++}$ and $\Sigma_c(2520)^{0/++}$ baryons, we have taken into account the recent Belle measurement \cite{Belle:2014} for average. The width of $\Xi_c(2645)^+$ is taken from \cite{Belle:dc}. For $\Xi_c(3055)^0$, we quote the preliminary result from Belle \cite{Kato}.}
\label{tab:spectrum}
\begin{center}
\begin{tabular}{|c|c ccc c c c|c|} \hline \hline
~~State~~ & ~~$J^P$~~ &~$S_\ell$~ & ~$L_\ell$~ &
~$J_\ell^{P_\ell}$~ &
~~~~~~~~~Mass~~~~~~~~~ & ~~~~Width~~~~ &~Decay modes~\\
\hline
 $\Lambda_c^+$ & ${1\over 2}^+$ & 0 & 0 & $0^+$ & $2286.46\pm0.14$ & & weak  \\
 \hline
 $\Lambda_c(2595)^+$ & ${1\over 2}^-$ & 0 & 1 & $1^-$ & $2592.25\pm0.28$ &
 $2.59\pm0.56$ & $\Lambda_c\pi\pi,\Sigma_c\pi$ \\
 \hline
 $\Lambda_c(2625)^+$ & ${3\over 2}^-$ & 0 & 1 & $1^-$ & $2628.11\pm0.19$ &
 $<0.97$ & $\Lambda_c\pi\pi,\Sigma_c\pi$ \\
 \hline
 $\Lambda_c(2765)^+$ & $?^?$ & ? & ? & $?$ & $2766.6\pm2.4$ & $50$ & $\Sigma_c\pi,\Lambda_c\pi\pi$ \\
 \hline
 $\Lambda_c(2880)^+$ & ${5\over 2}^+$ & ? & ? & ? & $2881.53\pm0.35$ & $5.8\pm1.1$
 & $\Sigma_c^{(*)}\pi,\Lambda_c\pi\pi,D^0p$ \\
 \hline
 $\Lambda_c(2940)^+$ & $?^?$ & ? & ? & $?$ & $2939.3^{+1.4}_{-1.5}$ & $17^{+8}_{-6}$ &
 $\Sigma_c^{(*)}\pi,\Lambda_c\pi\pi,D^0p$ \\ \hline
 $\Sigma_c(2455)^{++}$ & ${1\over 2}^+$ & 1 & 0 & $1^+$ & $2453.98\pm0.16$ &
 $1.94^{+0.08}_{-0.16}$ & $\Lambda_c\pi$ \\
 \hline
 $\Sigma_c(2455)^{+}$ & ${1\over 2}^+$ & 1 & 0 & $1^+$ & $2452.9\pm0.4$ &
 $<4.6$ & $\Lambda_c\pi$\\
 \hline
 $\Sigma_c(2455)^{0}$ & ${1\over 2}^+$ & 1 & 0 & $1^+$ & $2453.74\pm0.16$
 & $1.87^{+0.09}_{-0.17}$ & $\Lambda_c\pi$ \\
 \hline
 $\Sigma_c(2520)^{++}$ & ${3\over 2}^+$ & 1 & 0 & $1^+$ & $2517.9\pm0.6$
 & $14.8^{+0.3}_{-0.4}$ & $\Lambda_c\pi$\\
 \hline
 $\Sigma_c(2520)^{+}$ & ${3\over 2}^+$ & 1 & 0 & $1^+$ & $2517.5\pm2.3$
 & $<17$ & $\Lambda_c\pi$ \\
 \hline
 $\Sigma_c(2520)^{0}$ & ${3\over 2}^+$ & 1 & 0 & $1^+$ & $2518.8\pm0.6$
 & $15.3^{+0.3}_{-0.4}$ & $\Lambda_c\pi$ \\
 \hline
 $\Sigma_c(2800)^{++}$ & $?^?$ & ? & ? & ? & $2801^{+4}_{-6}$ & $75^{+22}_{-17}$ &
 $\Lambda_c\pi,\Sigma_c^{(*)}\pi,\Lambda_c\pi\pi$ \\
 \hline
 $\Sigma_c(2800)^{+}$ & $?^?$ & ? & ? & ? & $2792^{+14}_{-~5}$ & $62^{+60}_{-40}$ &
 $\Lambda_c\pi,\Sigma_c^{(*)}\pi,\Lambda_c\pi\pi$ \\
 \hline
 $\Sigma_c(2800)^{0}$ & $?^?$ & ? & ? & ? & $2806^{+5}_{-7}$ & $72^{+22}_{-15}$ &
 $\Lambda_c\pi,\Sigma_c^{(*)}\pi,\Lambda_c\pi\pi$\\
 \hline
 $\Xi_c^+$ & ${1\over 2}^+$ & 0 & 0 & $0^+$ & $2467.8^{+0.4}_{-0.6}$ & & weak \\ \hline
 $\Xi_c^0$ & ${1\over 2}^+$ & 0 & 0 & $0^+$ & $2470.88^{+0.34}_{-0.80}$ & & weak \\ \hline
 $\Xi'^+_c$ & ${1\over 2}^+$ & 1 & 0 & $1^+$ & $2575.6\pm3.1$ & & $\Xi_c\gamma$ \\ \hline
 $\Xi'^0_c$ & ${1\over 2}^+$ & 1 & 0 & $1^+$ & $2577.9\pm2.9$ & & $\Xi_c\gamma$ \\ \hline
 $\Xi_c(2645)^+$ & ${3\over 2}^+$ & 1 & 0 & $1^+$ & $2645.9^{+0.5}_{-0.6}$ & $2.6\pm0.5$ & $\Xi_c\pi$ \\
 \hline
 $\Xi_c(2645)^0$ & ${3\over 2}^+$ & 1 & 0 & $1^+$ & $2645.9\pm0.9$ & $<5.5$ & $\Xi_c\pi$ \\
 \hline
 $\Xi_c(2790)^+$ & ${1\over 2}^-$ & 0 & 1 & $1^-$ & $2789.9\pm3.2$ & $<15$ & $\Xi'_c\pi$\\
 \hline
 $\Xi_c(2790)^0$ & ${1\over 2}^-$ & 0 & 1 & $1^-$ & $2791.8\pm3.3$ & $<12$ & $\Xi'_c\pi$ \\
 \hline
 $\Xi_c(2815)^+$ & ${3\over 2}^-$ & 0 & 1 & $1^-$ & $2816.6\pm0.9$ & $<3.5$ & $\Xi^*_c\pi,\Xi_c\pi\pi,\Xi_c'\pi$ \\
 \hline
 $\Xi_c(2815)^0$ & ${3\over 2}^-$ & 0 & 1 & $1^-$ & $2819.6\pm1.2$ & $<6.5$ & $\Xi^*_c\pi,\Xi_c\pi\pi,\Xi_c'\pi$ \\
 \hline
$\Xi_c(2930)^0$ & $?^?$ & ? & ? & $?$ & $2931\pm6$ & $36\pm13$
 & $\Lambda_c \ov K$ \\
 \hline
 $\Xi_c(2980)^+$ & $?^?$ & ? & ? & $?$ & $2971.4\pm3.3$ & $26\pm7$
 & $\Sigma_c \ov K,\Lambda_c \ov K\pi,\Xi_c\pi\pi$  \\
 \hline
 $\Xi_c(2980)^0$ & $?^?$ & ? & ? & $?$ & $2968.0\pm2.6$ & $20\pm7$
 & ~$\Sigma_c \ov K,\Lambda_c \ov K\pi,\Xi_c\pi\pi$~ \\
 \hline
 $\Xi_c(3055)^+$ & $?^?$ & ? & ? & $?$ & $3054.2\pm1.3$ & $17\pm13$ &
 $\Sigma_c \ov K,\Lambda_c \ov K\pi,D\Lambda$  \\
 \hline
 $\Xi_c(3055)^0$ & $?^?$ & ? & ? & $?$ & $3059.7\pm0.8$ & $7.4\pm3.9$ &
 $\Sigma_c \ov K,\Lambda_c \ov K\pi,D\Lambda$  \\
 \hline
 $\Xi_c(3080)^+$ & $?^?$ & ? & ? & $?$ & $3077.0\pm0.4$ & $5.8\pm1.0$ &
 $\Sigma_c \ov K,\Lambda_c \ov K\pi,D\Lambda$  \\
 \hline
 $\Xi_c(3080)^0$ & $?^?$ & ? & ? & $?$ & $3079.9\pm1.4$ & $5.6\pm2.2$
 & $\Sigma_c \ov K,\Lambda_c \ov K\pi,D\Lambda$ \\
 \hline
$\Xi_c(3123)^+$ & $?^?$ & ? & ? & $?$ & $3122.9\pm1.3$ & $4.4\pm3.8$
 & $\Sigma_c^* \ov K,\Lambda_c \ov K\pi$ \\
 \hline
 $\Omega_c^0$ & ${1\over 2}^+$ & 1 & 0 & $1^+$ & $2695.2\pm1.7$ & & weak \\
 \hline
 $\Omega_c(2770)^0$ & ${3\over 2}^+$ & 1 & 0 & $1^+$ & $2765.9\pm2.0$ & & $\Omega_c\gamma$ \\
 \hline \hline
\end{tabular}
\end{center}
\end{table}

\vskip 0.3cm
In the following we discuss some of the excited charmed baryon
states:
\subsubsection{$\Lambda_c$ states}
The lowest-lying $p$-wave $\Lambda_c$ states are  $\tilde\Lambda_{c0}({1\over 2}^-), \Lambda_{c1}({1\over 2}^-,{3\over
2}^-),\tilde\Lambda_{c1}({1\over 2}^-,{3\over
2}^-)$ and $\tilde\Lambda_{c2}({3\over 2}^-,{5\over 2}^-)$.
A doublet $\Lambda_{c1}({1\over 2}^-,{3\over 2}^-)$ is formed by $\Lambda_c(2595)^+$ and $\Lambda_c(2625)^+$ \cite{Cho}.
The allowed strong decays are
$\Lambda_{c1}(1/2^-)\to[\Sigma_c\pi]_S,~[\Sigma_c^*\pi]_D$ and $\Lambda_{c1}(3/2^-)\to [\Sigma_c\pi]_D,~[\Sigma_c^*\pi]_{S,D},~[\Lambda_c\pi\pi]_P$.
This explains why the width of $\Lambda_c(2625)^+$ is narrower than that of
$\Lambda_c(2595)^+$. Because of isospin conservation in strong decays, $\Lambda_{c1}^+$ is not allowed to decay into $\Lambda_c^+\pi^0$.

$\Lambda_c(2765)^+$ is a broad state
first seen in $\Lambda_c^+\pi^+\pi^-$ by CLEO
\cite{CLEO:Lamc2880}. However, whether it is a
$\Lambda_c^+$ or a $\Sigma_c^+$ and whether the width might be due
to overlapping states are still not known. The Skyrme model \cite{Oh}
and the quark model \cite{Capstick} suggest a $J^P=\frac12^+$
$\Lambda_c$ state with a mass 2742 and 2775 MeV, respectively.
Therefore, $\Lambda_c(2765)^+$ could be a first positive-parity
excitation of $\Lambda_c$. It has also been proposed in the diquark model \cite{Ebert:2007} to be either the first radial ($2S$)
excitation of the $\Lambda_c$ with $J^P=\frac12^-$ containing the light scalar diquark or the first orbital
excitation ($1P$) of the $\Sigma_c$ with $J^P=\frac32^-$ containing the light axial vector diquark.

The state $\Lambda_c(2880)^+$ first observed by CLEO
\cite{CLEO:Lamc2880} in $\Lambda_c^+\pi^+\pi^-$ was also seen by
BaBar in the $D^0p$ spectrum \cite{BaBar:Lamc2940}.  Belle has studied the
experimental constraint on the $J^P$ quantum numbers of
$\Lambda_c(2880)^+$ \cite{Belle:Lamc2880} and found that $J^P=\frac52^+$ is favored by the angular analysis of
$\Lambda_c(2880)^+\to\Sigma_c^{0,++}\pi^\pm$ together with the ratio of $\Sigma^*\pi/\Sigma\pi$ measured to be
\begin{eqnarray} \label{eq:R}
 R\equiv {\Gamma(\Lambda_c(2880)\to\Sigma_c^*\pi^\pm)\over
 \Gamma(\Lambda_c(2880)\to\Sigma_c\pi^\pm)}=(24.1\pm6.4^{+1.1}_{-4.5})\%.
\end{eqnarray}
In the quark
model, the candidates for the parity-even spin-${5\over 2}$ state are
$\Lambda_{c2}(\frac52^+)$, $\hat\Lambda_{c2}(\frac52^+)$,
$\tilde\Lambda^1_{c2}(\frac52^+)$,
$\tilde\Lambda^2_{c2}(\frac52^+)$ and
$\tilde\Lambda^2_{c3}(\frac52^+)$ (see Table \ref{tab:pp}). The first four candidates with $J_\ell=2$ decay to $\Sigma_c\pi$ in a
$F$ wave and $\Sigma_c^*\pi$ in $F$ and $P$ waves. Neglecting the
$P$-wave contribution for the moment,
 \begin{eqnarray}
 {\Gamma\left(\Lambda_{c2}(5/2^+)\to
[\Sigma_c^*\pi]_F\right)\over \Gamma\left(\Lambda_{c2}(5/2^+)\to
[\Sigma_c\pi]_F\right)}&=& {4\over
5}\,{p_\pi^7(\Lambda_c(2880)\to\Sigma_c^*\pi)\over
p_\pi^7(\Lambda_c(2880)\to\Sigma_c\pi)} \nonumber \\
&=& {4\over 5}\times 0.29=0.23\,,
 \end{eqnarray}
where the factor of 4/5 follows from heavy quark symmetry.
At first glance, it appears that this is in good agreement with
experiment. However, the $\Sigma_c^*\pi$ channel is available via a
$P$-wave and is enhanced by a factor of $1/p_\pi^4$ relative to the
$F$-wave one. However, heavy quark symmetry cannot be applied
to calculate the contribution of the $[\Sigma_c^*\pi]_F$ channel to
the ratio $R$ as the reduced matrix elements are different for
$P$-wave and $F$-wave modes. In this case, one has to rely on a
phenomenological model to compute the ratio $R$.
As for $\tilde\Lambda^2_{c3}(\frac52^+)$, it decays to
$\Sigma_c^*\pi$, $\Sigma_c\pi$ and $\Lambda_c\pi$ all in $F$ waves.
It turns out that
 \begin{eqnarray}
 {\Gamma\left(\tilde\Lambda^2_{c3}(5/2^+)\to
[\Sigma_c^*\pi]_F\right)\over \Gamma\left(\tilde\Lambda^2_{c3}(5/2^+)\to
[\Sigma_c\pi]_F\right)}&=&{5\over
4}\,{p_\pi^7(\Lambda_c(2880)\to\Sigma_c^*\pi)\over
p_\pi^7(\Lambda_c(2880)\to\Sigma_c\pi)} \nonumber \\
&=& {5\over 4}\times 0.29=0.36\,.
 \end{eqnarray}
Although this deviates from the experimental measurement
(\ref{eq:R}) by $1\sigma$, it is a robust prediction. This has
motivated us to conjecture that that the first
positive-parity excited charmed baryon $\Lambda_c(2880)^+$ could be
an admixture of $\Lambda_{c2}(\frac52^+)$,
$\hat\Lambda_{c2}(\frac52^+)$ and $\tilde\Lambda^2_{c3}(\frac52^+)$
\cite{CC}.

It is worth mentioning that the Peking group
\cite{Zhu} has studied the strong decays of charmed baryons based
on the so-called $^3P_0$ recombination model. For the
$\Lambda_c(2880)$, Peking group found that (i) the possibility of
$\Lambda_c(2880)$ being a radial excitation is ruled out as its
decay into $D^0p$ is prohibited in the $^3P_0$ model if
$\Lambda_c(2880)$ is a first radial excitation of $\Lambda_c$, and
(ii) the states $\Lambda_{c2}(\frac52^+)$,
$\tilde\Lambda^1_{c2}(\frac52^+)$ and $\hat\Lambda_{c2}(\frac52^+)$ are excluded as they do not decay to  $D^0p$ according to the $^3P_0$ model. Moreover, the predicted ratios of $\Sigma_c^*\pi/\Sigma_c\pi$ are either too large or too small compared to experiment, for example,
 \begin{eqnarray}
 {\Gamma\left(\Lambda_{c2}(5/2^+)\to
\Sigma_c^*\pi\right)\over \Gamma\left(\Lambda_{c2}(5/2^+)\to
\Sigma_c\pi\right)} = 89\,, \qquad
 {\Gamma\left(\hat\Lambda_{c2}(5/2^+)\to
\Sigma_c^*\pi\right)\over \Gamma\left(\hat \Lambda_{c2}(5/2^+)\to
\Sigma_c\pi\right)} = 0.75\,.
 \end{eqnarray}
Both symmetric states $\Lambda_{c2}$ and $\hat\Lambda_{c2}$ are
thus ruled out. Hence, it appears that $\tilde\Lambda^2_{c3}(\frac52^+)$ dictates the inner structure of $\Lambda_c(2880)$.
\footnote{It has been argued in \cite{Zhong} that in the chiral quark model $\Lambda_c(2880)$ favors to be the state $|\Lambda_c\,^{2S+1}L_\sigma J^P\rangle=|\Lambda_c\,^2 D_{\lambda\lambda}{3\over 2}^+\rangle$ with $L_\rho=0$ and $L_\lambda=2$ rather than $|\Lambda_c\,^2 D_A{5\over 2}^+\rangle$ with $L_\rho=L_\lambda=1$ as the latter cannot decay into $D^0p$.
However, this is not our case as $\tilde\Lambda^2_{c3}(\frac52^+)$ does decay to $D^0p$ and can reproduce the measured value of $R$.}
However,  there are several issues with this assignment: (i)
the quark model indicates a $\Lambda_{c2}(\frac52^+)$
state around 2910 MeV which is close to the mass of
$\Lambda_c(2880)$, while the mass of $\tilde\Lambda^2_{c3}(\frac52^+)$ is
even higher \cite{Capstick}, (ii) $\tilde\Lambda^2_{c3}(\frac52^+)$ can decay to a $F$-wave $\Lambda_c\pi$
and this has not been seen by BaBar and Belle, and (iii) the calculated width 28.8 MeV is too large compared to the measured one $5.8\pm1.1$ MeV. One may argue that the $^3P_0$ model's prediction can be easily off by a factor of $2\sim 3$ from the experimental measurement due to its inherent uncertainties \cite{Zhu}.

It is interesting to notice that, based on the diquark idea, the
quantum numbers $J^P=\frac52^+$ have been correctly predicted in
\cite{Selem} for the $\Lambda_c(2880)$ before the Belle experiment.

The highest $\Lambda_c(2940)^+$ was first discovered by BaBar in
the $D^0p$ decay mode \cite{BaBar:Lamc2940} and  confirmed by
Belle in the decays $\Sigma_c^0\pi^+,\Sigma_c^{++}\pi^-$ which
subsequently decay into $\Lambda_c^+\pi^+\pi^-$
\cite{Belle:Lamc2880}. Its spin-parity assignment is quite diversified. For example, it has argued that  $\Lambda_c(2940)^+$ is the radial excitation of $\Lambda_c(2595)$ with $J^P={1\over 2}^-$, but the predicted mass is too large by of order 40 MeV  or it could be the first radial excitation of $\Sigma_c$ (not $\Lambda_c$!) with $J^P=3/2^+$ \cite{Ebert:2011}. The latter assignment has the advantage that the predicted mass is in better agreement with experiment.
Since the mass of $\Lambda_c(2940)^+$ is
barely below the threshold of $D^{*0}p$, this observation has
motivated the authors of \cite{He:2006is} to suggest an exotic molecular
state of $D^{*0}$ and $p$ with a binding energy of order 6 MeV and
$J^P={1\over 2}^-$ for $\Lambda_c(2940)^+$. The quark potential
model predicts a $\frac52^-$ $\Lambda_c$ state at 2900 MeV and a
$\frac32^+$ $\Lambda_c$ state at 2910 MeV \cite{Capstick}. A
similar result of 2906 MeV for $\frac32^+$ $\Lambda_c$ is also
obtained in the relativistic quark model \cite{Garcilazo}.


%

\subsubsection{$\Sigma_c$ states}
The highest isotriplet charmed baryons $\Sigma_c(2800)^{++,+,0}$
decaying to $\Lambda_c^+\pi$ were first measured by Belle
\cite{Belle:Sigc2800} with widths of order 70 MeV. The possible quark states are $\Sigma_{c0}({1\over 2}^-)$,  $\Sigma_{c1}({1\over 2}^-,{3\over 2}^-)$, $\tilde\Sigma_{c1}({1\over 2}^-,{3\over 2}^-)$ and $\Sigma_{c2}({3\over 2}^-,{5\over 2}^-)$.
The states $\Sigma_{c1}$ and $\tilde\Sigma_{c1}$ are ruled out
because their decays to $\Lambda_c^+\pi$ are prohibited in the heavy quark limit.  Now the
$\Sigma_{c2}({3\over 2}^-,{5\over 2}^-)$ baryon decays principally into the
$\Lambda_c\pi$ system in a $D$-wave, while $\Sigma_{c0}({1\over 2}^-)$  decays into $\Lambda_c\pi$ in an $S$-wave. Since HHChPT implies a very broad $\Sigma_{c0}$ with width of order 885 MeV (see Sec.III.B below), this  $p$-wave state is also excluded. Therefore, $\Sigma_c(2800)^{++,+,0}$ are likely to
be either $\Sigma_{c2}({3\over 2}^-)$ or $\Sigma_{c2}({5\over 2}^-)$ or their mixing. In the quark-diquark model \cite{Ebert:2011}, both of them have very close masses compatible with experiment.  Given that for light strange baryons, the first orbital excitation of the $\Sigma$ has also the quantum numbers $J^P=3/2^-$ (see Fig. \ref{fig:charmspect}), we will advocate a $\Sigma_{c2}(3/2^-)$ state for $\Sigma_c(2800)$.

\subsubsection{$\Xi_c$ states}

The states $\Xi_c(2790)$ and $\Xi_c(2815)$ form a doublet
$\Xi_{c1}({1\over 2}^-,{3\over 2}^-)$. Since the diquark transition
$1^-\to 0^++\pi$ is prohibited, $\Xi_{c1}({1\over 2}^-,{3\over
2}^-)$ cannot decay to $\Xi_c\pi$. The dominant decay mode is
$[\Xi'_c\pi]_S$ for $\Xi_{c1}({1\over 2}^-)$ and $[\Xi_c^*\pi]_S$
for $\Xi_{c1}({3\over 2}^-)$.

Many excited charmed baryon states  $\Xi_c(2980)$, $\Xi_c(3055)$, $\Xi_c(3080)$ and $\Xi_c(3123)$ have been seen at $B$ factories \cite{Belle:Xic2980,BaBar:Xic2980,Belle:dc}. Another state $\Xi_c(2930)^0$  omitted from the PDG summary table has been only seen by BaBar in the $\Lambda_c^+K^-$ mass projection of $B^-\to\Lambda_c^+\bar\Lambda_c^-K^-$ \cite{BaBar:Xic2930}. However, as we shall see below, it may form a sextet with $\Sigma_c(2800)$ and $\Omega_c(3050)$.
The states $\Xi_c(2980)$, $\Xi_c(3055)$, $\Xi_c(3080)$ and $\Xi_c(3123)$ could be the first positive-parity excitations of the $\Xi_c$.
The study of the Regge phenomenology is very useful for the $J^P$ assignment of charmed baryons \cite{Ebert:2011,Guo2008}. The Regge analysis suggests
$J^P=3/2^+$ for $\Xi_c(3055)$ and $5/2^+$ for $\Xi_c(3080)$ \cite{Ebert:2011}. From Table \ref{tab:3and6} below we shall see that $\Xi_c(3080)$ and $\Lambda_c(2880)$ form nicely a $J^P=5/2^+$ antitriplet.

In the relativistic quark-diquark model \cite{Ebert:2011},  $\Xi_c(2980)$ is a sextet $J^P={1\over 2}^+$ state. According to Table \ref{tab:pp}, possible candidates are $\Xi'_{c1}({1\over 2}^+),\hat\Xi'_{c1}({1\over 2}^+),\tilde\Xi'_{c0}({1\over 2}^+)$ and $\tilde\Xi'_{c1}({1\over 2}^+)$.
As pointed out in \cite{CC:2015}, strong decays of these four
states studied in \cite{Zhu} using the $^3P_0$ model show that $\tilde\Xi'_{c1}({1\over 2}^+)$ does not decay to $\Xi_c\pi$ and $\Lambda_c\ov K$ and has a width of 28 MeV consistent with experiment.
Therefore, the favored candidate for $\Xi_c(2980)$ is
$\tilde\Xi'_{c1}({1\over 2}^+)$ which has $J_\ell=L_\ell=1$.

The possible quark states for $J^P={5\over 2}^+$ $\Xi_c(3080)$ baryon in an antitriplet are $\Xi_{c2}({5\over 2}^+)$, $\hat\Xi_{c2}({5\over 2}^+)$, $\tilde \Xi^1_{c2}({5\over 2}^+)$, $\tilde \Xi^2_{c2}({5\over 2}^+)$ and $\tilde \Xi^2_{c3}({5\over 2}^+)$ (see Table \ref{tab:pp}).  Since
$\Xi_c(3080)$ is above the $D\Lambda$ threshold, the two-body mode $D\Lambda$ should exist though it has not been searched for in the $D\Lambda$ spectrum. Recall that the neutral $\Xi_c(3055)^0$ was observed recently by Belle in the $D^0\Lambda$ spectrum \cite{Kato}. According to the $^3P_0$ model, the first four states are excluded as they do not decay into $D\Lambda$ \cite{Zhu}. The only possibility left is $\tilde \Xi^2_{c3}({5\over 2}^+)$. This is the analog of $\tilde \Lambda^2_{c3}({5\over 2}^+)$ for $\Lambda_c(2880)$.
Nevertheless, the identification of $\tilde \Xi^2_{c3}({5\over 2}^+)$ with $\Xi_c(3080)$ encounters two potential problems:
(i) its width is dominated by $\Xi_c\pi$ and $\Lambda_c^+\ov K$ modes which have not been seen experimentally, and (ii) the predicted width of order 47 MeV \cite{Zhu} is too large compared to the measured one of order 5.7 MeV.


\subsubsection{$\Omega_c$ states}
Only two ground states have been observed thus far: $1/2^+$ $\Omega_c^0$ and $3/2^+$ $\Omega_c(2770)^0$. The latter was seen by BaBar in the electromagnetic decay $\Omega_c(2770)\to\Omega_c\gamma$ \cite{BaBar:Omegacst}.

\vskip 1.0 cm

\noindent\underline{\sl Molecule picture}
\vskip 0.2cm
Since $\Lambda_c(2940)^+$ and $\Sigma_c(2800)$ are barely below the $D^{*0}p$ and $DN$ thresholds, respectively, it is tempting to conjecture an
exotic molecular structure of $D^{*0}$ and $p$ for the former and a molecule state of $DN$ for the latter \cite{He:2006is,Dong:2010,He:2010,Yamaguchi:2013,Dong:2014,Zhang:2014}. Likewise, $\Xi_c(2980)$ could be a molecule state of $D\Lambda$.

The coupled-channel calculation of the baryon-meson $ND$ system has been performed to look for the isospin-spin channel which is attractive enough to form a molecule state \cite{Yamaguchi:2013,Carames}. It turns out that $(I)J^P=(0){1\over 2}^-$ is the most attractive one followed by $(I)J^P=(1){3\over 2}^-$. This suggests the possibility of $\Sigma_c(2800)$ being an $s$-wave $DN$ molecular state with $(I)J^P=(0){1\over 2}^-$ and $\Lambda_c(2940)$ an $s$-wave $D^*N$ molecular state with $(I)J^P=(1){3\over 2}^-$  (see Fig. 3 of \cite{Carames}). Another possibility is a $DN$ molecular state with $(I)J^P=(1){3\over 2}^-$ for $\Sigma_c(2800)$ and a $D^*N$ one with $(I)J^P=(0){1\over 2}^-$
for $\Lambda_c(2940)$.  Since $\Sigma_c(2800)$ has isospin 1 and moreover we have noted in passing that $\Sigma_c(2800)$ will be too broad if it is assigned to $J^P=1/2^-$, we conclude that the second possibility is more preferable (see also \cite{Zhang:2014}).

\vskip 0.5 cm

The possible spin-parity quantum numbers of the higher
excited charmed baryon resonances that have been suggested in the
literature are partially summarized in Table \ref{tab:qn}. Some of the
predictions are already ruled out by experiment. For example,
$\Lambda_c(2880)$ has $J^P={5\over 2}^+$ as seen by Belle.
Certainly, more experimental studies are needed in order to pin
down the quantum numbers.

{\squeezetable
\begin{table}[t]
\caption{Possible spin-parity quantum numbers for excited
charmed baryon resonances.}
\label{tab:qn}
\begin{center}
\begin{tabular}{|l|c|c| c| c| c| c| c| c| c|} \hline \hline
  &  $\Lambda_c(2765)$ & $\Lambda_c(2880)$ & $\Lambda_c(2940)$ & $\Sigma_c(2800)$ & $\Xi_c(2930)$
 & $\Xi_c(2980)$ & $\Xi_c(3055)$ &  $\Xi_c(3080)$ & $\Xi_c(3123)$ \\  \hline
 Capstick et al. \cite{Capstick} & ${1\over 2}^+$ & & ${3\over 2}^+,{5\over 2}^-$ & ${3\over 2}^-,{5\over 2}^-$ & & & & & \\
 B. Chen et al. \cite{Chen:2014} & ${1\over 2}^+(2S)$ & ${5\over 2}^+(1D)$ & ${1\over 2}^-(2P)$ &  & & ${1\over 2}^+(2S)$ & ${3\over 2}^+(1D)$ & ${5\over 2}^+(1D)$ & ${1\over 2}^-(2P)$ \\
 H. Chen et al. \cite{Chen:2015k} & & & ${1\over 2}^+,{1\over 2}^-$ & ${1\over 2}^-,{3\over 2}^-$ & & ${1\over 2}^-,{3\over 2}^-$ & & ${5\over 2}^-$ & \\
 Cheng et al. \cite{CC} & & & & ${3\over 2}^-$ & &  ${1\over 2}^+$ & & ${5\over 2}^+$ & \\
 Ebert et al. \cite{Ebert:2011} & ${1\over 2}^+(2S)$ & ${5\over 2}^+(1D)$ & ${1\over 2}^-(2P),{3\over 2}^+(2S)$ &  ${1\over 2}^-,{3\over 2}^-(1P)$ & ${1\over 2}^-,{3\over 2}^-,{5\over 2}^-$ & ${1\over 2}^+(2S)$ & ${3\over 2}^+(1D)$ & ${5\over 2}^+(1D)$ & ${7\over 2}^+(1D)$ \\
 Garcilazo et al. \cite{Garcilazo} & ${1\over 2}^+$ & ~${1\over 2}^-,~{3\over 2}^-$ & ${3\over 2}^+$ & ${1\over 2}^-,{3\over 2}^-$& & & & & \\
 Gerasyuata et al. \cite{Ger} & ${5\over 2}^-$ & ${1\over 2}^-$ & & ${5\over 2}^-$ & &  & & & \\
 Liu et al. \cite{Liu:2012} & & & & & ${1\over 2}^-(1P)$ & ${1\over 2}^-,{3\over 2}^-(1P)$ & ${3\over 2}^+(1D)$ & ${1\over 2}^+(2S)$ & ${3\over 2}^+,{5\over 2}^+(1D)$ \\
 Wilczek et al. \cite{Selem} & & ${5\over 2}^+$ & & & & & & & \\
 Zhong et al. \cite{Zhong} & ${1\over 2}^-(1P)$ & ${3\over 2}^+(1D)$ & ${5\over 2}^+(1D)$ & ${1\over 2}^-,{5\over 2}^-(1P)$ & & & & & \\
 \hline \hline
\end{tabular}
\end{center}
\end{table}
}

\begin{figure}[t]
\centerline{\psfig{file=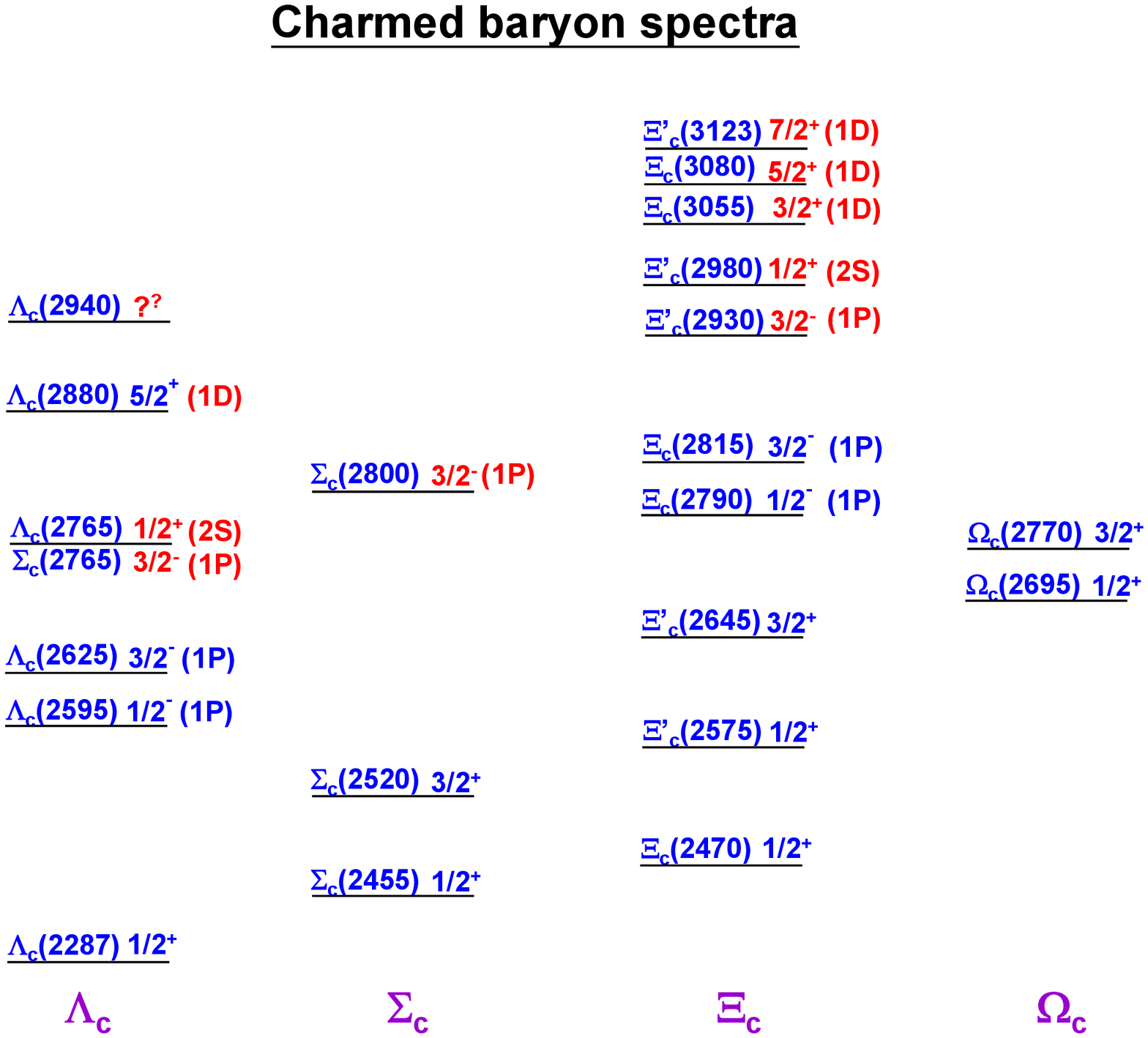,width=5.6in}}
\caption{Singly charmed baryon states where the spin-parity quantum numbers in red are taken from \cite{Ebert:2011}.} \label{fig:spectra}
\end{figure}

Charmed baryon spectroscopy has been studied extensively in
various models.
It appears that the spectroscopy is well described by  the heavy quark-light diquark picture elaborated by Ebert, Faustov and Galkin (EFG) \cite{Ebert:2011} (see also \cite{Chen:2014}). As noted in passing, the
quantum numbers $J^P=\frac52^+$ of $\Lambda_c(2880)$ have been correctly predicted in the model based on the diquark idea before the Belle experiment \cite{Selem}.
Moreover, EFG have shown that all available experimental data on heavy baryons fit nicely to the linear Regge trajectories, namely, the trajectories in the $(J,M^2)$ and $(n_r,M^2)$ planes for orbitally and radially excited heavy baryons, respectively:
 \begin{eqnarray}
J=\alpha M^2+\alpha_0, \qquad n_r=\beta M^2+\beta_0,
\end{eqnarray}
where $n_r$ is the radial excitation quantum number, $\alpha$, $\beta$ are the slopes and $\alpha_0$, $\beta_0$ are intercepts. The linearity, parallelism and equidistance of the Regge trajectories were verified. The predictions of the spin-parity quantum numbers of charmed baryons and their masses in \cite{Ebert:2011} can be regarded as a theoretical benchmark (see Fig. \ref{fig:spectra}).

\begin{table}[t]
\caption{Antitriplet and sextet states of charmed baryons. The $J^P$ quantum numbers of $\Xi_c(3080),\Xi'_c(2930),\Sigma_c(2800)$ are not yet established and the $\Omega_c(3/2^-)$ state has not been observed.
Mass differences $\Delta m_{\Xi_c\Lambda_c}\equiv m_{\Xi_c}-m_{\Lambda_c}$, $\Delta m_{\Xi'_c\Sigma_c}\equiv m_{\Xi'_c}-m_{\Sigma_c}$, $\Delta m_{\Omega_c\Xi'_c}\equiv m_{\Omega_c}-m_{\Xi'_c}$ are in units of MeV. } \label{tab:3and6}
\begin{center}
\begin{tabular}{|c|ccc c|} \hline\hline
  & $J^P$ & States & Mass difference & status \\
 \hline
 ~~${\bf \bar 3}$~~ & ~~$1/2^+$~~ &  $\Lambda_c(2287)^+$, $\Xi_c(2470)^+,\Xi_c(2470)^0$ & ~~$\Delta m_{\Xi_c\Lambda_c}=183$ ~~ & ~estab~ \\
 & ~~$1/2^-$~~ &  $\Lambda_c(2595)^+$, $\Xi_c(2790)^+,\Xi_c(2790)^0$ & $\Delta m_{\Xi_c\Lambda_c}=198$ & ~estab~ \\
 & ~~$3/2^-$~~ &  $\Lambda_c(2625)^+$, $\Xi_c(2815)^+,\Xi_c(2815)^0$ & $\Delta m_{\Xi_c\Lambda_c}=190$ & ~estab~ \\
 & ~~$5/2^+$~~ &  $\Lambda_c(2880)^+$, $\Xi_c(3080)^+,\Xi_c(3080)^0$ & $\Delta m_{\Xi_c\Lambda_c}=196$ & \cite{Ebert:2011} \\
 \hline
 ~~${\bf 6}$~~ & ~~$1/2^+$~~ &  $\Omega_c(2695)^0$, $\Xi'_c(2575)^{+,0},\Sigma_c(2455)^{++,+,0}$ & ~~~~$\Delta m_{\Xi'_c\Sigma_c}=124$, $\Delta m_{\Omega_c\Xi'_c}=119$~~ & ~estab~  \\
 & ~~$3/2^+$~~ &  $\Omega_c(2770)^0$, $\Xi'_c(2645)^{+,0},\Sigma_c(2520)^{++,+,0}$ & ~~~~$\Delta m_{\Xi'_c\Sigma_c}=128$, $\Delta m_{\Omega_c\Xi'_c}=120$~~ & ~estab~ \\
 & ~~$3/2^-$~~ &  $\Omega_c(3050)^0$, $\Xi'_c(2930)^{+,0},\Sigma_c(2800)^{++,+,0}$ & ~~~~$\Delta m_{\Xi'_c\Sigma_c}=131$, $\Delta m_{\Omega_c\Xi'_c}=119$~~ & \cite{Ebert:2011} \\
 \hline\hline
\end{tabular}
\end{center}
\end{table}

\vskip 0.4 cm
\noindent\underline{Antitriplet and  sextet states}

The antitriplet and sextet states of charmed baryons are listed in Table \ref{tab:3and6}. By now, the $J^P={1\over 2}^+$, ${1\over
2}^-$ and ${3\over 2}^-$ ${\bf \bar 3}$ states: ($\Lambda_c^+$, $\Xi_c^+,\Xi_c^0)$,
($\Lambda_c(2595)^+$, $\Xi_c(2790)^+,\Xi_c(2790)^0)$, ($\Lambda_c(2625)^+$, $\Xi_c(2815)^+,\Xi_c(2815)^0)$ respectively  and
$J^P={1\over 2}^+$ and ${3\over 2}^+$ ${\bf 6}$ states:
($\Omega_c,\Sigma_c,\Xi'_c$), ($\Omega_c^*,\Sigma_c^*,\Xi'^*_c$) respectively
are established.  It is clear that the mass difference $m_{\Xi_c}-m_{\Lambda_c}$ in the antitriplet states lies in the range of 180-200 MeV. We note in passing that $\Xi_c(3080)$ should carry the quantum numbers $J^P=5/2^+$. From Table \ref{tab:3and6} we see that $\Xi_c(3080)$ and $\Lambda_c(2880)$ form nicely a $J^P=5/2^+$ antitriplet as the mass difference between $\Xi_c(3080)$ and $\Lambda_c(2880)$ is consistent with that observed in other antitriplets. Likewise, the mass differences in $J^P=3/2^-$ sextet ($\Omega_c(3050)$, $\Xi'_c(2930),\Sigma_c(2800))$ predicted by the quark-diquark model are consistent with that measured in $J^P=1/2^+$ and $3/2^+$ sextets. Note that there is no $J^P={1\over 2}^-$ sextet as the $\Sigma_c(2800)$ with these spin-parity quantum numbers will be too broad to be observed.

On the basis of QCD sum rules, many charmed baryon multiplets classified according to $[{\bf 6}_F({\rm or~} {\bf \bar 3}_F),J_\ell,S_\ell, \rho/\lambda)]$ were recently studied in \cite{Chen:2015k}. Three sextets were proposed in this work: $(\Omega_c(3250),\Xi'_c(2980),\Sigma_c(2800))$ for $J^P=1/2^-,3/2^-$ and $(\Omega_c(3320),\Xi'_c(3080),\Sigma_c(2890))$ for $J^P=5/2^-$. Notice that
$\Xi'_c(2980)$ and $\Xi'_c(3080)$ were treated as $p$-wave baryons rather than the first positive-parity excitations as we have discussed before.
The results on the multiplet $[{\bf 6}_F,1,0,\rho]$ led the authors of \cite{Chen:2015k} to suggest that there are two $\Sigma_c(2800)$, $\Xi'_c(2980)$ and $\Omega_c(3250)$ states with $J^P=1/2^-$ and $J^P=3/2^-$. The mass splittings are $14\pm7$, $12\pm7$ and $10\pm6$ MeV, respectively. The predicted mass of $\Omega_c(1/2^-,3/2^-)$ is around $3250\pm200$ MeV. Using the central value of the predicted masses to label the states in the multiplet $[{\bf 6}_F,1,0,\rho]$ (see Table I of \cite{Chen:2015k}), one will have
$(\Omega_c(3250),\Xi'_c(2960),\Sigma_c(2730))$ for $J^P=1/2^-$ and $(\Omega_c(3260),\Xi'_c(2980),\Sigma_c(2750))$ for $J^P=3/2^-$. One can check that $\Delta m_{\Xi'_c\Sigma_c}=230\pm234$ MeV, and $\Delta m_{\Omega_c\Xi'_c}$ is of order $285\pm250$ MeV.  Due to the large theoretical uncertainties in masses, it is not clear if the QCD sum-rule calculations are compatible with the mass differences measured in $J^P=1/2^+$ and $3/2^+$ sextets. At any rate, it will be interesting to test these two different model predictions for $J^P=3/2^-$ and $1/2^-$ sextets in the future.

\begin{figure}[t]
\centerline{\psfig{file=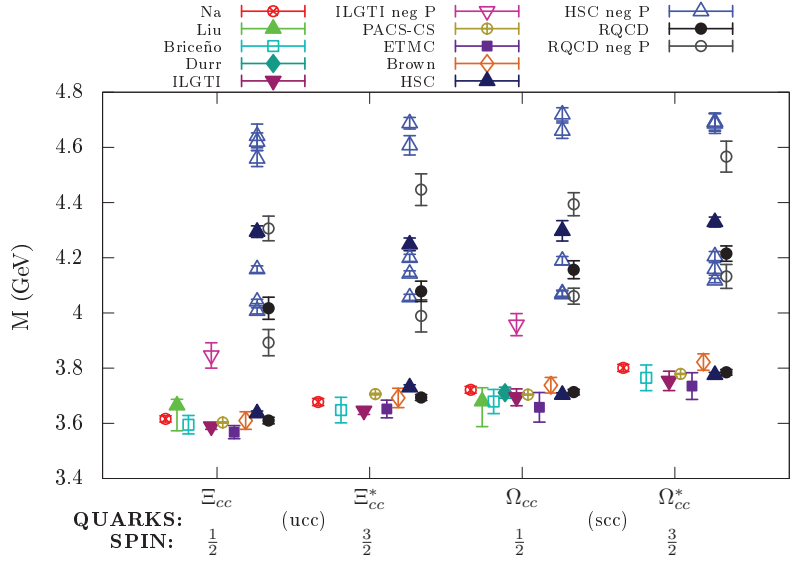,width=4.3in}}
\caption{Doubly charmed low-lying baryon spectra taken from \cite{Bali:2015}. } \label{fig:doubly}
\end{figure}

\subsection{Doubly charmed baryons}
Evidence of doubly charmed baryon states has been reported by SELEX
in $\Xi_{cc}(3520)^+\to\Lambda_c^+K^-\pi^+$ \cite{Selex02}.
Further observation of $\Xi_{cc}^+\to pD^+K^-$ was also announced by SELEX
\cite{Selex04}. However, none of the doubly charm states
discovered by SELEX has been confirmed by FOCUS \cite{FOCUS:dc}, BaBar
\cite{BaBar:dc}, Belle \cite{Belle:dc} and LHCb \cite{LHCb:dc} in spite of the $10^6$
$\Lambda_c$ events produced in $B$ factories, for example, versus 1630
$\Lambda_c$ events observed at SELEX.

The doubly charmed baryons $\Xi_{cc}^{(*)++},\Xi_{cc}^{(*)+},\Omega_{cc}^{(*)+}$ with the quark contents $ccu,ccd,ccs$ form an SU(3) triplet. They
have been studied extensively in many different approaches: quark model, light quark-heavy diquark model, QCD sum rules and lattice simulation. Tabulation of the predicted doubly charmed baryon masses calculated in various models can be found in \cite{Guo,Karliner:2014}. For recent QCD sum rule calculations, see e.g. \cite{Wang:dc}. Chiral corrections to the masses of doubly heavy baryons up to N$^3$LO were presented in \cite{Sun}.

There have been a number of recent lattice studies of doubly and triply charmed baryon spectra displayed in Fig. \ref{fig:doubly} by different groups: RQCD \cite{Bali:2015}, HSC \cite{HSC}, Brown et al. \cite{Brown},  ETMC \cite{ETMC},  ILGTI \cite{ILGTI}, PACS-CS \cite{PACS-CS}, Durr et al. \cite{Durr}, Briceno et al. \cite{Briceno},    Liu et al. \cite{Liu}  and Na et al. \cite{Na}. A new lattice calculation of $\Omega_{cc}^{(*)}$ and $\Omega_{ccc}$ was available in \cite{Can}.
Various lattice results are consist  with each other and they fall into the ranges
\begin{eqnarray} \label{eq:dcbmass}
M(\Xi_{cc})=3.54\sim 3.68\, {\rm GeV}, &\qquad&  M(\Xi^*_{cc})=3.61\sim 3.72\,{\rm GeV}, \nonumber \\
M(\Omega_{cc})=3.57\sim 3.76\, {\rm GeV}, &\qquad&  M(\Omega^*_{cc})=3.68\sim 3.85\,{\rm GeV},
\end{eqnarray}
and
\begin{eqnarray}
M(\Omega_{ccc})=4.70\sim 4.84\, {\rm MeV}.
\end{eqnarray}

Although lattice study suggests that the mass of the low-lying $\Xi_{cc}$ is larger than 3519 MeV, it is interesting to notice that  the authors of \cite{Guo} have calculated the masses of doubly and triply charmed baryons based on the Regge phenomenology and found $M(\Xi_{cc}^+)=3520.2^{+40.6}_{-39.8}$ MeV, in good agreement with SELEX.

\section{Strong decays}

Due to the rich mass spectrum and the relatively narrow widths of
the excited states, the charmed baryon system offers an excellent
ground for testing the ideas and predictions of heavy quark
symmetry and light flavor SU(3) symmetry. The pseudoscalar mesons
involved in the strong decays of charmed baryons such as
$\Sigma_c\to\Lambda_c\pi$ are soft. Therefore, heavy quark
symmetry of the heavy quark and chiral symmetry of the light
quarks will have interesting implications for the low-energy
dynamics of heavy baryons interacting with the Goldstone bosons.

The strong decays of charmed baryons are most conveniently
described by the heavy hadron chiral perturbation theory (HHChPT) in which heavy quark symmetry and chiral symmetry are incorporated
\cite{Yan,Wise}. Heavy baryon chiral Lagrangians  were first constructed in \cite{Yan} for strong decays of $s$-wave charmed baryons and in \cite{Cho,Pirjol} for $p$-wave ones. Previous phenomenological studies of the strong decays of $p$-wave charmed baryons based on HHChPT can be found in \cite{Cho,Pirjol,Chiladze,Falk03,CC}.
The chiral Lagrangian involves two coupling constants
$g_1$ and $g_2$ for $P$-wave transitions between $s$-wave and
$s$-wave baryons \cite{Yan}, six couplings $h_{2}-h_7$ for the
$S$-wave transitions between $s$-wave and $p$-wave baryons, and
eight couplings $h_{8}-h_{15}$ for the $D$-wave transitions
between $s$-wave and $p$-wave baryons \cite{Pirjol}. The general
chiral Lagrangian for heavy baryons coupling to the pseudoscalar
mesons can be expressed compactly in terms of superfields. We will
not write down the relevant Lagrangians here; instead the reader
is referred to Eqs. (3.1) and (3.3) of \cite{Pirjol}.
The partial widths relevant for our purposes are \cite{Pirjol}:
 \begin{eqnarray} \label{eq:swavecoupling}
 \Gamma(\Sigma_c^*\to \Sigma_c\pi)={g_1^2\over 2\pi
 f_\pi^2}\,{m_{\Sigma_c}\over m_{\Sigma_c^*}}p_\pi^3, &&
 \Gamma(\Sigma_c\to \Lambda_c\pi)={g_2^2\over 2\pi
 f_\pi^2}\,{m_{\Lambda_c}\over m_{\Sigma_c}}p_\pi^3, \nonumber \\
 \Gamma(\Lambda_{c1}(1/2^-)\to \Sigma_c\pi)={h_2^2\over 2\pi
 f_\pi^2}\,{m_{\Sigma_c}\over m_{\Lambda_{c1}}}E_\pi^2p_\pi, &&
 \Gamma(\Sigma_{c0}(1/2^-)\to \Lambda_c\pi)={h_3^2\over 2\pi
 f_\pi^2}\,{m_{\Lambda_c}\over m_{\Sigma_{c0}}}E_\pi^2p_\pi,
 \nonumber \\
  \Gamma(\Lambda_{c1}(3/2^-)\to\Sigma_c\pi)={2h_8^2\over 9\pi
 f_\pi^2}\,{m_{\Sigma_c}\over m_{\Lambda_{c1}(3/2)}}\,p_\pi^5, &&
  \Gamma\left(\Sigma_{c1}({3/2}^-)\to\Sigma_c^{(*)}\pi\right) =
 {h_{9}^2\over 9\pi f_\pi^2}\,{m_{\Sigma_c^{(*)}}\over
 m_{\Sigma_{c1}(3/2)}}p_\pi^5,
 \nonumber \\
 \Gamma\left(\Sigma_{c2}({3/ 2}^-)\to\Lambda_c\pi\right)
 = {4h_{10}^2\over 15\pi f_\pi^2}\,{m_{\Lambda_c}\over
 m_{\Sigma_{c2}}}p_\pi^5, &&
 \Gamma\left(\Sigma_{c2}({3/2}^-)\to\Sigma_c^{(*)}\pi\right) =
 {h_{11}^2\over 10\pi f_\pi^2}\,{m_{\Sigma_c^{(*)}}\over
 m_{\Sigma_{c2}}}p_\pi^5,
  \\
 \Gamma\left(\Sigma_{c2}({5/ 2}^-)\to\Sigma_c\pi\right)
 = {2h_{11}^2\over 45\pi f_\pi^2}\,{m_{\Sigma_c}\over
 m_{\Sigma_{c2}}}p_\pi^5, &&
 \Gamma\left(\Sigma_{c2}({5/2}^-)\to\Sigma_c^{*}\pi\right) =
 {7h_{11}^2\over 45\pi f_\pi^2}\,{m_{\Sigma_c^{*}}\over
 m_{\Sigma_{c2}}}p_\pi^5,
 \nonumber \\ \nonumber
 \end{eqnarray}
where $f_\pi=132$ MeV. The dependence on the pion momentum is proportional to $p_\pi$, $p_\pi^3$ and $p_\pi^5$ for $S$-wave, $P$-wave and $D$-wave transitions, respectively. It is obvious that the couplings $g_1,g_2,h_2,\cdots,h_7$ are dimensionless, while $h_8,\cdots,h_{15}$ have canonical dimension $E^{-1}$.

\subsection{Strong decays of $s$-wave charmed baryons}
Since the strong decay $\Sigma_c^*\to\Sigma_c\pi$ is
kinematically prohibited,
the coupling $g_1$ cannot be extracted directly from the strong
decays of heavy baryons.
In the framework of HHChPT, one can use some measurements as input to fix the
coupling $g_2$ which, in turn, can be used to predict the rates of
other strong decays. Among the strong decays $\Sigma_c^{(*)}\to\Lambda_c\pi$, $\Sigma_c^{++}\to\Lambda_c^+\pi^+$ is the most well measured. Hence, we shall use this mode to extract the coupling $g_2$. Based on the 2006 data \cite{PDG2006} of $\Gamma(\Sigma_c^{++})=\Gamma(\Sigma_c^{++}\to\Lambda_c^+\pi^+)=
2.23\pm0.30\,{\rm MeV}$, the coupling $g_2$ is extracted to be
\begin{eqnarray} \label{eq:g2}
 |g_2|_{_{2006}}=0.605^{+0.039}_{-0.043}\,.
\end{eqnarray}
The predicted rates of other modes are shown in Table \ref{tab:strongdecayS2006}.
It is clear that the agreement between theory and experiment is excellent except the predicted width for $\Sigma_c^{*++}\to\Lambda_c^+\pi^+$ is a bit too large.

Using the new data from 2014 Particle Data Group \cite{PDG} in conjunction with the new measurements of $\Sigma_c$ and $\Sigma_c^*$ widths by Belle \cite{Belle:2014}, we obtain the new average $\Gamma(\Sigma_c^{++}\to\Lambda_c^+\pi^+)=
1.94^{+0.08}_{-0.16}\,{\rm MeV}$ (see Table \ref{tab:spectrum}). Therefore, the coupling $g_2$ is reduced to
\begin{eqnarray} \label{eq:newg2}
 |g_2|_{_{2015}}=0.565^{+0.011}_{-0.024}\,.
\end{eqnarray}
From Table \ref{tab:strongdecayS} we see that the agreement between theory and experiment is further improved: The predicted $\Xi_c(2645)^+$ width is consistent with the first new measurement by Belle \cite{Belle:dc} and the new calculated width for $\Sigma_c^{*++}\to\Lambda_c^+\pi^+$ is now in agreement with experiment.
It is also clear that the $\Sigma_c$ width is
smaller than that of $\Sigma_c^*$ by a factor of $\sim 7$,
although they will become the same in the limit of heavy quark
symmetry.

\begin{table}[t]
\caption{Decay widths (in units of MeV) of $s$-wave charmed
baryons where the measured rates are taken from 2006 PDG \cite{PDG2006}. } \label{tab:strongdecayS2006}
\begin{center}
\begin{tabular}{|c|c c|} \hline \hline
~~~~~~~Decay~~~~~~~ & Expt. & ~~HHChPT~~  \\
\hline
 $\Sigma_c^{++}\to\Lambda_c^+\pi^+$ & ~~$2.23\pm0.30$~~ & input   \\ \hline
 $\Sigma_c^{+}\to\Lambda_c^+\pi^0$ & $<4.6$ & $2.6\pm0.4$  \\ \hline
 $\Sigma_c^{0}\to\Lambda_c^+\pi^-$ & $2.2\pm0.4$ & $2.2\pm0.3$  \\ \hline
 $\Sigma_c(2520)^{++}\to\Lambda_c^+\pi^+$ & $14.9\pm1.9$ & $16.7\pm2.3$  \\  \hline
 $\Sigma_c(2520)^{+}\to\Lambda_c^+\pi^0$ & $<17$ & $17.4\pm2.3$   \\  \hline
 $\Sigma_c(2520)^{0}\to\Lambda_c^+\pi^-$ & $16.1\pm2.1$ & $16.6\pm2.2$    \\  \hline
 $\Xi_c(2645)^+\to\Xi_c^{0,+}\pi^{+,0}$ & $<3.1$ & $2.8\pm0.4$  \\  \hline
 $\Xi_c(2645)^0\to\Xi_c^{+,0}\pi^{-,0}$ & $<5.5$ & $2.9\pm0.4$  \\ \hline \hline
\end{tabular}
\end{center}
\end{table}

\begin{table}[t]
\caption{Decay widths (in units of MeV) of $s$-wave charmed
baryons. Data are taken from 2014 PDG \cite{PDG} together with the new measurements of $\Sigma_c$, $\Sigma_c^*$ \cite{Belle:2014}
and $\Xi_c(2645)^+$ widths \cite{Belle:dc}.
Theoretical predictions of \cite{Tawfiq} are taken from
Table IV of \cite{Ivanov}.} \label{tab:strongdecayS}
\begin{center}
\begin{tabular}{|c|c c c c c c|} \hline \hline
~~~~~~~Decay~~~~~~~ & Expt. & ~HHChPT~ & Tawfiq & Ivanov & Huang &
Albertus
 \\
& ~~\cite{PDG}~~ &  & et al. \cite{Tawfiq} &
 et al. \cite{Ivanov} &  et al. \cite{Huang95} & ~~et al. \cite{Albertus}~~  \\
\hline
 $\Sigma_c^{++}\to\Lambda_c^+\pi^+$ & $1.94^{+0.08}_{-0.16}$ & input & $1.51\pm0.17$ & $2.85\pm0.19$
 &  2.5  & $2.41\pm0.07$  \\ \hline
 $\Sigma_c^{+}\to\Lambda_c^+\pi^0$ & $<4.6$ & $2.3^{+0.1}_{-0.2}$ & $1.56\pm0.17$ & $3.63\pm0.27$ &
 3.2 & $2.79\pm0.08$ \\ \hline
 $\Sigma_c^{0}\to\Lambda_c^+\pi^-$ & $1.9^{+0.1}_{-0.2}$ & $1.9^{+0.1}_{-0.2}$  & $1.44\pm0.16$ & $2.65\pm0.19$ &
 2.4 & $2.37\pm0.07$ \\ \hline
 $\Sigma_c(2520)^{++}\to\Lambda_c^+\pi^+$ & $14.8^{+0.3}_{-0.4}$ & $14.5^{+0.5}_{-0.8}$  & $11.77\pm1.27$ & $21.99\pm0.87$ &
 8.2 & $17.52\pm0.75$   \\  \hline
 $\Sigma_c(2520)^{+}\to\Lambda_c^+\pi^0$ & $<17$ & $15.2^{+0.6}_{-1.3}$ & $$ & $$
 &8.6 & $17.31\pm0.74$  \\  \hline
 $\Sigma_c(2520)^{0}\to\Lambda_c^+\pi^-$ & $15.3^{+0.4}_{-0.5}$ & $14.7^{+0.6}_{-1.2}$ & $11.37\pm1.22$ & $21.21\pm0.81$ &
 8.2 & $16.90\pm0.72$   \\  \hline
 $\Xi_c(2645)^+\to\Xi_c^{0,+}\pi^{+,0}$ & $2.6\pm0.5$ & $2.4^{+0.1}_{-0.2}$ & $1.76\pm0.14$ & $3.04\pm0.37$ &
 & $3.18\pm0.10$  \\  \hline
 $\Xi_c(2645)^0\to\Xi_c^{+,0}\pi^{-,0}$ & $<5.5$ & $2.5^{+0.1}_{-0.2}$ & $1.83\pm0.06$ & $3.12\pm0.33$ &
 & $3.03\pm0.10$  \\ \hline \hline
\end{tabular}
\end{center}
\end{table}

\subsection{Strong decays of $p$-wave charmed baryons}

Since $\Lambda_c(2595)^+$ and $\Lambda_c(2625)^+$ form a doublet
$\Lambda_{c1}({1\over 2}^-,{3\over 2}^-)$, it appears from Eq. (\ref{eq:swavecoupling}) that the couplings $h_2$ and $h_8$ in principle can be extracted from $\Lambda_c(2595)\to
\Sigma_c\pi$ and from $\Lambda_c(2625)\to
\Sigma_c\pi$, respectively. Likewise, the information on the couplings $h_{10}$ and $h_{11}$ can be inferred from the strong decays of $\Sigma_c(2800)$ identified with $\Sigma_{c2}(3/2^-)$.
Couplings other than $h_2$, $h_8$ and $h_{10}$ can be related to each
other via the quark model \cite{Pirjol}.

Although the coupling $h_2$ can be inferred from the two-body decay $\Lambda_c(2595)\to\Sigma_c\pi$, this method is less accurate because this decay is kinematically barely allowed or even prohibited depending on the mass of $\Lambda_c(2595)^+$. For the old mass measurement $m(\Lambda_c(2595))=2595.4\pm0.6$ MeV \cite{PDG2006}, $\Lambda_c(2595)^+\to\Sigma_c^{++}\pi^-,\Sigma_c^0\pi^+$ and $\Lambda_c(2595)^+\to\Sigma^+\pi^0$ are kinematically barely allowed. But for the new measurement $m(\Lambda_c(2595))=2592.25\pm0.28$ MeV by CDF \cite{CDF:2595}, only the last mode is allowed. Moreover, the finite width effect of the intermediate resonant states could become important \cite{Chiladze}.

\begin{table}[t]
\caption{Decay widths (in units of MeV) of $p$-wave charmed
baryons where the measured rates are taken from 2006 PDG \cite{PDG2006}. } \label{tab:strongdecayP2006}
\begin{center}
\begin{tabular}{|c|c c|} \hline \hline
~~~~~~~Decay~~~~~~~ & Expt. & ~~HHChPT~~  \\
& ~~\cite{PDG2006}~~& ~\cite{CC}~   \\
\hline
 $\Lambda_c(2595)^+\to (\Lambda_c^{+}\pi\pi)_R$ & ~~$2.63^{+1.56}_{-1.09}$~~ & input  \\ \hline
 $\Lambda_c(2595)^+\to \Sigma_c^{++}\pi^-$ & $0.65^{+0.41}_{-0.31}$ & $0.72^{+0.43}_{-0.30}$  \\ \hline
 $\Lambda_c(2593)^+\to \Sigma_c^{0}\pi^+$ & $0.67^{+0.41}_{-0.31}$ & $0.77^{+0.46}_{-0.32}$ \\ \hline
 $\Lambda_c(2593)^+\to \Sigma_c^{+}\pi^0$ & & $1.57^{+0.93}_{-0.65}$
  \\ \hline
 $\Lambda_c(2625)^+\to \Sigma_c^{++}\pi^-$ & $<0.10$ & $ 0.029$  \\ \hline
 $\Lambda_c(2625)^+\to \Sigma_c^{0}\pi^+$ & $<0.09$ & $0.029$  \\ \hline
 $\Lambda_c(2625)^+\to \Sigma_c^{+}\pi^0$ & & $0.041$  \\ \hline
 $\Lambda_c(2625)^+\to \Lambda_c^+\pi\pi$ & $<1.9$ & $0.21$  \\ \hline
 $\Sigma_c(2800)^{++}\to\Lambda_c\pi,\Sigma_c^{(*)}\pi$ & $75^{+22}_{-17}$ & input  \\ \hline
 $\Sigma_c(2800)^{+}\to\Lambda_c\pi,\Sigma_c^{(*)}\pi$ & $62^{+60}_{-40}$ & input   \\ \hline
 $\Sigma_c(2800)^0\to\Lambda_c\pi,\Sigma_c^{(*)}\pi$ & $61^{+28}_{-18}$ & input \\ \hline
 $\Xi_c(2790)^+\to\Xi'^{0,+}_c\pi^{+,0}$ & $<15$ & $8.0^{+4.7}_{-3.3}$  \\ \hline
 $\Xi_c(2790)^0\to\Xi'^{+,0}_c\pi^{-,0}$  & $<12$ & $8.5^{+5.0}_{-3.5}$ \\ \hline
 $\Xi_c(2815)^+\to\Xi^{*+,0}_c\pi^{0,+}$ & $<3.5$ & $3.4^{+2.0}_{-1.4}$  \\ \hline
 $\Xi_c(2815)^0\to\Xi^{*+,0}_c\pi^{-,0}$ & $<6.5$ & $3.6^{+2.1}_{-1.5}$  \\ \hline \hline
\end{tabular}
\end{center}
\end{table}

We next turn to the three-body decays $\Lambda_c^+\pi\pi$ of  $\Lambda_c(2595)^+$ and $\Lambda_c(2625)^+$ to extract $h_2$ and $h_8$.
Aa shown in \cite{CC},  the 2006 data  for $\Gamma(\Lambda_c(2595))=3.6^{+2.0}_{-1.3}$ MeV \cite{PDG2006} and for the $\Lambda_c(2595)$ mass lead to the resonant rate \cite{CC}
\begin{eqnarray}
\Gamma(\Lambda_c(2593)^+\to\Lambda_c^+\pi\pi)_R &=& (2.63^{+1.56}_{-1.09})\,{\rm
 MeV},
\end{eqnarray}
as shown in Table \ref{tab:strongdecayP}.
Assuming the pole contributions to $\Lambda_c(2595)^+\to \Lambda_c^+\pi\pi$ due to the intermediate states $\Sigma_c$ and $\Sigma_c^*$, the resonant rate
for the process
$\Lambda_{c_1}^+(2595)\to \Lambda_c^+\pi^+\pi^-$ can be calculated
in the framework of heavy hadron chiral perturbation theory
\cite{Pirjol}.
Numerically, we found
\begin{eqnarray} \label{eq:h2h8}
 \Gamma(\Lambda_c(2595)^+\to\Lambda_c^+\pi\pi)_R&=& 13.82h_2^2
 +26.28h_8^2-2.97h_2h_8, \nonumber \\
 \Gamma(\Lambda_c(2625)^+\to\Lambda_c^+\pi\pi)_R&=& 0.617h_2^2+0.136\times
 10^6h_8^2-27h_2h_8,
\end{eqnarray}
where $\Lambda_c^+\pi\pi=\Lambda_c^+\pi^+\pi^- +\Lambda_c^+\pi^0\pi^0$.
It is clear that the limit on $\Gamma(\Lambda_c(2625))$ gives an
upper bound on $h_8$ of order $10^{-3}$ (in units of MeV$^{-1}$),
whereas the decay width of $\Lambda_c(2595)$ is entirely governed
by the coupling $h_2$. Specifically, we have \cite{CC}
\begin{eqnarray} \label{eq:h2fw}
 |h_2|_{_{\rm 2006}}=0.437^{+0.114}_{-0.102}\,, \qquad\quad |h_8|_{_{\rm 2006}}< 3.65\times
 10^{-3}\,{\rm MeV}^{-1}\,.
\end{eqnarray}

\begin{figure}[t]
\centerline{\psfig{file=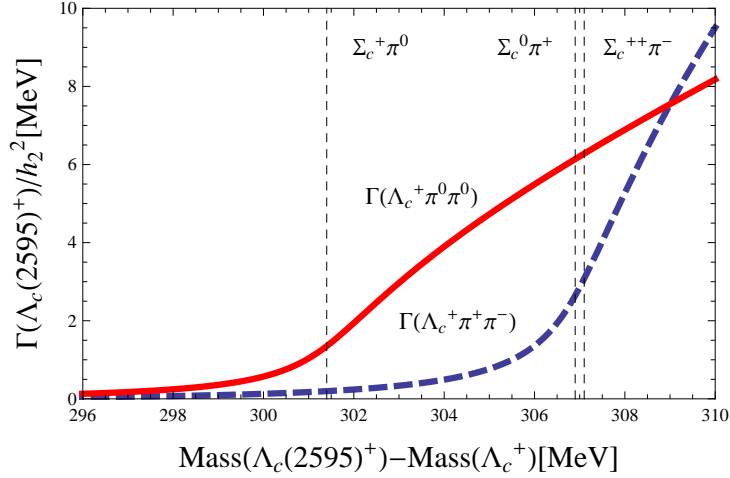,width=3.8in}}
\caption{Calculated dependence of $\Gamma(\Lambda_c^+\pi^0\pi^0)/h^2_2$ (full curve) and $\Gamma(\Lambda_c^+\pi^+\pi^-)/h^2_2$ (dashed curve) on $m(\Lambda_c(2595)^+)-m(\Lambda_c^+)$, where we have used the parameters $g_2=0.565$, $h_2=0.63$ and $h_8=0.85\times 10^{-3}\,{\rm MeV}^{-1}$. } \label{fig:Lambdac2595}
\end{figure}

It was pointed out in \cite{Falk03} that the proximity of the $\Lambda_c(2595)^+$ mass to the sum of the masses of its decay products will lead to an important threshold effect which will lower the $\Lambda_c(2595)^+$ mass by $2-3$ MeV than the one observed. A more sophisticated treatment of the mass lineshape of $\Lambda_c(2595)^+\to\Lambda_c^+\pi^+\pi^-$ by CDF yields $m(\Lambda_c(2595))=2592.25\pm0.28$ MeV \cite{CDF:2595}, which is 3.1 MeV smaller than the 2006 world average.
Therefore, the strong decay $\Lambda_c(2595)\to\Lambda_c\pi\pi$ is very close to the threshold. With the new measurement of  $m(\Lambda_c(2595))$, we have (in units of MeV) \cite{CC:2015}
\begin{eqnarray} \label{eq:h2h8new}
 \Gamma(\Lambda_c(2595)^+\to\Lambda_c^+\pi\pi)_R &=& g_2^2(20.45h_2^2
 +43.92h_8^2-8.95h_2h_8), \nonumber \\
 \Gamma(\Lambda_c(2625)^+\to\Lambda_c^+\pi\pi)_R &=& g_2^2(1.78h_2^2+4.557\times
 10^6h_8^2-79.75h_2h_8).
\end{eqnarray}

By performing a fit to the measured $M(pK^-\pi^+\pi^+)-M(pK^-\pi^+)$  mass difference distributions and using $g_2^2=0.365$, CDF obtained $h_2^2=0.36\pm0.08$ or $|h_2|=0.60\pm0.07$ \cite{CDF:2595}. This corresponds to a decay width $\Gamma(\Lambda_c(2595)^+)=2.59\pm0.30\pm0.47$ MeV \cite{CDF:2595}.
For the width of $\Lambda_c(2625)^+$, CDF observed a value consistent with zero and therefore calculated an upper limit 0.97 MeV using a Bayesian approach.
From the CDF measurements $\Gamma(\Lambda_c(2595)^+)=2.59\pm0.56$ MeV and $\Gamma(\Lambda_c(2625)^+)<0.97$ MeV, we obtain
\begin{eqnarray} \label{eq:h2h8,2014}
 |h_2|_{_{\rm 2015}}=0.63\pm0.07\,, \qquad\quad |h_8|_{_{\rm 2015}}< 2.32\times  10^{-3}\,{\rm MeV}^{-1}\,.
\end{eqnarray}
Hence, the magnitude of the coupling $h_2$ is greatly enhanced from 0.437 to 0.63\,. Our $h_2$ is slightly different from the value of 0.60 obtained by CDF. This is because CDF used $|g_2|=0.604$ to calculate the mass dependence of $\Gamma(\Lambda_c^+\pi\pi)$, while we used $|g_2|=0.565$.

\begin{table}[t]
\caption{Decay widths (in units of MeV) of $p$-wave charmed
baryons. Data are taken from 2014 PDG \cite{PDG} together with the new measurements of $\Sigma_c$, $\Sigma_c^*$ \cite{Belle:2014}
and $\Xi_c(2645)^+$ widths \cite{Belle:dc}.
Theoretical predictions of \cite{Tawfiq} are taken from
Table IV of \cite{Ivanov}.} \label{tab:strongdecayP}
\begin{center}
\begin{tabular}{|c|c c c c c c|} \hline \hline
~~~~~~~Decay~~~~~~~ & Expt. & ~HHChPT~ & Tawfiq & Ivanov & Huang & Zhu   \\
& ~~\cite{PDG}~~&  & ~~et al. \cite{Tawfiq}~~ &
 ~~~et al. \cite{Ivanov}~~~ &  ~~et al. \cite{Huang95}~~ & ~~\cite{Zhu}~~  \\
\hline
 $\Lambda_c(2595)^+\to (\Lambda_c^{+}\pi\pi)_R$ & $2.59\pm0.56$ & input & $$
 & $$ & $2.5$ &  \\ \hline
 $\Lambda_c(2595)^+\to \Sigma_c^{++}\pi^-$ & $$ & $$ & $1.47\pm0.57$ &
 $0.79\pm0.09$ & $0.55^{+1.3}_{-0.55}$ & 0.64  \\ \hline
 $\Lambda_c(2595)^+\to \Sigma_c^{0}\pi^+$ & $$ & $$ & $1.78\pm0.70$
 & $0.83\pm0.09$ & $0.89\pm0.86$ & 0.86 \\ \hline
 $\Lambda_c(2595)^+\to \Sigma_c^{+}\pi^0$ & & $2.74^{+0.57}_{-0.60}$ & $1.18\pm0.46$
 & $0.98\pm0.12$ & $1.7\pm0.49$ & 1.2  \\ \hline
 $\Lambda_c(2625)^+\to \Sigma_c^{++}\pi^-$ & $<0.10$ & $\lsim 0.028$ & $0.44\pm0.23$ & $0.076\pm0.009$ &
 $0.013$ & 0.011  \\ \hline
 $\Lambda_c(2625)^+\to \Sigma_c^{0}\pi^+$ & $<0.09$ & $\lsim 0.040$ & $0.47\pm0.25$ & $0.080\pm0.009$
 & 0.013 & 0.011  \\ \hline
 $\Lambda_c(2625)^+\to \Sigma_c^{+}\pi^0$ & & $\lsim 0.029$ & $0.42\pm0.22$ & $0.095\pm0.012$
 & 0.013 & 0.011  \\ \hline
 $\Lambda_c(2625)^+\to \Lambda_c^+\pi\pi$ & $<0.97$ & $\lsim 0.35$ & $$ & $$
 & 0.11 &   \\ \hline
 $\Sigma_c(2800)^{++}\to\Lambda_c\pi,\Sigma_c^{(*)}\pi$ & $75^{+22}_{-17}$ & input &
 & & &  \\ \hline
 $\Sigma_c(2800)^{+}\to\Lambda_c\pi,\Sigma_c^{(*)}\pi$ & $62^{+60}_{-40}$ & input &
 & & &  \\ \hline
 $\Sigma_c(2800)^0\to\Lambda_c\pi,\Sigma_c^{(*)}\pi$ & $72^{+22}_{-15}$ & input &
 & & & \\ \hline
 $\Xi_c(2790)^+\to\Xi'^{0,+}_c\pi^{+,0}$ & $<15$ & $16.7^{+3.6}_{-3.6}$ &
 $$ & $$ & &  \\ \hline
 $\Xi_c(2790)^0\to\Xi'^{+,0}_c\pi^{-,0}$  & $<12$ & $17.7^{+2.9}_{-3.8}$ &
 $$ & $$ & & \\ \hline
 $\Xi_c(2815)^+\to\Xi^{*+,0}_c\pi^{0,+}$ & $<3.5$ & $7.1^{+1.5}_{-1.5}$ &
 $2.35\pm0.93$ & $0.70\pm0.04$ & &  \\ \hline
 $\Xi_c(2815)^0\to\Xi^{*+,0}_c\pi^{-,0}$ & $<6.5$ & $7.7^{+1.7}_{-1.7}$ &
 $$ & $$ & &  \\ \hline \hline
\end{tabular}
\end{center}
\end{table}

The fact that the coupling $h_2$ obtained in 2006 and 2015 is so different is ascribed to the fact that the mass of $\Lambda_c(2595)^+$ is 3.1 MeV lower than the previous world average due to the threshold effect. To illustrate this, we consider the dependence of $\Gamma(\Lambda_c^+\pi^+\pi^-)/h^2_2$ and $\Gamma(\Lambda_c^+\pi^0\pi^0)/h^2_2$ on $\Delta M(\Lambda_c(2595))\equiv M(\Lambda_c(2595)^+)-M(\Lambda_c^+)$ as depicted in Fig. \ref{fig:Lambdac2595}. It is evident that $\Gamma(\Lambda_c^+\pi\pi)/h^2_2$ at $\Delta M(\Lambda_c(2595))=305.79$ MeV is smaller than that at 308.9 MeV. This explains why $h_2$ should become larger when $\Delta M(\Lambda_c(2595))$ becomes smaller.

The $\Xi_c(2790)$ and $\Xi_c(2815)$ baryons form a doublet
$\Xi_{c1}({1\over 2}^-,{3\over 2}^-)$.  Using the
coupling $h_2$ obtained from (\ref{eq:h2h8,2014}) and  assuming SU(3) flavor symmetry, the predicted $\Xi_c(2790)$ and $\Xi_c(2815)$
widths are shown in Table \ref{tab:strongdecayP}. It is evident that the predicted two-body decay rates of $\Xi_c(2790)^0$ and $\Xi_c(2815)^+$ exceed the current experimental limits because of the enhancement of $h_2$ (see Table \ref{tab:strongdecayP}). Hence, there is a tension for the coupling $h_2$ as its value extracted from from $\Lambda_c(2595)^+\to \Lambda_c^+\pi\pi$ will imply $\Xi_c(2790)^0\to\Xi'_c\pi$ and $\Xi_c(2815)^+\to\Xi_c^*\pi$ rates slightly above current limits. It is conceivable that SU(3) flavor symmetry breaking can help account for the discrepancy.

Some information on the coupling $h_{10}$ can be inferred from
the strong decays of $\Sigma_c(2800)$.  From Eq. (\ref{eq:swavecoupling}) and the quark model relation $|h_3|=\sqrt{3}|h_2|$ from \cite{Pirjol}, we obtain, for example, $\Gamma(\Sigma_{c0}^{++}\to\Lambda_c^+\pi^+)\approx 885$ MeV. Hence, $\Sigma_c(2800)$ cannot be identified with $\Sigma_{c0}(1/2^-)$.
Using the quark model relation $h_{11}^2=2h_{10}^2$ and the measured widths of
$\Sigma_c(2800)^{++,+,0}$ (Table \ref{tab:spectrum}), we obtain
 \begin{eqnarray}
|h_{10}|=(0.85^{+0.11}_{-0.08})\times 10^{-3}\,{\rm MeV}^{-1}\,.
 \end{eqnarray}
The quark model relation $|h_8|=|h_{10}|$ then leads to \begin{eqnarray}
|h_{8}|\approx (0.85^{+0.11}_{-0.08})\times 10^{-3}\,{\rm MeV}^{-1}\,,
\end{eqnarray}
which improves the previous limit (\ref{eq:h2h8,2014}) by a factor of
3. The calculated partial widths of $\Lambda_c(2625)^+$
shown in Table \ref{tab:strongdecayP} are consistent with experimental limits.

\section{Lifetimes}

\subsection{Singly charmed baryons}

Among singly charmed baryons, the antitriplet states $\Lambda_c^+,~\Xi_c^+,~\Xi_c^0$ and the $\Omega_c^0$ baryon in the sextet decay weakly. The world averages of their lifetimes in 2006 were \cite{PDG2006}
  \begin{eqnarray} \label{eq:exptlifetime}
&& \tau(\Lambda^+_c)= (200\pm6)\times 10^{-15}s, \qquad
\tau(\Xi^+_c)= (442\pm26)\times 10^{-15}s,
  \nonumber \\
&& \tau(\Xi^0_c)= (112^{+13}_{-10})\times 10^{-15}s, \qquad~~
\tau(\Omega^0_c)= (69\pm12)\times 10^{-15}s.
  \end{eqnarray}
These results remain the same even in 2014 \cite{PDG}.
As we shall see below, the lifetime hierarchy
$\tau(\Xi_c^+)>\tau(\Lambda_c^+)>\tau(\Xi_c^0)
>\tau(\Omega_c^0)$ is qualitatively understood in the OPE (operator product expansion) approach but not
quantitatively.

Based on the OPE approach for the analysis of inclusive weak
decays, the inclusive rate of the charmed baryon is schematically
represented by
 \begin{eqnarray}
 \Gamma({\cal B}_c\to f) = {G_F^2m_c^5\over
192\pi^3}V_{\rm CKM}\left(A_0+{A_2\over m_c^2}+{A_3\over
m_c^3}+{\cal O}({1\over m_c^4})\right),
 \end{eqnarray}
with $V_{\rm CKM}$ being the relevant CKM matrix element.
The $A_0$ term comes from the $c$ quark decay and is common to all
charmed hadrons. There is no linear $1/m_Q$ corrections to the
inclusive decay rate due to the lack of gauge-invariant
dimension-four operators \cite{Chay,Bigi92}, a consequence known
as Luke's theorem \cite{Luke}. Nonperturbative corrections start
at order $1/m_Q^2$ and they are model independent. Spectator
effects in inclusive decays due to the Pauli interference and
$W$-exchange contributions account for $1/m_c^3$ corrections and
they have two eminent features: First, the estimate of spectator
effects is model dependent; the hadronic four-quark matrix
elements are usually evaluated by assuming the factorization
approximation for mesons and the quark model for baryons. Second,
there is a two-body phase-space enhancement factor of $16\pi^2$
for spectator effects relative to the three-body phase space for
heavy quark decay. This implies that spectator effects, being of
order $1/m_c^3$, are comparable to and even exceed the $1/m_c^2$
terms.

\begin{figure}[t]
\centerline{\psfig{file=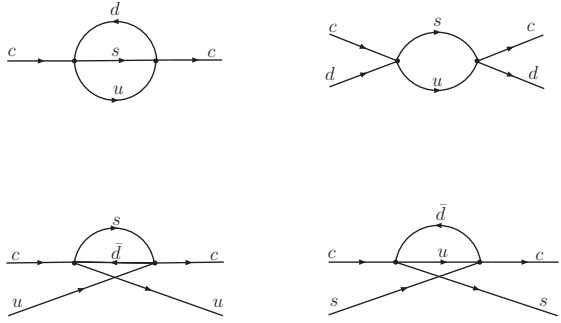,width=3.0in}}
\caption{Contributions to nonleptonic decay rates of charmed
baryons from four-quark operators: (a) $c$-quark decay, (b)
$W$-exchange, (c) destructive Pauli interference and (d)
constructive interference.} \label{fig:fourquarkNL}
\end{figure}

In general, the total width of the charmed baryon ${\cal B}_c$
receives contributions from inclusive nonleptonic and semileptonic
decays: $\Gamma({\cal B}_c)=\Gamma_{\rm NL}({\cal
B}_c)+\Gamma_{\rm SL}({\cal B}_c)$. The nonleptonic contribution
can be decomposed into
 \begin{eqnarray}
 \Gamma_{\rm NL}({\cal B}_c)=\Gamma^{\rm dec}({\cal B}_c)+\Gamma^{\rm ann}
 ({\cal B}_c)+\Gamma^{\rm
 int}_-({\cal B}_c)+\Gamma^{\rm int}_+({\cal B}_c),
 \end{eqnarray}
corresponding to the $c$-quark decay, the $W$-exchange
contribution, destructive and constructive Pauli interferences. It
is known that the inclusive decay rate is governed by the
imaginary part of an effective nonlocal forward transition
operator $T$. Therefore, $\Gamma^{\rm dec}$ corresponds to the
imaginary part of Fig. \ref{fig:fourquarkNL}(a) sandwiched between
the same ${\cal B}_c$ states. At the Cabibbo-allowed level,
$\Gamma^{\rm dec}$ represents the decay rate of $c\to su\bar d$,
and $\Gamma^{\rm ann}$ denotes the contribution due to the
$W$-exchange diagram $cd\to us$. The interference $\Gamma^{\rm
int}_-$ ($\Gamma^{\rm int}_+$) arises from the destructive
(constructive) interference between the $u$ ($s$) quark produced
in the $c$-quark decay and the spectator $u$ ($s$) quark in the
charmed baryon ${\cal B}_c$. Notice that the constructive Pauli
interference is unique to the charmed baryon sector as it does not
occur in the bottom sector. From the quark content of the
charmed baryons, it is clear that
at the Cabibbo-allowed level, the destructive interference occurs
in $\Lambda_c^+$ and $\Xi_c^+$ decays (Fig. \ref{fig:fourquarkNL}(c)), while $\Xi_c^+,\Xi_c^0$ and
$\Omega_c^0$ can have constructive interference $\Gamma^{\rm int}_+$ (Fig. \ref{fig:fourquarkNL}(d)).
Since $\Omega_c^0$
contains two $s$ quarks, it is natural to expect that $\Gamma^{\rm
int}_+(\Omega_c^0)\gg \Gamma^{\rm int}_+(\Xi_c)$. The $W$-exchange contribution (Fig. \ref{fig:fourquarkNL}(b))
occurs  only for $\Xi_c^0$ and $\Lambda_c^+$ at the same
Cabibbo-allowed level. In the heavy quark expansion approach, the
above-mentioned spectator effects can be described in terms of the
matrix elements of local four-quark operators.

\begin{table}
\caption{Various contributions to the decay rates (in units of
$10^{-12}$ GeV) of singly charmed baryons \cite{Cheng:1997}.  Experimental values are taken from \cite{PDG}.}
\label{tab:lifetime}
\begin{center}
\begin{tabular}{|c|c c c c l l  l  l|} \hline \hline
 & $\Gamma^{\rm dec}$ & $\Gamma^{\rm ann}$ & $\Gamma^{\rm int}_-$ &
$\Gamma^{\rm int}_+$ & ~ $\Gamma_{\rm SL}$ & ~$\Gamma^{\rm tot}$ &
~$\tau(10^{-13}s)$~ & ~ $\tau_{\rm expt}(10^{-13}s)$~ \\
\hline
 $\Lambda_c^+$~ & ~1.006~ & ~1.342~ & ~$-0.196$~ & & ~0.323~ &
~2.492~ & ~  2.64~ &  ~$2.00\pm 0.06$~   \\
 $\Xi_c^+$ & 1.006 & 0.071 & $-0.203$ & 0.364 & ~0.547 &
~1.785~ & ~ 3.68~ & ~$4.42\pm0.26$  \\
 $\Xi_c^0$ & 1.006 & 1.466 & & 0.385 & ~0.547 &
~3.404 & ~ 1.93 & ~$1.12^{+0.13}_{-0.10}$ \\
 $\Omega_c^0$ & 1.132 & 0.439 & & 1.241 & ~1.039 &
~3.851 & ~ 1.71  & ~$0.69\pm 0.12$  \\
\hline \hline
\end{tabular}
\end{center}
\end{table}
\vskip 0.4cm

The inclusive
nonleptonic rates of charmed baryons in the valence quark
approximation and in the limit of $m_s/m_c=0$ have the expressions \cite{Cheng:1997}:
 \begin{eqnarray} \label{eq:lifetimes}
 \Gamma_{\rm NL}(\Lambda_c^+) &=& \Gamma^{\rm
 dec}(\Lambda_c^+)+\cos\theta_C^2\Gamma^{\rm ann}+\Gamma^{\rm
 int}_-+\sin\theta_C^2\Gamma^{\rm int}_+,  \nonumber \\
 \Gamma_{\rm NL}(\Xi_c^+) &=& \Gamma^{\rm
 dec}(\Xi_c^+)+\sin\theta_C^2\Gamma^{\rm ann}+\Gamma^{\rm
 int}_-+\cos\theta_C^2\Gamma^{\rm int}_+,  \nonumber \\
 \Gamma_{\rm NL}(\Xi_c^0) &=& \Gamma^{\rm
 dec}(\Xi_c^0)+\Gamma^{\rm ann}+\Gamma^{\rm int}_+,  \nonumber \\
 \Gamma_{\rm NL}(\Omega_c^0) &=& \Gamma^{\rm
 dec}(\Omega_c^0)+6\sin\theta_C^2\Gamma^{\rm ann}+{10\over 3}\cos\theta_C^2\Gamma^{\rm int}_+,
 \end{eqnarray}
with $\theta_C$ being the Cabibbo angle.

The results of a model calculation in \cite{Cheng:1997} are shown in Table
\ref{tab:lifetime}. It is clear that the lifetime pattern
 \begin{eqnarray} \label{eq:lifepattern}
\tau(\Xi_c^+)>\tau(\Lambda_c^+)>\tau(\Xi_c^0)>\tau(\Omega_c^0)
 \end{eqnarray}
is in accordance with experiment.  This lifetime hierarchy is
qualitatively understandable. The $\Xi_c^+$ baryon is
longest-lived among charmed baryons because of the smallness of
$W$-exchange and partial cancellation between constructive and
destructive Pauli interferences, while $\Omega_c$ is
shortest-lived due to the presence of two $s$ quarks in the
$\Omega_c$ that renders the contribution of $\Gamma^{\rm int}_+$
largely enhanced. Since
$\Gamma^{\rm int}_+$ is always positive, $\Gamma^{\rm int}_-$ is
negative and that the constructive interference is larger than the
magnitude of the destructive one, this explains why
$\tau(\Xi_c^+)>\tau(\Lambda_c^+)$. It is also clear from Table
\ref{tab:lifetime} that, although the qualitative feature of the
lifetime pattern is comprehensive, the quantitative estimates of
charmed baryon lifetimes and their ratios are still rather poor.

\subsection{Doubly charmed baryons}

\begin{table}[t]
\caption{Predicted lifetimes of doubly charmed baryons
in units of $10^{-13}s$. } \label{tab:lifetimes_dc}
\begin{center}
\begin{tabular}{|c|c c c c |} \hline \hline
 & ~~~Kiselev et al.~~ & ~~Guberina et al.~~ & ~~Chang et al.~~ & ~~Karliner et al.~~  \\
&  \cite{Kiselev:2001} & \cite{Guberina} & \cite{Chang} & \cite{Karliner} \\
\hline
 ~~$\Xi_{cc}^{++}$~~~ & $4.6\pm0.5$ & 10.5 & 6.7 & 1.85 \\
 ~~$\Xi_{cc}^{+}$~~~ & $1.6\pm0.5$ & 2.0 & 2.5 & 0.53 \\
 ~~$\Omega_{cc}^{+}$~~~ & $2.7\pm0.6$ & 3.0 & 2.1 &  \\ \hline
 \hline
\end{tabular}
\end{center}
\end{table}

The inclusive
nonleptonic rates of doubly charmed baryons in the valence quark
approximation and in the limit of $m_s/m_c=0$ have the expressions:
 \begin{eqnarray} \label{eq:lifetimes_dc}
 \Gamma_{\rm NL}(\Xi_{cc}^{+}) &=& \Gamma^{\rm
 dec}(\Xi_{cc}^{+})+\cos\theta_C^2\Gamma^{\rm ann}+\sin\theta_C^2\Gamma^{\rm int}_+,  \nonumber \\
 \Gamma_{\rm NL}(\Xi_{cc}^{++}) &=& \Gamma^{\rm
 dec}(\Xi_{cc}^{++})+\Gamma^{\rm
 int}_-,  \nonumber \\
 \Gamma_{\rm NL}(\Omega_{cc}^+) &=& \Gamma^{\rm
 dec}(\Omega_{cc}^+)+\sin\theta_C^2\Gamma^{\rm ann}+\cos\theta_C^2\Gamma^{\rm int}_+.
 \end{eqnarray}
Since $\Gamma^{\rm int}_+$ is positive and $\Gamma^{\rm int}_-$ is
negative,  it is obvious that $\Xi_{cc}^{++}$ is longest-lived, while $\Xi_{cc}^+$ ($\Omega_{cc}^+$) is shortest-lived if $\Gamma^{\rm int}_+>\Gamma^{\rm ann}$ ($\Gamma^{\rm int}_+<\Gamma^{\rm ann}$). In general, we have
\begin{eqnarray}
\tau(\Xi_{cc}^{++})\gg\tau(\Omega_{cc}^+)\sim\tau(\Xi_{cc}^+).
\end{eqnarray}
The predictions available in the literature are summarized in Table \ref{tab:lifetimes_dc}. Note that the lifetime of $\Xi_{cc}^+$ was measured by SELEX to be $\tau(\Xi_{cc}^+)<0.33\times 10^{-13}s$ \cite{Selex02}.

Since the mass splitting between $\Xi_{cc}^*$ and $\Xi_{cc}$ and between $\Omega_{cc}^*$ and $\Omega_{cc}$ is less than 100 MeV (see also Eq. ({\ref{eq:dcbmass}) for lattice calculations)
\begin{eqnarray}
m_{\Xi_{cc}^*}-m_{\Xi_{cc}}=m_{\Sigma_c^*}-m_{\Sigma_c}\approx 65\,{\rm MeV}, \qquad
m_{\Omega_{cc}^*}-m_{\Omega_{cc}}=m_{\Omega_c^*}-m_{\Omega_c}\approx 71\,{\rm MeV},
\end{eqnarray}
it is clear that only electromagnetic decays are allowed for $\Omega_{cc}^*$ and $\Xi_{cc}^*$.

\section{Hadronic weak decays}
\subsection{Nonleptonic decays}

Contrary to the significant progress made over the last 10 years
or so in the studies of hadronic weak decays in the bottom baryon sector, advancement in
the arena of charmed baryons, both theoretical and experimental, has
been very slow.

In the naive factorization approach, the coefficients $a_1$ for
the external $W$-emission amplitude and $a_2$ for internal
$W$-emission are given by $(c_1+{c_2\over N_c})$ and
$(c_2+{c_1\over N_c})$, respectively. However, we have learned
from charmed meson decays that the naive factorization approach
never works for the decay rate of color-suppressed decay modes,
though it usually operates for color-allowed decays. Empirically,
it was learned in the 1980s that if the Fierz-transformed terms
characterized by $1/N_c$ are dropped, the discrepancy between
theory and experiment is greatly improved \cite{Fuk}. This leads
to the so-called large-$N_c$ approach for describing hadronic $D$
decays \cite{Buras}. Theoretically, explicit calculations based on
the QCD sum-rule analysis \cite{BS} indicate that the Fierz terms
are indeed largely compensated by the nonfactorizable corrections.

As the discrepancy between theory and experiment for charmed meson
decays gets much improved in the $1/N_c$ expansion method, it is
natural to ask if this scenario also works in the baryon sector?
This issue can be settled down by the experimental measurement of
the Cabibbo-suppressed mode $\Lambda_c^+\to p\phi$, which receives
contributions only from the factorizable diagrams. As pointed out
in \cite{CT92}, the large-$N_c$ predicted rate is in good
agreement with the measured value. By contrast, its decay rate
prdicted by the naive factorization approximation will be too small by
a factor of 15. Therefore, the $1/N_c$ approach also works for the
factorizable amplitude of charmed baryon decays. This also implies
that the inclusion of nonfactorizable contributions is inevitable
and necessary. If nonfactorizable effects amount to a redefinition
of the effective parameters $a_1$, $a_2$ and are universal (i.e.,
channel-independent) in charm decays, then we still have a new
factorization scheme with the universal parameters $a_1,~a_2$ to
be determined from experiment.

It is known in the heavy
meson case that nonfactorizable contributions will render the
color suppression of internal $W$-emission ineffective. However,
the $W$-exchange in baryon decays is not subject to color
suppression even in the absence of nonfactorizable terms. A simple
way to see this is to consider the large-$N_c$ limit. Although the
$W$-exchange diagram is down by a factor of $1/N_c$ relative to
the external $W$-emission one, it is compensated by the fact that
the baryon contains $N_c$ quarks in the limit of large $N_c$, thus
allowing $N_c$ different possibilities for $W$ exchange between
heavy and light quarks \cite{Korner}. That is, the pole
contribution can be as important as the factorizable one. The
experimental measurement of the decay modes
$\Lambda^+_c\to\Xi^0K^+, \Delta^{++}K^-$, which proceed only through the $W$-exchange contributions, indicates that $W$-exchange indeed plays an
essential role in charmed baryon decays.

On the theoretical side, various approaches had been made to investigate weak decays of heavy baryons, including the current algebra approach \cite{Korner79,Uppal}, the factorization scheme, the pole model technique \cite{CT92,CT93,XK92,XK92b,Verma98,Zen}, the relativistic quark model \cite{Korner,Ivanov98} and the quark-diagram scheme \cite{CCT,Kohara}.
Various model predictions of the branching fractions and decay
asymmetries  can be found in Tables VI-VII of \cite{Cheng:2009} for ${\cal B}_c\to{\cal B}+P$
decays, Table VIII for ${\cal B}_c\to{\cal B}+V$ decays and Table IX for ${\cal B}_c\to{\cal B}({3\over 2}^+)+P(V)$ decays.

\subsection{Discussions}

\subsubsection{$\Lambda_c^+$ decays}
Experimentally, nearly all the branching fractions of the
$\Lambda_c^+$ are measured relative to the $pK^-\pi^+$ mode.
Based on ARGUS and CLEO data, PDG has made a model-dependent determination of the absolute branching fraction ${\cal B}(\Lambda_c^+\to pK^-\pi^+)=(5.0\pm1.3)\%$ \cite{PDG}. Recently, Belle has reported a value of $(6.84\pm0.24^{+0.21}_{-0.27})\%$ \cite{Zupanc} from the reconstruction of $D^*p\pi$ recoiling against the $\Lambda_c^+$ production in $e^+ e^-$ annihilation. Hence, uncertainties are much reduced and, most importantly, this measurement is model independent! More recently, BESIII has also measured this mode directly with the preliminary result  ${\cal B}(\Lambda_c^+\to pK^-\pi^+)=(5.77\pm0.27)\%$ (statistical error only) \cite{Lyu}. Its precision is comparable to the Belle's result.
Another approach is to exploit a particular decay $B^+\to p\pi^+\pi^+\overline{\Sigma}_c^{--}$ and its charge conjugate to measure
${\cal B}(\Lambda_c^+\to pK^-\pi^+)$ also in a model independent manner \cite{Contu}.

\begin{table}
\caption{Branching fractions of the Cabibbo-allowed two-body decays of $\Lambda_c^+$ in units of \%. Data are taken from PDG \cite{PDG} except that the absolute branching fraction ${\cal B}(\Lambda_c^+\to pK^-\pi^+)=(5.0\pm1.3)\%$ is replaced by the new measurement of $(6.84\pm0.24^{+0.21}_{-0.27})\%$ by Belle \cite{Zupanc}. } \label{tab:BRs}
\begin{center}
\begin{tabular}{|lc | lc|lc|}
\hline\hline ~~~Decay & $\B$ & ~~~Decay & $\B$ & ~~~Decay & $\B$  \\
\hline
~~$\Lambda^+_c\to \Lambda \pi^+$~~ & 1.46$\pm$ 0.13 & ~~$\Lambda^+_c\to \Lambda \rho^+$~~ & $<6.5$ & ~~$\Lambda^+_c\to \Delta^{++}K^-$  &  $1.16\pm0.07$\\
\hline
~~$\Lambda^+_c\to \Sigma^0 \pi^+$~~ & 1.44$\pm$ 0.14 & ~~$\Lambda^+_c\to \Sigma^0 \rho^+$  & & ~~$\Lambda^+_c\to \Sigma^{*0} \pi^+$ & \\
\hline
~~$\Lambda^+_c\to \Sigma^+ \pi^0$~~ & 1.37$\pm$ 0.30 & ~~$\Lambda^+_c\to \Sigma^+ \rho^0$~~  & $<1.9$ & ~~$\Lambda^+_c\to \Sigma^{*+} \pi^0$& \\
\hline
~~$\Lambda^+_c\to \Sigma^+ \eta$~~ & 0.75$\pm$ 0.11 & ~~$\Lambda^+_c\to \Sigma^+ \omega$~~  &  3.7$\pm$ 1.0 & ~~$\Lambda^+_c\to \Sigma^{*+}\eta$~~ & $1.16\pm0.35$ \\
\hline
~~$\Lambda^+_c\to \Sigma^+ \eta^\prime$~~ & & ~~$\Lambda^+_c\to \Sigma^+ \phi$~~ & 0.42$\pm$ 0.07 & ~~$\Lambda^+_c\to \Sigma^{*+} \eta^\prime$ &\\
\hline
~~$\Lambda^+_c\to \Xi^0 K^+$~~ & 0.53$\pm$ 0.13 & ~~$\Lambda^+_c\to \Xi^0 K^{*+}$~~ & 0.53$\pm$ 0.19 & ~~$\Lambda^+_c\to \Xi^{*0}K^+$~~  & $0.36\pm0.10$\\
\hline
~~$\Lambda^+_c\to p \bar K^0$~~ &  3.2$\pm$ 0.3 & ~~$\Lambda^+_c\to p \bar K^{*0}$~~  & 2.1$\pm$ 0.3 & ~~$\Lambda^+_c\to \Delta^+\bar K^0$~~ & 1.36$\pm$ 0.44 \\
\hline \hline
\end{tabular}
\end{center}
\end{table}

Branching fractions of the Cabibbo-allowed two-body decays of $\Lambda_c^+$ are displayed in Table \ref{tab:BRs}. Data taken from PDG \cite{PDG} are scaled up by a factor of 1.37 for the central values due to the new measurement of ${\cal B}(\Lambda_c^+\to pK^-\pi^+)$ by Belle \cite{Zupanc}. BESIII has recently measured 2-body, 3-body and 4-body decay modes of $\Lambda_c^+$ with precision significantly improved \cite{Lyu}. For example, $\B(\Lambda_c^+\to \Lambda\pi^+)=(1.20\pm0.07)\%$ obtained by BESIII has a much better precision than the value of $(1.07\pm0.28)\%$ quoted by PDG \cite{PDG}.

Many of the $\Lambda_c^+$ decay modes such as $\Sigma^+K^+K^-$, $\Sigma^+\phi$, $\Xi^{(*)}K^{(*)+}$ and $\Delta ^{++}K^-$ can only proceed through $W$-exchange.  The
experimental measurement of them implies the importance of $W$-exchange, which is not subject to color suppression in charmed baryon decays.

Some Cabibbo-suppressed modes such as $\Lambda_c^+\to\Lambda K^+$ and
$\Lambda_c^+\to\Sigma^0K^+$ have been measured by Belle \cite{Belle:LamcCS} and BaBar \cite{BaBar:LamcCS}, respectively. Their branching fractions are of order $10^{-3}-10^{-4}$.
The first measured Cabibbo-suppressed mode $\Lambda_c^+\to p\phi$
is of particular interest because it receives contributions only
from the factorizable diagram and is expected to be color
suppressed in the naive factorization approach. A
calculation in \cite{CT96} yields
 \begin{eqnarray}
 {\cal B}(\Lambda_c^+\to p\phi)=2.26\times 10^{-3}a_2^2, \qquad
 \alpha(\Lambda_c^+\to p\phi)=-0.10\,.
 \end{eqnarray}
From the experimental measurement ${\cal B}(\Lambda_c^+\to
p\phi)=(11.2\pm2.3)\times 10^{-4}$ \cite{PDG},
\footnote{We have scaled up the PDG number $(8.7\pm 2.7)\times 10^{-4}$ \cite{PDG} by a factor of 1.37 for its central value.}
it follows that
 \begin{eqnarray} \label{eq:a2}
 |a_2|_{\rm expt}=0.70\pm0.07\,.
 \end{eqnarray}
This is consistent with the $1/N_c$ approach where
$a_2=c_2(m_c)\approx -0.59$\,.

All the models
except the model of \cite{Verma98} predict a positive decay
asymmetry $\alpha$ for the decay $\Lambda_c^+\to\Sigma^+\pi^0$ (see Table VII of \cite{Cheng:2009}). Therefore,
the measurement of $\alpha=-0.45\pm0.31\pm0.06$ by CLEO
\cite{CLEO:alpha} is a big surprise. If the negative sign of
$\alpha$ is confirmed in the future, this will imply an opposite
sign between $s$-wave and $p$-wave amplitudes for this decay,
contrary to the model expectation. The implication of this has
been discussed in detail in \cite{CT92}. Since the error of the
previous CLEO measurement is very large, it is crucial to carry out
more accurate measurements of the decay asymmetry for
$\Lambda_c^+\to\Sigma^+\pi^0$.

\subsubsection{$\Xi_c^+$ decays}
No absolute branching fractions have been measured. The branching
ratios listed in Tables VI and VIII of \cite{Cheng:2009} are the
ones relative to $\Xi_c^+\to\Xi^-\pi^+\pi^+$. Several
Cabibbo-suppressed decay modes such as $p\bar K^{*0}$,
$\Sigma^+\phi$, $\Sigma^+\pi^+\pi^-$, $\Sigma^-\pi^+\pi^+$  and $\Xi(1690)K^+$ have been observed \cite{PDG}.

The Cabibbo-allowed decays $\Xi_c^+\to {\cal B}(3/2^+)+P$ have
been studied and they are believed to be forbidden as they do not
receive factorizable and $1/2^\pm$ pole contributions
\cite{XK92b,Korner}. However, the $\Sigma^{*+}\bar K^0$ mode was
seen by FOCUS before \cite{FOCUS:SigK} and this may indicate the
importance of pole contributions beyond low-lying $1/2^\pm$
intermediate states.

\subsubsection{$\Xi_c^0$ decays}
No absolute branching fractions have been measured so far. However,
there are several measurements of the ratios of branching fractions,
for example \cite{PDG},
 \begin{eqnarray}
 R_1={\Gamma(\Xi_c^0\to\Lambda
K_S^0)\over\Gamma(\Xi_c^0\to\Xi^-\pi^+)}=0.21\pm0.02\pm0.02, \quad
R_2=
  {\Gamma(\Xi_c^0\to\Omega^-
K^+)\over\Gamma(\Xi_c^0\to\Xi^-\pi^+)}=0.297\pm0.024\,.
 \end{eqnarray}
The decay modes $\Xi_c^0\to\Omega^-K^+$ and $\Sigma^+K^-$ and $\Sigma^+\pi^-$ proceed only through $W$-exchange.  The measured branching ratio of $\Omega^-K^+$ relative to $\Xi^-\pi^+$ implies the significant role played by the $W$-exchange mechanism. The model of K\"orner and Kr\"amer \cite{Korner}
predicts $R_2=0.33$ (see Table IX of \cite{Cheng:2009}), in agreement with
experiment, but its prediction $R_1=0.06$ is too small compared to
the data.


\subsubsection{$\Omega_c^0$ decays}
One of the unique features of the $\Omega_c^0$ decay is that the
decay $\Omega_c^0\to\Omega^-\pi^+$ proceeds only via external
$W$-emission, while $\Omega_c^0\to \Xi^{*0}\bar K^0$  via
the factorizable internal $W$-emission diagram.  Various model predictions of Cabibbo-allowed $\Omega_c^0\to {\cal B}(3/2^+)+P(V)$ are displayed
in Table IX of \cite{Cheng:2009} with the unknown parameters $a_1$ and
$a_2$. From the decay $\Lambda_c^+\to p\phi$ we learn that
$|a_2|=0.70\pm0.07$. Recently, the hadronic weak decays of the $\Omega_c^0$ have been studied in \cite{Dhir} in great details with the finding that most of the decay channels in $\Omega_c^0$ decays proceed only through the
$W$-exchange diagram; moreover, the $W$-exchange contributions dominate in the rest of processes with some exception. Observation of such decays would shed some light on the mechanism of $W$-exchange effects in these decay modes.

\subsection{Charm-flavor-conserving nonleptonic decays}
There is a special class of weak decays of charmed baryons which
can be studied in a reliable way, namely, heavy-flavor-conserving
nonleptonic decays. Some examples are the singly
Cabibbo-suppressed decays $\Xi_c\to\Lambda_c\pi$ and
$\Omega_c\to\Xi'_c\pi$. The idea is simple: In these decays only
the light quarks inside the heavy baryon will participate in weak
interactions; that is, while the two light quarks undergo weak
transitions, the heavy quark behaves as a ``spectator". As the
emitted light mesons are soft, the $\Delta S=1$ weak interactions
among light quarks can be handled by the well known short-distance
effective Hamiltonian.
This special class
of weak decays usually can be calculated more reliably than the
conventional charmed baryon weak decays.
The synthesis of heavy-quark and chiral
symmetries provides a natural setting for investigating these
reactions \cite{ChengHFC}. The weak decays $\Xi_Q\to\Lambda_Q\pi$
with $Q=c,b$  were also studied in \cite{Voloshin00,Faller}.

The combined symmetries of heavy and light quarks severely
restricts the weak interactions allowed. In the symmetry limit, it
is found that there cannot be ${\cal B}_{\bar 3}-{\cal B}_6$ and
${\cal B}^*_6-{\cal B}_6$ nonleptonic weak transitions \cite{ChengHFC}. Symmetries alone permit three types of transitions:  ${\cal B}_{\bar 3}-{\cal
B}_{\bar 3}$,  ${\cal B}_{6}-{\cal B}_6$ and ${\cal B}^*_6-{\cal
B}_6$ transitions. However, in both the MIT bag and diquark
models, only ${\cal B}_{\bar 3}-{\cal B}_{\bar 3}$ transitions
have nonzero amplitudes. The general amplitude for $\B_i\to \B_f+P$ reads
\begin{eqnarray}
M(\B_i\to \B_f+P)=i\bar u_f(A-B\gamma_5)u_i,
\end{eqnarray}
where $A$ and $B$ are the $S$- and $P$-wave amplitudes, respectively. The $S$-wave amplitude can be evaluated using current algebra in terms of the parity-violating commutator term. For example, the $S$-wave amplitude of $\Xi_c^+\to\Lambda_c^+\pi^0$ is given by
\begin{eqnarray}
A(\Xi_c^+\to\Lambda_c^+\pi^0)=-{1\over \sqrt{2}f_\pi}\langle \Lambda_c^+\uparrow|{\cal H}_{\rm eff}^{\rm PC}|\Xi_c^+\uparrow\rangle,
\end{eqnarray}
while the $P$-wave amplitude arises from the ground-state baryon poles \cite{ChengHFC}
\begin{eqnarray}
B(\Xi_c^+\to\Lambda_c^+\pi^0)={g_2\over 2f_\pi}\,{m_{\Xi_c}+m_{\Xi'_c}\over  m_{\Lambda_c}-m_{\Xi_{c2}} }\langle \Lambda_c^+\uparrow|{\cal H}_{\rm eff}^{\rm PC}|\Xi_c^+\uparrow\rangle \sin\phi,
\end{eqnarray}
where $\phi$ is mixing angle of $\Xi_c$ with $\Xi'_c$ and $\Xi_{c1}$, $\Xi_{c2}$ being their mass eigenstates. The matrix element $\langle \Lambda_c^+\uparrow|{\cal H}_{\rm eff}^{\rm PC}|\Xi_c^+\uparrow\rangle$ was evaluated in \cite{ChengHFC} using two different models: the MIT bag model \cite{MIT} and the diquark model.

The predicted rates are \cite{ChengHFC}
 \begin{eqnarray}
&& \Gamma(\Xi_c^0\to\Lambda_c^+\pi^-) = 1.7\times 10^{-15}\,{\rm GeV},
 \qquad
 \Gamma(\Xi_c^+\to\Lambda_c^+\pi^0) = 1.0\times 10^{-15}\,{\rm GeV},
 \nonumber \\
 && \Gamma(\Omega_c^0\to\Xi'^+_c\pi^-) = 4.3\times 10^{-17}\,{\rm
 GeV},
 \end{eqnarray}
and the corresponding branching fractions are
 \begin{eqnarray} \label{eq:HFC}
&& {\cal B}(\Xi_c^0\to\Lambda_c^+\pi^-) = 2.9\times 10^{-4},
 \qquad
 {\cal B}(\Xi_c^+\to\Lambda_c^+\pi^0) = 6.7\times 10^{-4},
 \nonumber \\
 &&  {\cal B}(\Omega_c^0\to\Xi'^+_c\pi^-) = 4.5\times 10^{-6}.
 \end{eqnarray}
As stated above, the ${\cal B}_{6}-{\cal B}_6$ transition
$\Omega_c^0\to\Xi'^+_c\pi^-$ vanishes in the chiral limit. It
receives a finite factorizable contribution as a result of
symmetry-breaking effect. At any rate, the predicted branching
fractions for the charm-flavor-conserving decays
$\Xi_c^0\to\Lambda_c^+\pi^-$ and $\Xi_c^+\to\Lambda_c^+\pi^0$ are
of order $10^{-3}\sim 10^{-4}$ and  should be readily accessible
in the near future.

\subsection{Semileptonic decays}

{\squeezetable
\begin{table}
\caption{Predicted semileptonic decay rates (in units of
$10^{10}s^{-1}$) and decay asymmetries (second entry) in various models. The absolute branching fraction ${\cal B}(\Lambda_c^+\to pK^-\pi^+)=(5.0\pm1.3)\%$ is replaced by the new measurement of $(6.84\pm0.24^{+0.21}_{-0.27})\%$ by Belle \cite{Zupanc} for the data of $\Gamma(\Lambda_c^+\to\Lambda^0\ell^+\nu_\ell)$ taken from PDG \cite{PDG}.  Predictions of \cite{Marcial} are obtained in the
non-relativistic quark model and the MIT bag model (in
parentheses).  } \label{tab:SL}
\begin{center}
\begin{tabular}{|l|c|c|c|c|c|c|c|c| c|c|c|} \hline \hline
 Process
 & ~\cite{CT96}~ & \cite{Marcial} & ~\cite{Singleton}~  & \cite{Ivanov96}
 & ~\cite{Luo}~ & \cite{Carvalho} & \cite{Huang} & \cite{Gutsche} & \cite{Azizi} & \cite{Pervin}  & Expt. \cite{PDG} \\ \hline
 $\Lambda_c^+\to\Lambda^0 e^+\nu_e$  & 7.1 & 11.2 (7.7)  & 9.8  &
 7.22 & 7.0 &  $13.2\pm1.8$ & $10.9\pm3.0$ & & & & $14.4\pm 2.6$  \\
 & & & & $-0.812$ & & $-1$ & $-0.88\pm0.03$ & & & & $-0.86\pm0.04$ \\ \hline
 $\Lambda_c^+\to\Lambda^0 \mu^+\nu_e$  & 7.1 & 11.2 (7.7)  & 9.8  &
 7.22 & 7.0 &  $13.2\pm1.8$ & $10.9\pm3.0$ & & & & $13.3\pm 2.8$  \\
 \hline
 $\Lambda_c^+\to n e^+\nu_e$  &  &
 &  & 1.32 &  & & & 1.01 & & ~0.96,~1.37~ & \\ \hline
 $\Xi_c\to\Xi e^+\nu_e$  & 7.4 & 18.1 (12.5) &
 8.5  & 8.16 & 9.7 & & & & $64.8\pm22.6$ & & seen \\ \hline
 $\Xi_c\to\Sigma e^+\nu_e$  &  &
 & & & & & & &  $3.3\pm1.7$ & &  \\ \hline
\end{tabular}
\end{center}
\end{table}}

The exclusive semileptonic decays of charmed baryons:
$\Lambda_c^+\to\Lambda e^+(\mu^+)\nu_e$, $\Xi_c^+\to \Xi^0
e^+\nu_e$ and $\Xi_c^0\to \Xi^-e^+\nu_e$ have been observed
experimentally. Their rates depend on the ${\cal B}_c\to{\cal B}$
form factors $f_i(q^2)$ and $g_i(q^2)$ ($i=1,2,3$) defined by
\begin{eqnarray} \label{eq:FF}
 \langle {\cal B}_f(p_f)|V_\mu-A_\mu|{\cal B}_c(p_i)\rangle &=& \bar{u}_f(p_f)
[f_1(q^2)\gamma_\mu+if_2(q^2)\sigma_{\mu\nu}q^\nu+f_3(q^2)q_\mu   \nonumber \\
&&
-(g_1(q^2)\gamma_\mu+ig_2(q^2)\sigma_{\mu\nu}q^\nu+g_3(q^2)q_\mu)\gamma_5]
u_i(p_i).
\end{eqnarray}
These form factors have been evaluated in the
non-relativistic quark model \cite{Marcial,Singleton,CT96,Pervin}, the
MIT bag model \cite{Marcial}, the relativistic quark model
\cite{Ivanov96,Gutsche}, the light-front quark model \cite{Luo} and QCD
sum rules \cite{Carvalho,Huang,Azizi}. Experimentally, the only
information available so far is the form-factor ratio measured in
the semileptonic decay $\Lambda_c\to\Lambda e\bar{\nu}$. In the
heavy quark limit, the six $\Lambda_c\to\Lambda$ form factors are
reduced to two:
 \begin{eqnarray}
\langle
\Lambda(p)|\bar{s}\gamma_\mu(1-\gamma_5)c|\Lambda_c(v)\rangle=\,\bar{u}
_{_\Lambda}\left(F_1^{\Lambda_c\Lambda}(v\cdot p)+v\!\!\!/
F_2^{\Lambda_c \Lambda}(v\cdot
p)\right)\gamma_\mu(1-\gamma_5)u_{_{\Lambda_c}}.
 \end{eqnarray}
Assuming a dipole $q^2$ behavior for form factors, the ratio
$R=\tilde{F}_2^{\Lambda_c\Lambda}/\tilde{F}_1^{\Lambda_c\Lambda}$
is measured by CLEO to be \cite{CLEO:sl}
 \begin{eqnarray} R=-0.31\pm 0.05\pm 0.04\,.
 \end{eqnarray}

Various model predictions of the charmed baryon semileptonic decay
rates and decay asymmetries are shown in Table \ref{tab:SL}. Dipole $q^2$ dependence for form factors is assumed whenever the
form factor momentum dependence is not available in the model. Four different sets of predictions for $\Lambda_c^+\to n e^+\nu_e$ not listed in Table \ref{tab:SL} were presented in the sum rule calculations of \cite{Azizi2}. The semileptonic decays of $\Omega_c$ have been treated in \cite{Pervin:omegac} within the framework of a constituent quark model.
From Table \ref{tab:SL} we see that the computed branching fractions of
$\Lambda_c^+\to\Lambda e^+\nu$ falling in the range $1.4\%\sim 2.6\%$
are slightly smaller than experiment, $(2.9\pm0.5)\%$ [$(2.1\pm0.6)\%$ in PDG \cite{PDG}].
Branching fractions of $\Xi_c^0\to\Xi^-e^+\nu$ and $\Xi_c^+\to\Xi^0
e^+\nu$ are predicted to lie in the ranges $(0.8\sim 2.0)\%$
and $(3.3\sim 8.1)\%$, respectively, except that the QCD sum rule calculation in \cite{Azizi} predicts a much large rate for $\Xi_c\to\Xi e^+\nu_e$.
Experimentally, only the
ratios of the branching fractions are available so far \cite{PDG}
 \begin{eqnarray}
 {\Gamma(\Xi_c^+\to\Xi^0
 e^+\nu)\over \Gamma(\Xi_c^+\to\Xi^-\pi^+\pi^+)}= 2.3\pm0.6^{+0.3}_{-0.6},
 \qquad
 {\Gamma(\Xi_c^0\to\Xi^-
 e^+\nu)\over\Gamma(\Xi_c^0\to\Xi^-\pi^+)} =
 3.1\pm1.0^{+0.3}_{-0.5}\,.
 \end{eqnarray}

There has been active studies in semileptonic decays of doubly charmed baryons. The interested reader can consult to \cite{Albertus:2011,Faessler,Ebert:2004,Guo:doubly} for further references.

Just as the hadronic decays discussed in the last subsection, there are also heavy-flavor-conserving semileptonic processes, for example, $\Xi_c^0\to \Lambda_c^+(\Sigma_c^+)e^-\bar\nu_e$ and $\Omega_c^0\to\Xi_c^+ e^-\bar\nu_e$.
In these decays only
the light quarks inside the heavy baryon will participate in weak
interactions, while the heavy quark behaves as a spectator. This topic has been recently investigated in \cite{Faller}. Due to the severe phase-space suppression, the branching fractions are of order $10^{-6}$ in the best cases, typically $10^{-7}$ to $10^{-8}$.

\section{Electromagnetic and weak radiative decays}

Although radiative decays are well measured in the charmed meson
sector, e.g. $D^*\to D\gamma$ and $D_s^{*+}\to D_s^+\gamma$, only
three of the radiative modes in the charmed baryon sector have
been seen, namely, $\Xi'^0_c\to\Xi_c^0\gamma$, $\Xi'^+_c\to
\Xi^+_c\gamma$ and $\Omega_c^{*0}\to\Omega_c^0\gamma$. This is
understandable  because $m_{\Xi'_c}-m_{\Xi_c}\approx 108$ MeV and
$m_{\Omega_c^*}-m_{\Omega_c}\approx 71$ MeV. Hence, $\Xi'_c$ and
$\Omega_c^*$ are governed by the electromagnetic decays. However,
it will be difficult to measure the rates of these decays because
these states are too narrow to be experimentally resolvable.
Nevertheless, we shall systematically study the two-body
electromagnetic decays of charmed baryons and also weak radiative
decays.

\subsection{Electromagnetic decays}

In the charmed baryon sector, the following two-body electromagnetic
decays are of interest:
 \begin{eqnarray}
\B_6 \to \B_{\overline{3}} + \gamma & : & \Sigma_c \rightarrow
\Lambda_c + \gamma,  \quad \Xi^\prime_c \rightarrow \Xi_c + \gamma , \nonumber \\
 \B^\ast_6 \rightarrow \B_{\overline{3}} + \gamma & : &
\Sigma^\ast_c \rightarrow \Lambda_c + \gamma ,  \quad \Xi^{\prime
\ast}_c \rightarrow
\Xi_c + \gamma ,  \nonumber \\
\B^\ast_6 \rightarrow \B_6 + \gamma & : & \Sigma^\ast_c \rightarrow
\Sigma_c + \gamma , \quad \Xi^{\prime \ast}_c \rightarrow
\Xi^\prime_c + \gamma ,  \quad \Omega^\ast_c \rightarrow \Omega_c
+ \gamma ,
 \end{eqnarray}
where we have denoted the spin $\frac{1}{2}$ baryons as $\B_6$ and
$\B_{\overline{3}}$ for a symmetric sextet {\bf 6}  and
antisymmetric antitriplet {\bf \={3}}, respectively, and the spin
$\frac{3}{2}$ baryon by $\B^\ast_6$.

An ideal theoretical framework for studying the above-mentioned
electromagnetic decays is provided by the formalism in which the
heavy quark symmetry and the chiral symmetry of light quarks are
combined \cite{Yan,Wise}. When supplemented by the nonrelativistic
quark model, the formalism determines completely the low energy
dynamics of heavy hadrons. The electromagnetic interactions of
heavy hadrons consist of two distinct contributions: one from
gauging electromagnetically the chirally invariant strong
interaction Lagrangians for heavy mesons and baryons given in
\cite{Yan,Wise}, and the other from the anomalous magnetic moment
couplings of the heavy particles.  The heavy quark symmetry
reduces the number of free parameters needed to describe the
magnetic couplings to the photon.   There are two undetermined
parameters for the ground-state heavy baryons. All these
parameters are related simply to the magnetic moments of the light
quarks in the nonrelativistic quark model.  However, the charmed
quark is not particularly heavy ($m_c \simeq 1.6$ GeV), and it
carries a charge of $\frac{2}{3} e$.  Consequently, the
contribution from its magnetic moment cannot be neglected. The
chiral and electromagnetic gauge-invariant Lagrangian for heavy
baryons can be found in Eqs. (3.8) and (3.9) of \cite{Cheng93}, denoted by ${\cal L}_B^{(1)}$ and ${\cal L}_B^{(2)}$, respectively.

The most general gauge-invariant Lagrangian Eq. (3.9) of \cite{Cheng93} for magnetic transitions of heavy baryons can be recast in terms of superfields \cite{ChengSU3}
\begin{eqnarray} \label{eq:emL}
{\cal L}_B^{(2)} &=& -i3a_1 {\rm tr}(\bar S^\mu QF_{\mu\nu}S^\nu)+\sqrt{3}a_2\epsilon_{\mu\nu\alpha\beta}{\rm tr}(\bar S^\mu Qv^\nu F^{\alpha\beta}T)+h.c.  \nonumber \\
&& +3a'_1{\rm tr}(\bar S^\mu Q'\sigma\cdot F S_\mu)-{3\over 2}a'_1{\rm tr}(\bar TQ'\sigma\cdot F T),
\end{eqnarray}
where $\sigma\cdot F\equiv \sigma_{\mu\nu}F^{\mu\nu}$, $Q={\rm diag}(2/3, -1/3,-1/3)$ is the charge matrix for the light $u,d,$ and $s$ quarks, $Q'=e_Q$ is the charge of the heavy quark. In the above equation,
\begin{eqnarray}
T=\B_{\bar 3}, \qquad \quad S^\mu=\B_6^{*\mu}-{1\over\sqrt{3}}(\gamma^\mu+v^\mu)\gamma_5 \B_6\,.
\end{eqnarray}
It follows that \cite{ChengSU3}
\begin{eqnarray}
A[S^\mu_{ij}(v)\to S^\nu_{ij}+\gamma(\varepsilon,k)] &=& i{3\over 2}a_1\overline {\cal U}^\nu(Q_{ii}+Q_{jj})(k_\nu\varepsilon_\mu-k_\mu\varepsilon_\nu){\cal U}^\mu-i6a'_1Q'\overline {\cal U}^\mu k\!\!\!/ \varepsilon\!\!\!/ {\cal U}_\mu, \nonumber \\
A[S^\mu_{ij}(v)\to T_{ij}+\gamma(\varepsilon,k)] &=& -2\sqrt{3/2}\,a_2\epsilon_{\mu\nu\alpha\beta}\bar u_{\bar 3}v^\nu k^\alpha\varepsilon^\beta(Q_{ii}-Q_{jj}){\cal U}^\mu~(i<j),
\end{eqnarray}
where $k_\mu$ is the photon 4-momentum and $\varepsilon_\mu$ is
the polarization 4-vector. As stressed in \cite{Cheng93}, SU(3) breaking effects due to light-quark mass differences can be incorporated by replacing the charge matrix $Q$ by
\begin{eqnarray}
Q\to\tilde Q={\rm diag}\left( {2\over 3}, -{\alpha\over 3}, -{\beta\over 3}\right)
\end{eqnarray}
with $\alpha={M_u/M_d}$ and $\beta=M_u/M_s$. To avoid any confusion with the current quark mass $m_q$, we have used capital letters to denote the constituent quark masses. In the quark model, the coefficients $a_1$ and $a_2$ are simply related to the Dirac magnetic moments of the light quarks
\begin{eqnarray}
a_1=-{e\over 3}{1\over M_u}, \qquad a_2={e\over 2\sqrt{6}}{1\over M_u},
\end{eqnarray}
whereas $a'_1$ is connected to those of heavy quarks. Explicitly, $a'_1$ is fixed by heavy quark symmetry to be
\begin{eqnarray}
a'_1={e\over 12}{1\over M_Q}.
\end{eqnarray}

Within the framework of HHChPT, the authors of \cite{Jiang} proceeded to construct chiral Lagrangians at the level ${\cal O}(p^2)$ and ${\cal O}(p^3)$ and then calculated the electromagnetic decay amplitudes of charmed baryons up to ${\cal O}(p^3)$. It is not clear if their ${\cal O}(p^2)$ Lagrangian (see Eq. (12) of \cite{Jiang}) characterized by the four couplings $f_2,f_3,\tilde{f}_3$ and $f_4$ are equivalent to the first two terms of the ${\cal O}(p)$ Lagrangian given by Eq. (\ref{eq:emL}). The unknown couplings there were also estimated using the quark model.

The general amplitudes of electromagnetic decays are given by
\cite{Cheng93}
 \begin{eqnarray}
 A(\B_6\to \B_{\bar 3}+\gamma) &=& i\eta_1\bar u_{\bar
3}\sigma_{\mu\nu}k^\mu \varepsilon^\nu u_6,   \nonumber \\
A(\B^*_6\to \B_{\bar 3}+\gamma) &=&
i\eta_2\epsilon_{\mu\nu\alpha\beta}
\bar u_{\bar 3}\gamma^\nu k^\alpha\varepsilon^\beta u^\mu, \nonumber \\
A(\B^*_6\to \B_6+\gamma) &=& i\eta_3\epsilon_{\mu\nu\alpha\beta}
\bar u_6\gamma^\nu k^\alpha\varepsilon^\beta u^\mu.
 \end{eqnarray}
The corresponding decay rates are
\cite{Cheng93}
 \begin{eqnarray}
\Gamma(\B_6\to \B_{\bar 3}+\gamma) &=& \eta_1^2\,{k^3\over \pi},  \nonumber \\
\Gamma(\B_6^*\to \B_{\bar 3}+\gamma) &=& \eta_2^2\,{k^3\over
3\pi}\,{ 3m_i^2+m_f^2\over 4m_i^2},   \nonumber \\
\Gamma(\B_6^*\to \B_6+\gamma) &=& \eta_3^2\,{k^3\over 3\pi}\,{3m_i^2
+m_f^2\over 4m_i^2},
 \end{eqnarray}
where $m_i$ ($m_f$) is the mass of the parent (daughter) baryon.
The coupling constants $\eta_i$ can be calculated using the quark
model for $a_1$, $a_2$ and $a'_1$ \cite{Cheng93,Cheng97}:
 \begin{eqnarray} \label{eq:eta}
\eta_1(\Sigma_c^+\to\Lambda_c^+)=\,{e\over 6\sqrt{3}}\left({2\over
M_u}+{1 \over M_d}\right),&&
\eta_1(\Xi_c^{'+}\to\Xi_c^+)=\,{e\over 6\sqrt{3}}
\left({2\over M_u}+{1\over M_s}\right),   \nonumber  \\
\eta_1(\Xi_c^{'0}\to\Xi_c^0)=\,{e\over 6\sqrt{3}}\left({1\over
M_s}-{1 \over M_d}\right),&&
\eta_2(\Sigma_c^{*+}\to\Lambda_c^+)=\,{e\over 3\sqrt{6}}
\left({2\over M_u}+{1\over M_d}\right),   \nonumber  \\
\eta_2(\Xi_c^{'*+}\to\Xi_c^+)=\,{e\over 3\sqrt{6}}
\left({2\over M_u}+{1\over M_s}\right), && \eta_2(\Xi_c^{'*0}\to\Xi_c^0)=\,{e\over 3\sqrt{6}}\left(-{1\over
M_d}+{1\over M_s}\right),
  \\
\eta_3(\Sigma_c^{*++}\to\Sigma_c^{++})=\,{2\sqrt{2}e\over
9}\left({1\over M_u} -{1\over
M_c}\right),&&\eta_3(\Sigma_c^{*0}\to\Sigma_c^0)=\,{2\sqrt{2}e\over
9}\left(-{1\over 2M_d}-{1\over M_c}\right),   \nonumber  \\
\eta_3(\Sigma_c^{*+}\to\Sigma_c^+)=\,{\sqrt{2}e\over
9}\left({1\over M_u} -{1\over 2M_d}-{2\over M_c}\right),
&&
\eta_3(\Omega_c^{*0}\to\Omega_c^0)=\,{2\sqrt{2}e\over
9}\left(-{1\over 2M_s}-{1\over M_c}\right), \nonumber \\
\eta_3(\Xi_c^{'*+}\to\Xi_c^{'+})=\,{\sqrt{2}e\over
9}\left({1\over M_u} -{1\over 2M_s}-{2\over M_c}\right), && \eta_3(\Xi_c^{'*0}\to\Xi_c^{'0})=\,{\sqrt{2}e\over
9}\left(-{1\over 2M_d}-{1\over 2M_s}-{2\over M_c}\right). \nonumber
\end{eqnarray}
Using the constituent quark masses, $M_u=338$ MeV, $M_d=322$ MeV, $M_s=510$ MeV \cite{PDG}, and $M_c=1.6$ GeV, the calculated results are
summarized in the second column of Table \ref{tab:em}. Some other model predictions are also listed there for comparison.

Radiative decays of $s$-wave charmed baryons are considered in \cite{Dey} in the quark model with predictions similar to ours.
A similar
procedure is followed in \cite{Tawfiq01} where the heavy quark
symmetry is supplemented with light-diquark symmetries to
calculate the widths of $\Sigma_c^+\to\Lambda_c^+\gamma$ and
$\Sigma_c^*\to\Sigma_c\gamma$. The authors of \cite{Ivanov}
apply the relativistic quark model to predict various
electromagnetic decays of charmed baryons. Besides the magnetic
dipole (M1) transition, the author of \cite{Savage95} also
considered and estimated the electric quadrupole (E2) amplitude
for $\Sigma_c^{*+}\to\Lambda_c^+\gamma$ arising from the chiral
loop correction. A detailed analysis of the E2
contributions was presented in \cite{Pich}. The E2 amplitudes
appear at different higher orders for the three kinds of decays:
${\cal O}(1/\Lambda_\chi^2)$ for $\B^*_6\to \B_6+\gamma$, ${\cal
O}(1/m_Q\Lambda_\chi^2)$ for $\B^*_6\to \B_{\bar 3}+\gamma$ and
${\cal O}(1/m_Q^3\Lambda_\chi^2)$ for $\B_6\to \B_{\bar 3}+\gamma$.
Therefore, the E2 contribution to $\B_6\to \B_{\bar 3}+\gamma$ is
completely negligible. The electromagnetic decays were calculated in \cite{Aliev,ZGWang} using the QCD sum rule method, while they were studied within the framework of the modified bag model in \cite{Bernotas}.

{\squeezetable
\begin{table}[t]
\caption{Electromagnetic decay rates (in units of keV) of $s$-wave charmed
baryons. Among the four different results listed in \cite{Dey} and \cite{Majethiya}, we quote those denoted by $\Gamma_\gamma^{(0)}$ and ``Present (ecqm)", respectively.  } \label{tab:em}
\begin{center}
\begin{tabular}{|l|c|c|c|c|c|c|c| c|c|c|}
\hline\hline ~~~~Decay & HHChPT & HHChPT & Dey & Ivanov & Tawfiq
 & Ba\~nuls  & Aliev & Wang & Bernotas & Majethiya  \\
 & \cite{Cheng97,Cheng93}  & \cite{Jiang} & et al. \cite{Dey} & et al. \cite{Ivanov}  & et al. \cite{Tawfiq01} & et al. \cite{Pich} & et al. \cite{Aliev} &  \cite{ZGWang} & et al. \cite{Bernotas} & et al. \cite{Majethiya}
   \\
\hline
 $\Sigma^+_c\to \Lambda_c^+\gamma$ & 91.5 & 164.16 & 120 & $60.7\pm1.5$  & 87 & & & & 46.1 & 60.55  \\ \hline
 $\Sigma_c^{*+}\to\Lambda_c^+\gamma$ & 150.3 & 892.97 & 310 & $151\pm4$ & & &$130\pm45$ & & 126 & 154.48  \\ \hline
 $\Sigma_c^{*++}\to\Sigma_c^{++}\gamma$ & 1.3  & 11.60 & 1.6 & & 3.04 &  & $2.65\pm1.20$ & $6.36^{+6.79}_{-3.31}$ & 0.826 & 1.15  \\ \hline
 $\Sigma_c^{*+}\to\Sigma_c^+\gamma$ & 0.002 & 0.85 & 0.001 & $0.14\pm0.004$ & 0.19 & & $0.40\pm0.16$ & $0.40^{+0.43}_{-0.21}$ & 0.004 & $<10^{-4}$  \\ \hline
 $\Sigma_c^{*0}\to\Sigma_c^0\gamma$ & 1.2 &  2.92 & 1.2 & $$ & 0.76 &  & $0.08\pm0.03$ & $1.58^{+1.68}_{-0.82}$ & 1.08 & 1.12  \\ \hline
 $\Xi'^+_c\to\Xi_c^+\gamma$ & 19.7  & 54.31 & 14 & $12.7\pm1.5$ & & & & & 10.2 &  \\ \hline
 $\Xi'^{*+}_c\to\Xi_c^+\gamma$ & 63.5  & 502.11 & 71 & $54\pm3$ & & & $52\pm25$ & & 44.3 & 63.32  \\ \hline
 $\Xi'^{*+}_c\to\Xi_c^{'+}\gamma$ & 0.06  & 1.10 & 0.10 & & & & & $0.96^{+1.47}_{-0.67}$ & 0.011 & \\ \hline
 $\Xi'^0_c\to\Xi_c^0\gamma$ & 0.4  & 0.02 &  0.33 & $0.17\pm0.02$ & & $1.2\pm0.7$  & & & 0.0015 & \\ \hline
 $\Xi'^{*0}_c\to\Xi_c^0\gamma$ & 1.1 &   0.36 & 1.7 & $0.68\pm0.04$ & & $5.1\pm2.7$ & $0.66\pm0.32$ & & 0.908 & 0.30  \\ \hline
 $\Xi'^{*0}_c\to\Xi_c^{'0}\gamma$ & 1.0 &   3.83 & 1.6 & &  &  &  & $1.26^{+0.80}_{-0.46}$ & 1.03 & \\ \hline
 $\Omega_c^{*0}\to\Omega_c^0\gamma$ & 0.9  & 4.82 & 0.71 & & &  &  & $1.16^{+1.12}_{-0.54}$  & 1.07& 2.02 \\ \hline
 \hline
\end{tabular}
\end{center}
\end{table}
}

\begin{table}[t]
\caption{Electromagnetic decay rates (in units of keV) of $p$-wave charmed
baryons.} \label{tab:empwave}
\begin{center}
\begin{tabular}{|l|c c c c c c|}
\hline\hline ~~~~Decay &  ~~~Ivanov~~~ & ~~Tawfiq~~ & Aziza Baccouche & ~~Zhu~~ & ~~Chow~~  & Gamermann  \\
 & et al. \cite{Ivanov}  & et al. \cite{Tawfiq01}  & et al. \cite{Cohen}
 & \cite{Zhu:2000} & \cite{Chow} & et al. \cite{Gamermann}  \\
\hline
 $1/2^-\to 1/2^+(3/2^+)\gamma$ & & & & & & \\ \hline
 $\Lambda_c(2595)^+\to\Lambda_c^+\gamma$ & $115\pm1$ & & 25 & 36 & 16 & $274\pm52$ \\
 $\Lambda_c(2595)^+\to\Sigma_c^+\gamma$ & $77\pm1$ & 71 & & 11 & & $2.1\pm0.4$  \\
 $\Lambda_c(2595)^+\to\Sigma_c^{*+}\gamma$ & $6\pm0.1$ & 11 & & 1 & $$ &  \\ \hline
 $3/2^-\to 1/2^+(3/2^+)\gamma$ & & & & & &  \\ \hline
 $\Lambda_c(2625)^+\to\Lambda_c^+\gamma$ & $151\pm2$ & & & 48 & 21 & \\
 $\Lambda_c(2625)^+\to\Sigma_c^+\gamma$ & $35\pm0.5$ & 130 & & 5 & & \\
 $\Lambda_c(2625)^+\to\Sigma_c^{*+}\gamma$ & $46\pm0.6$ & 32 & & 6 & & \\
 $\Xi_c(2815)^+\to\Xi_c^+\gamma$ & $190\pm5$ & & & & & \\
 $\Xi_c(2815)^0\to\Xi_c^0\gamma$ & $497\pm14$ & & & & & \\ \hline
 \hline
\end{tabular}
\end{center}
\end{table}

It is evident from Table \ref{tab:em} that the predictions in \cite{Cheng97,Cheng93} and \cite{Jiang} all based on HHChPT are quite different for the following three modes:  $\Sigma_c^{*++}\to \Sigma_c^{++}\gamma$, $\Sigma_c^{*+}\to\Lambda_c^+\gamma$ and $\Xi'^{*+}_c\to\Xi_c^+\gamma$.
Indeed, the results for the last two modes in \cite{Jiang} are larger than all other existing predictions by one order of magnitude! It is naively expected that all HHChPT approaches should agree with each other to the lowest order of chiral expansion provided that the coefficients are inferred from the nonrelativistic quark model. The lowest order predictions $\Gamma(\Sigma_c^{*+}\to\Lambda_c^+\gamma)=756$ keV and  $\Gamma(\Xi'^{*+}_c\to\Xi_c^+\gamma)=403$ keV obtained in \cite{Jiang} are still very large. Note that a recent lattice calculation in  \cite{Bahtiyar} yields $\Gamma(\Omega_c^*\to \Omega_c\gamma)=0.074\pm0.008$ keV which is much smaller than $\Gamma(\Omega_c^*)=4.82$ keV predicted in \cite{Jiang}.

Chiral-loop corrections to the M1 electromagnetic decays and to
the strong decays of heavy baryons have been computed at the one
loop order in \cite{ChengSU3}. The leading chiral-loop effects we
found are nonanalytic in the forms of $m/ \Lambda_\chi$ and
$(m^2/\Lambda^2_\chi)\ln(\Lambda^2/ m^2)$ (or $m_q^{1/2}$ and
$m_q\ln m_q$, with $m_q$ being the light quark mass). Some results
are \cite{ChengSU3}
 \begin{eqnarray}
\Gamma(\Sigma_c^+\to\Lambda_c^+\gamma)= \,112\,{\rm keV}, \quad
\Gamma({\Xi'}_c^+\to\Xi_c^+\gamma)= \,29\,{\rm keV}, \quad
\Gamma({\Xi'}_c^0\to\Xi_c^0\gamma)= \,0.15\,{\rm keV},
 \end{eqnarray}
which should be compared with the corresponding quark-model
results: 92 keV, 20 keV and 0.4 keV (Table \ref{tab:em}).

The electromagnetic decays $\Xi'^{*0}_c\to\Xi_c^0\gamma$ and $\Xi'^{0}_c\to\Xi_c^0\gamma$ are of
special interest. It has been advocated in \cite{Lu} that a
measurement of their branching fractions will allow us to determine one of the
coupling constants in HHChPT, namely, $g_1$. They are forbidden at tree level in SU(3)
limit [see Eq. (\ref{eq:eta})]. In heavy baryon chiral
perturbation theory, this radiative decay is induced via chiral
loops where SU(3) symmetry is broken by the light current quark
masses. By identifying the chiral loop contribution to
$\Xi'^{*0}_c\to\Xi_c^0\gamma$ with the quark model prediction
given in Eq. (\ref{eq:eta}), it was found in \cite{Cheng97} that
one of the two possible solutions is in accord with the quark
model expectation for $g_1$.

For the electromagnetic decays of $p$-wave charmed baryons, the
search for $\Lambda_c(2593)^+\to\Lambda_c^+\gamma$ and
$\Lambda_c(2625)^+\to\Lambda_c^+\gamma$ has been failed so far.
On the theoretical side, the interested reader is referred to
\cite{Tawfiq01,Ivanov,Lu,Zhu:2000,Cho,Cohen,Chow,Gamermann,Dong:rad} for more details. Some predictions are collected in Table \ref{tab:empwave} and they are more diversified than the $s$-wave case. For the electromagnetic decays of doubly charmed baryons, see e.g. \cite{Branz,Bernotas}.

The electromagnetic decays considered so far do not test
critically the heavy quark symmetry nor the chiral symmetry.  The
results follow simply from the quark model.  There are examples in
which both the heavy quark symmetry and the chiral symmetry enter
in a crucial way. These are the radiative decays of heavy baryons
involving an emitted pion. Some examples which are kinematically
allowed are
 \begin{eqnarray}
\Sigma_c \rightarrow \Lambda_c \pi \gamma,~~ \Sigma^\ast_c
\rightarrow \Lambda_c \pi \gamma,~~ \Sigma^\ast_c \rightarrow
\Sigma_c \pi \gamma,~~  \Xi^\ast_c \rightarrow \Xi_c \pi \gamma.
 \end{eqnarray}
It turns out that the contact interaction dictated by the Lagrangian  ${\cal L}_B^{(1)}$ can be nicely tested by the decay $\Sigma_c^0\to\Lambda_c^+\pi^-\gamma$, whereas a test on the chiral structure of ${\cal L}_B^{(2)}$ is provided by the process $\Sigma_c^+\to\Lambda_c^+\pi^0\gamma$; see \cite{Cheng93} for the analysis.

\subsection{Weak radiative decays}

At the quark level, there are three different types of processes
which can contribute to the weak radiative decays of heavy
hadrons, namely, single-, two- and three-quark transitions
\cite{Kamal82}. The single-quark transition mechanism comes from
the so-called electromagnetic penguin diagram. Unfortunately, the
penguin process $c\to u\gamma$ is very suppressed and hence it
plays no role in charmed hadron radiative decays. There are two
contributions from the two-quark transitions: one from the
$W$-exchange diagram accompanied by a photon emission from the
external quark, and the other from the same $W$-exchange diagram
but with a photon radiated from the $W$ boson. The latter is
typically suppressed by a factor of $m_qk/M_W^2$ ($k$ being the
photon energy) as compared to the former bremsstrahlung process
\cite{Kamal83}. For charmed baryons, the Cabibbo-allowed decay
modes via $c\bar{u}\to s\bar{d}\gamma$ (Fig. \ref{fig:rad}) or
$cd\to us\gamma$ are
 \begin{eqnarray} \label{eq:rad}
  \Lambda_c^+\to\Sigma^+\gamma,\qquad \Xi_c^0\to\Xi^0\gamma.
 \end{eqnarray}
Finally, the three-quark transition involving $W$-exchange between
two quarks and a photon emission by the third quark is quite
suppressed because of very small probability of finding three
quarks in adequate kinematic matching with the baryons
\cite{Kamal82,Hua}.

\begin{figure}[t]
\centerline{\psfig{file=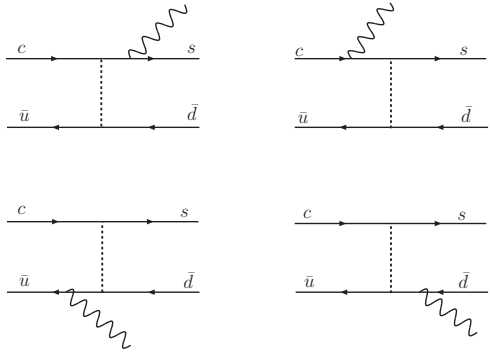,width=2.6in}}
 \caption{$W$-exchange diagrams contributing to the quark-quark
bremsstrahlung process $c+\bar{u}\to s+\bar{d}+\gamma$. The
$W$-annihilation type diagrams are not shown here.}
\label{fig:rad}
\end{figure}

The general amplitude of the weak radiative baryon decay reads
 \begin{eqnarray}
 A({\cal B}_i\to {\cal B}_f\gamma)=\,i\bar{u}_f(a+b\gamma_5)\sigma_{\mu\nu}\varepsilon^\mu
k^ \nu u_i,
 \end{eqnarray}
where $a$ and $b$ are parity-conserving and -violating amplitudes,
respectively. The corresponding decay rate is
 \begin{eqnarray}
 \Gamma({\cal B}_i\to {\cal B}_f\gamma)=\,{1\over 8\pi}\left( {m^2_i-m^2_f\over
m_i}\right) ^3(|a|^2+|b|^2).
 \end{eqnarray}

Nonpenguin weak radiative decays of charmed baryons such as those
in (\ref{eq:rad}) are characterized by emission of a hard photon
and the presence of a highly virtual intermediate quark between
the electromagnetic and weak vertices. It has been shown in
\cite{Chengweakrad} that these features should make possible to
analyze these processes by perturbative QCD; that is, these
processes are describable by an effective local and gauge
invariant Lagrangian:
 \begin{eqnarray}
 {\cal H}_{\rm eff}(c\bar{u}\to s\bar{d}\gamma)=\,{G_F\over
2\sqrt{2}}V_{cs} V_{ud}^*(c_+{O}^F_++c_-{O}^F_-),
 \end{eqnarray}
with
 \begin{eqnarray}
{O}^F_\pm(c\bar{u}\to s\bar{d}\gamma)=\,{e\over m_i^2-m_f^2}
&\Bigg\{& \left( e_s{m_f\over m_s}+e_u{m_i\over
m_u}\right)\left(\tilde{F}_{\mu\nu}+iF_{\mu\nu}
\right)O_\pm^{\mu\nu}   \\
&-&\left(e_d{m_f\over m_d}+e_c{m_i\over
m_c}\right)\left(\tilde{F}_{\mu\nu}
-iF_{\mu\nu}\right)O_\mp^{\mu\nu}\Bigg\},
 \end{eqnarray}
where $m_i=m_c+m_u$, $m_f=m_s+m_d$, $\tilde{F}_{\mu\nu}\equiv
{1\over 2}\epsilon_{\mu\nu\alpha\beta}F^{ \alpha\beta}$ and
  \begin{eqnarray}
 O_\pm^{\mu\nu}=\,\bar{s}\gamma^\mu(1-\gamma_5)c\bar{u}\gamma^\nu(1-\gamma_5)d
\pm \bar{s}\gamma^\mu(1-\gamma_5)d\bar{u}\gamma^\nu(1-\gamma_5)c.
 \end{eqnarray}

For the charmed baryon radiative decays, one needs to evaluate the
matrix element $\langle {\cal B}_f|O_\pm^{\mu\nu}|{\cal
B}_i\rangle$.  Since the quark-model wave functions best resemble
the hadronic states in the frame where both baryons are static,
the static MIT bag model was thus adopted in
\cite{Chengweakrad} for the calculation. The predictions are
\footnote{The branching fraction of $\Xi_c^0\to\Xi^0\gamma$ has been updated using the current lifetime of $\Xi_c^0$.}
 \begin{eqnarray}
 && {\cal B}(\Lambda_c^+\to
\Sigma^+\gamma)=\,4.9\times 10^{-5},\qquad \alpha
(\Lambda_c^+\to\Sigma^+\gamma)=-0.86\,,   \nonumber \\
&& {\cal B}(\Xi_c^0\to\Xi^0\gamma)=3.5\times 10^{-5}, \qquad~~
\alpha(\Xi_c^0 \to\Xi^0\gamma)=-0.86\,.
 \end{eqnarray}
A different analysis of the same decays was carried out in
\cite{Uppalrad} with the results
 \begin{eqnarray}
&& {\cal B}(\Lambda_c^+\to \Sigma^+\gamma)=\,2.8\times
10^{-4},~~~~\alpha (\Lambda_c^+\to\Sigma^+\gamma)=0.02\,,
\nonumber \\ && {\cal B}(\Xi_c^0\to\Xi^0\gamma)=1.5\times
10^{-4}, \qquad \alpha(\Xi_c^0 \to\Xi^0\gamma)=-0.01\,.
 \end{eqnarray}
Evidently, these predictions (especially the decay asymmetry) are
very different from the ones obtained in \cite{Chengweakrad}.

Finally, it is worth remarking that, in analog to the
heavy-flavor-conserving nonleptonic weak decays as discussed in
Sec. VI.C, there is a special class of weak radiative decays in
which heavy flavor is conserved, for example, $\Xi_c \to
\Lambda_c \gamma$ and $\Omega_c \to \Xi_c \gamma$.  In these
decays, weak radiative transitions arise from the light quark sector
of the heavy baryon whereas the heavy quark behaves as a
spectator. However, the dynamics of these radiative decays is more
complicated than their counterpart in nonleptonic weak decays,
e.g., $\Xi_c \to \Lambda_c \pi$. In any event, it deserves an investigation.

\section{Conclusions}
In this report
we began with a brief overview of the spectroscopy of charmed baryons and discussed their possible structure and spin-parity assignments in the quark model. For the $p$-wave baryons,
We have assigned $\Sigma_{c2}({3\over 2}^-)$ to $\Sigma_c(2800)$. As for first positive-parity excitations,
with the help of the relativistic quark-diquark model and the $^3P_0$ model, we have identified $\tilde\Lambda_{c3}^2({5\over 2}^+)$ with $\Lambda_c(2800)$, $\tilde\Xi_{c}'({1\over 2}^+)$ with $\Xi_c(2980)$, and $\tilde\Xi_{c3}^2({5\over 2}^+)$ with $\Xi_c(3080)$, though the first and last assignments may encounter some potential problems.

It should be stressed that the mass analysis alone is usually not adequate to
pin down the spin-parity quantum numbers of higher excited charmed baryon states, a study of their strong decays is necessary. for example, $\Sigma_{c0}({1\over 2}^-)$,  $\Sigma_{c1}({1\over 2}^-,{3\over 2}^-)$ and $\Sigma_{c2}({3\over 2}^-,{5\over 2}^-)$ for $\Sigma_c(2800)$ all have similar masses. The analysis of strong decays allows us to exclude the first two possibilities. It should be stressed that the Regge phenomenology and the mass relations for antitriplet and sextet multiplets also provide very useful guidance for the spin-parity quantum numbers.

Based on various theoretical tools such as lattice QCD and the QCD sum rule method, there are a lot of theoretical activities on charmed baryon spectroscopy, especially for doubly and triply charmed baryons. However, progress in the hadronic decays, radiative decays and lifetimes has been very slow. Experimentally, nearly all the branching fractions of the
$\Lambda_c^+$ are measured relative to the $pK^-\pi^+$ mode. The recent measurements ${\cal B}(\Lambda_c^+\to pK^-\pi^+)=(6.84\pm0.24^{+0.21}_{-0.27})\%$ by Belle and
$(5.77\pm0.27)\%$ (statistical error only) by BESIII are very encouraging. Moreover, BESIII has recently measured 2-body, 3-body and 4-body nonleptonic decay modes of $\Lambda_c^+$ with precision significantly improved. It is conceivable that many new data emerged from LHCb and BESIII in the immediate future and from the experiments at J-PARC and ${\rm \bar P}$ANDA in the future can be used to test the underlying mechanism for hadronic weak decays.

\section{Acknowledgments}

I am grateful to Hai-Bo Li for inviting me to attend the International Workshop on Physics at Future High Intensity Collider @ 2-7 GeV in China held at University of Science and Technology of China on January 13-16, 2015.
He urged me to write up the review on charmed baryons which will appear in the special issue of {\sl Frontiers of Physics}, ``Potential Physics at a Super Tau-Charm Factory". This research was supported in part by the Ministry of Science and Technology of R.O.C. under Grant No. 104-2112-M-001-022.

\newcommand{\bi}{\bibitem}


\begin{thebibliography}{199}

\bibitem{Cheng:2009}
  H.~Y.~Cheng,
  Int.\ J.\ Mod.\ Phys.\ A {\bf 24}, Suppl. 1 (2009) 593
  [arXiv:0809.1869 [hep-ex]], Chap. 24\,.

\bi{Korner94} J. G. K\"orner, M. Kr\"amer, and D. Pirjol, Prog.
Part. Nucl. Phys. {\bf 33}, 787 (1994) [hep-ph/9406359].

\bi{Bigireview} S. Bianco, F.L. Fabbri, D. Benson, and I.I. Bigi,
Riv. Nuovo Cim. {\bf 26}, \#7-8 (2003) [hep-ex/0309021].

\bibitem{Roberts}
  W.~Roberts and M.~Pervin,
  Int.\ J.\ Mod.\ Phys.\ A {\bf 23}, 2817 (2008)
  [arXiv:0711.2492 [nucl-th]].

\bibitem{Klempt}
  E.~Klempt and J.~M.~Richard,
  Rev.\ Mod.\ Phys.\  {\bf 82}, 1095 (2010)
  [arXiv:0901.2055 [hep-ph]].

\bibitem{Crede}
  V.~Crede and W.~Roberts,
  Rept.\ Prog.\ Phys.\  {\bf 76}, 076301 (2013)
  [arXiv:1302.7299 [nucl-ex]].

\bibitem{Bphysics}
  A.~J.~Bevan {\it et al.}  [BaBar and Belle Collaborations],
  Eur.\ Phys.\ J.\ C {\bf 74}, no. 11, 3026 (2014)
  [arXiv:1406.6311 [hep-ex]], Chap. 19.4.

\bibitem{PDG}
K.A. Olive {\it et al.} [Particle Data Group], Chin. Phys. C {\bf 38}, 090001 (2014).

\bi{Pirjol}
  D.~Pirjol and T.~M.~Yan,
  Phys.\ Rev.\ D {\bf 56}, 5483 (1997)
  [hep-ph/9701291].


\bibitem{CC}
  H.~Y.~Cheng and C.~K.~Chua,
  Phys.\ Rev.\ D {\bf 75}, 014006 (2007)
  [hep-ph/0610283].

\bibitem{Belle:2014}
  S.~H.~Lee {\it et al.}  [Belle Collaboration],
  Phys.\ Rev.\ D {\bf 89}, no. 9, 091102 (2014)
  [arXiv:1404.5389 [hep-ex]].

\bibitem{Belle:dc}
  Y.~Kato {\it et al.}  [Belle Collaboration],
  Phys.\ Rev.\ D {\bf 89},  052003 (2014)
  [arXiv:1312.1026 [hep-ex]].

\bibitem{Kato}
  Y.~Kato [Belle Collaboration],
  PoS DIS {\bf 2014}, 195 (2014).

\bi{Cho} P. Cho, Phys. Rev. D {\bf 50}, 3295 (1994).

\bi{CLEO:Lamc2880} M. Artuso {\it et al.} [CLEO collaboration],
Phys. Rev. Lett. {\bf 86}, 4479 (2001).

\bi{Oh} Y. Oh and B.Y. Park, Phys. Rev. D {\bf 53}, 1605 (1996).

\bibitem{Capstick} S. Capstick and N. Isgur, Phys. Rev.D {\bf 34}, 2809
(1986); L.A. Copley, N. Isgur, and G. Karl, Phys. Rev. D {\bf 20},
768 (1979).

\bibitem{Ebert:2007}
  D.~Ebert, R.~N.~Faustov and V.~O.~Galkin,
  Phys.\ Lett.\ B {\bf 659}, 612 (2008)
  [arXiv:0705.2957 [hep-ph]].

\bi{BaBar:Lamc2940}
  B.~Aubert {\it et al.}  [BaBar Collaboration],
  Phys.\ Rev.\ Lett.\  {\bf 98}, 012001 (2007)
  [hep-ex/0603052].

\bi{Belle:Lamc2880}
  K.~Abe {\it et al.}  [Belle Collaboration],
  Phys.\ Rev.\ Lett.\  {\bf 98}, 262001 (2007)
  [hep-ex/0608043].

\bibitem{Zhu}
  C.~Chen, X.~L.~Chen, X.~Liu, W.~Z.~Deng and S.~L.~Zhu,
  Phys.\ Rev.\ D {\bf 75}, 094017 (2007)
  [arXiv:0704.0075 [hep-ph]].

\bibitem{Zhong}
  X.~H.~Zhong and Q.~Zhao,
  Phys.\ Rev.\ D {\bf 77}, 074008 (2008)
  [arXiv:0711.4645 [hep-ph]].

\bi{Selem} A. Selem, {\it A Diquark Interpretation of the
Structure and Energies of Hadrons}, Senior thesis, M.I.T. (2005);
A. Selem and F. Wilczek, hep-ph/0602128.

\bibitem{Ebert:2011}
  D.~Ebert, R.~N.~Faustov and V.~O.~Galkin,
  Phys.\ Rev.\ D {\bf 84}, 014025 (2011)
  [arXiv:1105.0583 [hep-ph]].

\bibitem{He:2006is}
  X.~G.~He, X.~Q.~Li, X.~Liu and X.~Q.~Zeng,
  Eur.\ Phys.\ J.\ C {\bf 51}, 883 (2007)
  [hep-ph/0606015].

\bibitem{Garcilazo} H. Garcilazo, J. Vijande, and A. Valcarce, J.
Phys. G {\bf 34}, 961 (2007).

\bi{Belle:Sigc2800}  R. Mizuk {\it et al.} [Belle Collaboration],
Phys. Rev. Lett. {\bf 94}, 122002 (2005).

\bi{Belle:Xic2980}
  R.~Chistov {\it et al.}  [Belle Collaboration],
  Phys.\ Rev.\ Lett.\  {\bf 97}, 162001 (2006)
  [hep-ex/0606051].

\bi{BaBar:Xic2980}
   B.~Aubert {\it et al.}  [BaBar Collaboration],
  Phys.\ Rev.\ D {\bf 77}, 012002 (2008)
  [arXiv:0710.5763 [hep-ex]].

\bibitem{CC:2015}
  H.~Y.~Cheng and C.~K.~Chua,
  arXiv:1508.05653 [hep-ph].

\bibitem{BaBar:Xic2930}
  B.~Aubert {\it et al.} [BaBar Collaboration],
  Phys.\ Rev.\ D {\bf 77}, 031101 (2008)
  [arXiv:0710.5775 [hep-ex]].

\bibitem{Guo2008}
  X.~H.~Guo, K.~W.~Wei and X.~H.~Wu,
  Phys.\ Rev.\ D {\bf 78}, 056005 (2008)
  [arXiv:0809.1702 [hep-ph]].

\bibitem{Chen:2014}
  B.~Chen, K.~W.~Wei and A.~Zhang,
  Eur.\ Phys.\ J.\ A {\bf 51}, 82 (2015)
  [arXiv:1406.6561 [hep-ph]].

\bibitem{Chen:2015k}
  H.~X.~Chen, W.~Chen, Q.~Mao, A.~Hosaka, X.~Liu and S.~L.~Zhu,
  Phys.\ Rev.\ D {\bf 91}, no. 5, 054034 (2015)
  [arXiv:1502.01103 [hep-ph]].

\bibitem{Ger}
  S.~M.~Gerasyuta and E.~E.~Matskevich,
  Int.\ J.\ Mod.\ Phys.\ E {\bf 17}, 585 (2008)
  [arXiv:0709.0397 [hep-ph]].

\bibitem{Liu:2012}
  L.~H.~Liu, L.~Y.~Xiao and X.~H.~Zhong,
  Phys.\ Rev.\ D {\bf 86}, 034024 (2012)
  [arXiv:1205.2943 [hep-ph]].

\bi{BaBar:Omegacst}  B.~Aubert {\it et al.}  [BaBar Collaboration],
  Phys.\ Rev.\ Lett.\  {\bf 97}, 232001 (2006)
  [hep-ex/0608055].

\bibitem{Dong:2010}
  Y.~B. Dong, A.~Faessler, T.~Gutsche and V.~E.~Lyubovitskij,
  Phys.\ Rev.\ D {\bf 81}, 074011 (2010)
  [arXiv:1002.0218 [hep-ph]].

\bibitem{He:2010}
  J.~He and X.~Liu,
  Phys.\ Rev.\ D {\bf 82}, 114029 (2010)
  [arXiv:1008.1500 [hep-ph]].

\bibitem{Yamaguchi:2013}
  Y.~Yamaguchi, S.~Ohkoda, S.~Yasui and A.~Hosaka,
  Phys.\ Rev.\ D {\bf 87},  074019 (2013)
  [arXiv:1301.4557 [hep-ph]].

\bibitem{Dong:2014}
  Y.~B. Dong, A.~Faessler, T.~Gutsche and V.~E.~Lyubovitskij,
  Phys.\ Rev.\ D {\bf 90}, no. 9, 094001 (2014)
  [arXiv:1407.3949 [hep-ph]].

\bibitem{Zhang:2014}
  J.~R.~Zhang,
  Phys.\ Rev.\ D {\bf 89}, no. 9, 096006 (2014)
  [arXiv:1212.5325 [hep-ph]].

\bibitem{Carames}
  T.~F.~Carames and A.~Valcarce,
  Phys.\ Rev.\ D {\bf 90}, no. 1, 014042 (2014)
  [arXiv:1507.08046 [hep-ph]].


\bibitem{Selex02}
M.~Mattson {\it et al.}  [SELEX Collaboration],
  Phys.\ Rev.\ Lett.\  {\bf 89}, 112001 (2002)
  [hep-ex/0208014].

\bi{Selex04}
 A.~Ocherashvili {\it et al.}  [SELEX Collaboration],
  Phys.\ Lett.\ B {\bf 628}, 18 (2005)
  [hep-ex/0406033].

\bi{FOCUS:dc} S. P. Ratti [FOCUS Collaboration], Nucl. Phys. Proc. Suppl. 115, 33 (2003).

\bi{BaBar:dc}
  B.~Aubert {\it et al.}  [BaBar Collaboration],
  Phys.\ Rev.\ D {\bf 74}, 011103 (2006)
  [hep-ex/0605075].

\bibitem{LHCb:dc}
  R.~Aaij {\it et al.}  [LHCb Collaboration],
  JHEP {\bf 1312}, 090 (2013)
  [arXiv:1310.2538 [hep-ex]].

\bibitem{Guo}
  K.~W.~Wei, B.~Chen and X.~H.~Guo,
  arXiv:1503.05184 [hep-ph].

\bibitem{Karliner:2014}
  M.~Karliner and J.~L.~Rosner,
  Phys.\ Rev.\ D {\bf 90}, no. 9, 094007 (2014)
  [arXiv:1408.5877 [hep-ph]].

\bibitem{Wang:dc}
  Z.~G.~Wang,
  Eur.\ Phys.\ J.\ A {\bf 47}, 81 (2011)
  [arXiv:1003.2838 [hep-ph]];
  Eur.\ Phys.\ J.\ C {\bf 68}, 459 (2010)
  [arXiv:1002.2471 [hep-ph]];
  Eur.\ Phys.\ J.\ A {\bf 45}, 267 (2010)
  [arXiv:1001.4693 [hep-ph]].
  
\bibitem{Sun}
  Z.~F.~Sun, Z.~W.~Liu, X.~Liu and S.~L.~Zhu,
   Phys.\ Rev.\ D {\bf 91}, no. 9, 094030 (2015)
  [arXiv:1411.2117 [hep-ph]].

\bibitem{Bali:2015}
  P.~Perez-Rubio, S.~Collins and G. Bali,
  Phys.\ Rev.\ D {\bf 92}, no. 3, 034504 (2015)
  [arXiv:1503.08440 [hep-lat]].

\bibitem{HSC}
M.~Padmanath, R.~G.~Edwards, N.~Mathur and M.~Peardon,
  Phys.\ Rev.\ D {\bf 91}, no. 9, 094502 (2015)
  [arXiv:1502.01845 [hep-lat]].

\bibitem{Brown}
  Z.~S.~Brown, W.~Detmold, S.~Meinel and K.~Orginos,
  Phys.\ Rev.\ D {\bf 90}, no. 9, 094507 (2014)
  [arXiv:1409.0497 [hep-lat]].

\bibitem{ETMC}
  C.~Alexandrou, V.~Drach, K.~Jansen, C.~Kallidonis and G.~Koutsou [ETM Collaboration],
  Phys.\ Rev.\ D {\bf 90}, no. 7, 074501 (2014)
  [arXiv:1406.4310 [hep-lat]].

\bibitem{ILGTI}
  S.~Basak, S.~Datta, A.~T.~Lytle, M.~Padmanath, P.~Majumdar and N.~Mathur [ILGTI Collaboration],
  PoS LATTICE {\bf 2013}, 243 (2014)
  [arXiv:1312.3050 [hep-lat]].

\bibitem{PACS-CS}
  Y.~Namekawa {\it et al.}  [PACS-CS Collaboration],
  Phys.\ Rev.\ D {\bf 87}, no. 9, 094512 (2013)
  [arXiv:1301.4743 [hep-lat]].

\bibitem{Durr}
  S.~Durr, G.~Koutsou and T.~Lippert,
  Phys.\ Rev.\ D {\bf 86}, 114514 (2012)
  [arXiv:1208.6270 [hep-lat]].

\bibitem{Briceno}
  R.~A.~Briceno, H.~W.~Lin and D.~R.~Bolton,
  Phys.\ Rev.\ D {\bf 86}, 094504 (2012)
  [arXiv:1207.3536 [hep-lat]].

\bibitem{Liu}
  L.~Liu, H.~W.~Lin, K.~Orginos and A.~Walker-Loud,
  Phys.\ Rev.\ D {\bf 81}, 094505 (2010)
  [arXiv:0909.3294 [hep-lat]].

\bibitem{Na}
  H.~Na and S.~A.~Gottlieb,
  PoS LAT {\bf 2007}, 124 (2007)
  [arXiv:0710.1422 [hep-lat]].

\bi{Can}
  K. U. Can, G. Erkol, M. Oka, T. T. Takahashi,
  arXiv:1508.03048.

\bi{Yan}
  T.~M.~Yan, H.~Y.~Cheng, C.~Y.~Cheung, G.~L.~Lin, Y.~C.~Lin and H.~L.~Yu,
  Phys.\ Rev.\ D {\bf 46}, 1148 (1992)
  [Erratum-ibid.\ D {\bf 55}, 5851 (1997)].

\bi{Wise} M.B. Wise, Phys. Rev. D {\bf 45}, 2188 (1992); G.
Burdman and J. Donoghue, Phys. Lett. B {\bf 280}, 287 (1992).

\bibitem{Chiladze}
  G.~Chiladze and A.~F.~Falk,
  Phys.\ Rev.\ D {\bf 56}, 6738 (1997)
  [hep-ph/9707507].

\bi{Falk03}
  A.~E.~Blechman, A.~F.~Falk, D.~Pirjol and J.~M.~Yelton,
  Phys.\ Rev.\ D {\bf 67}, 074033 (2003)
  [hep-ph/0302040].

\bibitem{PDG2006} Y.M. Yao {\it et al.} [Particle Data Group], J. Phys. G
{\bf 33}, 1 (2006).


\bi{Tawfiq} S. Tawfiq, P.J. O'Donnell, and J.G. K\"orner, Phys.
Rev. D {\bf 58}, 054010 (1998).

\bi{Ivanov}
  M.~A.~Ivanov, J.~G.~Korner, V.~E.~Lyubovitskij and A.~G.~Rusetsky,
  Phys.\ Rev.\ D {\bf 60}, 094002 (1999)
  [hep-ph/9904421].

\bi{Huang95} M.Q. Huang, Y.B. Dai, and C.S. Huang, Phys. Rev. D
{\bf 52}, 3986 (1995); Phys. Rev. D {\bf 55}, 7317(E) (1997).

\bi{Albertus} C. Albertus, Hern\'andez, J. Nieves, and J.M.
Verde-Velasco, Phys. Rev. D {\bf 72}, 094022 (2005).


\bibitem{CDF:2595}
  T.~Aaltonen {\it et al.}  [CDF Collaboration],
  Phys.\ Rev.\ D {\bf 84}, 012003 (2011)
  [arXiv:1105.5995 [hep-ex]].

\bi{CLEO:Xic2815}  J.P. Alexander {\it et al.} [CLEO Collaboration],
Phys. Rev. Lett. {\bf 83}, 3390 (1999).


\bi{Bigi92} I.I. Bigi, N.G. Uraltsev, and A.I. Vainshtein, Phys.
Lett. B {\bf 293}, 430 (1992); B {\bf 297}, 477(E) (1992).

\bi{Chay} J. Chay, H. Georgi, and B. Grinstein, Phys. Lett. B {\bf
247}, 399 (1990); J. Chay and S.J. Rey, Z. Phys. C {\bf 68}, 431
(1995).

\bi{Luke} M.E. Luke, Phys. Lett.  B {\bf 252}, 447 (1990).

\bibitem{Cheng:1997}
  H.~Y.~Cheng,
  Phys.\ Rev.\ D {\bf 56}, 2783 (1997)
  [hep-ph/9704260].

\bibitem{Kiselev:2001}
  V.~V.~Kiselev and A.~K.~Likhoded,
  Phys.\ Usp.\  {\bf 45}, 455 (2002)
  [Usp.\ Fiz.\ Nauk {\bf 172}, 497 (2002)]
  [hep-ph/0103169];
  V.~V.~Kiselev, A.~K.~Likhoded and A.~I.~Onishchenko,
  Phys.\ Rev.\ D {\bf 60}, 014007 (1999)
  [hep-ph/9807354].

\bibitem{Guberina}
  B.~Guberina, B.~Melic and H.~Stefancic,
  Eur.\ Phys.\ J.\ C {\bf 9}, 213 (1999)
  [Erratum: Eur.\ Phys.\ J.\ C {\bf 13}, 551 (2000)]
  [hep-ph/9901323].

\bibitem{Chang}
  C.~H.~Chang, T.~Li, X.~Q.~Li and Y.~M.~Wang,
  Commun.\ Theor.\ Phys.\  {\bf 49}, 993 (2008)
  [arXiv:0704.0016 [hep-ph]].

\bibitem{Karliner}
  M.~Karliner and J.~L.~Rosner,
  Phys.\ Rev.\ D {\bf 90}, no. 9, 094007 (2014)
  [arXiv:1408.5877 [hep-ph]].

\bibitem{Fuk}
M. Fukugita, T. Inami, N. Sakai, and S. Yazaki, Phys. Lett. B {\bf
72}, 237 (1977); D. Tadi\'c and J. Trampeti\'c, ibid. B{\bf 114},
179 (1982); M. Bauer and B. Stech, ibid. B {\bf 152}, 380 (1985).

\bibitem{Buras}
A.J. Buras, J.-M. G\'erard, and R. R\"uckl, Nucl. Phys. {\bf
B268}, 16 (1986).

\bi{BS} B. Blok and M. Shifman, Sov. J. Nucl. Phys. {\bf 45}, 35,
301, 522 (1987).

\bi{CT92} H.Y. Cheng and B. Tseng, Phys. Rev. D {\bf 46}, 1042
(1992); D {\bf 55}, 1697(E) (1997).

\bi{Korner} J.G. K\"orner and M. Kr\"amer, Z. Phys. C {\bf 55},
659 (1992).

\bi{Korner79} J.G. K\"orner, G. Kramer, and J. Willrodt, Phys.
Lett. B {\bf 78}, 492 (1978); Z. Phys. C {\bf 2}, 117 (1979); B.
Guberina, D. Tadi\'c, and J. Trampeti\'c, Z. Phys. C {\bf 13}, 251
(1982); F. Hussain and M.D. Scadron, Nuovo Cimento, A {\bf 79},
248 (1984); F. Hussain and K. Khan, ibid. A {\bf 88}, 213 (1985);
R.E. Karlsen and M.D. Scadron, Europhys. Lett. {\bf 14}, 319
(1991); D. Ebert and W. Kallies, Phys. Lett. B {\bf 131}, 183
(1983); B {\bf 148}, 502(E) (1984); Yad. Fiz. {\bf 40}, 1250
(1984); Z. Phys. C {\bf 29}, 643 (1985); H.Y. Cheng, Z. Phys. C
{\bf 29}, 453 (1985); Yu.L. Kalinovsky, V.N. Pervushin, G.G.
Takhtamyshev, and N.A. Sarikov, Sov. J. Part. Nucl. {\bf 19}, 47
(1988); S. Pakvasa, S.F. Tuan, and S.P. Rosen, Phys. Rev. D {\bf
42}, 3746 (1990); G. Kaur and M.P. Khanna, Phys. Rev. D {\bf 44},
182 (1991); ibid. D {\bf 45}, 3024 (1992); G. Turan and J.O. Eeg,
Z. Phys. C {\bf 51}, 599 (1991).

\bi{Uppal} T. Uppal, R.C. Verma, and M.P. Khana, Phys. Rev. D {\bf
49}, 3417 (1994).

\bi{CT93} H.Y. Cheng and B. Tseng, Phys. Rev. D {\bf 48}, 4188
(1993).

\bi{XK92} Q.P. Xu and A.N. Kamal, Phys. Rev. D {\bf 46}, 270
(1992).

\bi{XK92b} Q.P. Xu and A.N. Kamal, Phys. Rev. D {\bf 46}, 3836
(1992).

\bi{Verma98} K.K. Sharma and R.C. Verma, Eur. Phys. J. C {\bf 7},
217 (1999).

\bi{Zen} P. \.Zenczykowski, Phys. Rev. D {\bf 50}, 5787, 3285, 402
(1994).

\bi{Ivanov98} M.A. Ivanov, J.G. K\"orner, V.E. Lyubovitskij, and
A.G. Tusetsky, Phys. Rev. D {\bf 57}, 1 (1998).

\bi{CCT}
  L.~L.~Chau, H.~Y.~Cheng and B.~Tseng,
  Phys.\ Rev.\ D {\bf 54}, 2132 (1996)
  [hep-ph/9508382].

\bi{Kohara} Y. Kohara, Phys. Rev. D {\bf 44}, 2799 (1991).

\bibitem{Zupanc}
  A.~Zupanc {\it et al.}  [Belle Collaboration],
  Phys.\ Rev.\ Lett.\  {\bf 113}, no. 4, 042002 (2014)
  [arXiv:1312.7826 [hep-ex]].

\bibitem{Lyu} X. R. Lyu, talk presented at The 7th International Conference on Charm Physics, May 18-22, 2015, Wayne State University, Detroit, USA.

\bibitem{Contu}
  A.~Contu, D.~Fonnesu, R.~G.~C.~Oldeman, B.~Saitta and C.~Vacca,
  Eur.\ Phys.\ J.\ C {\bf 74}, no. 12, 3194 (2014)
  [arXiv:1408.6802 [hep-ex]].

\bi{Belle:LamcCS}
 K. Abe et al. [Belle Collaboration], Phys. Lett. B 524, 33
 (2002).

\bi{BaBar:LamcCS}
 B.~Aubert {\it et al.} [BaBar Collaboration],
  Phys.\ Rev.\ D {\bf 75}, 052002 (2007)
  [hep-ex/0601017].

\bi{CT96} H.Y. Cheng and B. Tseng, Phys. Rev. D {\bf 53}, 1457
(1996); Phys. Rev. D {\bf 55}, 1697(E) (1997).

\bi{CLEO:alpha}  M. Bishai {\it et al.,} [CLEO Collaboration], Phys.
Lett. B {\bf 350}, 256 (1995).

\bi{FOCUS:SigK}  J.M. Link {\it et al.,} [FOCUS Collaboration], Phys.
Lett. B {\bf 571}, 139 (2003).

\bibitem{Dhir}
  R.~Dhir and C.~S.~Kim,
  Phys.\ Rev.\ D {\bf 91}, no. 11, 114008 (2015)
  [arXiv:1501.04259 [hep-ph]].


\bi{ChengHFC}
  H.~Y.~Cheng, C.~Y.~Cheung, G.~L.~Lin, Y.~C.~Lin, T.~M.~Yan and H.~L.~Yu,
  Phys.\ Rev.\ D {\bf 46}, 5060 (1992).

\bi{Voloshin00}  M.~B.~Voloshin,
  Phys.\ Lett.\ B {\bf 476}, 297 (2000)
  [hep-ph/0001057];
  X.~Li and M.~B.~Voloshin,
  Phys.\ Rev.\ D {\bf 90}, no. 3, 033016 (2014)
  [arXiv:1407.2556 [hep-ph]].

\bibitem{Faller}
  S.~Faller and T.~Mannel,
  arXiv:1503.06088 [hep-ph].

\bi{MIT} A. Chodos, R.L. Jaffe, K. Johnson, and C.B. Thorn, Phys.
Rev. D {\bf 10}, 2599 (1974); T. DeGrand, R.L. Jaffe, K. Johnson,
and J. Kiskis, Phys. Rev. D {\bf 12}, 2060 (1975).

\bi{Marcial}
 R. P\'erez-Marcial, R. Huerta, A. Garcia, and M.
 Avila-Aoki, Phys. Rev. D {\bf 40}, 2955 (1989); ibid. D {\bf 44},
 2203(E) (1991).

\bi{Singleton} R. Singleton, Phys. Rev. D  {\bf 43}, 2939 (1991).

\bi{Ivanov96}
  M.~A.~Ivanov, V.~E.~Lyubovitskij, J.~G.~K\"orner and P.~Kroll,
  Phys.\ Rev.\ D {\bf 56}, 348 (1997)
  [hep-ph/9612463].

\bi{Luo} C.W. Luo, Eur. Phys. J. C {\bf 1}, 235 (1998).

\bi{Carvalho} R.S. Marques de Carvalho {\it et al.,} Phys. Rev. D
{\bf 60}, 034009 (1999).

\bi{Huang}
  M.~Q.~Huang and D.~W.~Wang,
  hep-ph/0608170.

\bibitem{Gutsche}
  T.~Gutsche, M.~A.~Ivanov, J.~G.~K\"orner, V.~E.~Lyubovitskij and P.~Santorelli,
  Phys.\ Rev.\ D {\bf 90}, no. 11, 114033 (2014)
  [arXiv:1410.6043 [hep-ph]].

\bibitem{Azizi}
  K.~Azizi, Y.~Sarac and H.~Sundu,
  Eur.\ Phys.\ J.\ A {\bf 48}, 2 (2012)
  [arXiv:1107.5925 [hep-ph]].

\bibitem{Pervin}
  M.~Pervin, W.~Roberts and S.~Capstick,
  Phys.\ Rev.\ C {\bf 72}, 035201 (2005)
  [nucl-th/0503030].

\bibitem{CLEO:sl}
  J.~W.~Hinson {\it et al.} [CLEO Collaboration],
  Phys.\ Rev.\ Lett.\  {\bf 94}, 191801 (2005)
  [hep-ex/0501002].

\bibitem{Azizi2}
  K.~Azizi, M.~Bayar, Y.~Sarac and H.~Sundu,
  Phys.\ Rev.\ D {\bf 80}, 096007 (2009)
  [arXiv:0908.1758 [hep-ph]].

\bibitem{Pervin:omegac}
  M.~Pervin, W.~Roberts and S.~Capstick,
  Phys.\ Rev.\ C {\bf 74}, 025205 (2006)
  [nucl-th/0603061].

\bibitem{Albertus:2011}
  C.~Albertus, E.~Hernandez and J.~Nieves,
  Phys.\ Lett.\ B {\bf 704}, 499 (2011)
  [arXiv:1108.1296 [hep-ph]].

\bibitem{Faessler}
  A.~Faessler, T.~Gutsche, M.~A.~Ivanov, J.~G.~K\"orner and V.~E.~Lyubovitskij,
  Phys.\ Rev.\ D {\bf 80}, 034025 (2009)
  [arXiv:0907.0563 [hep-ph]].

\bibitem{Ebert:2004}
  D.~Ebert, R.~N.~Faustov, V.~O.~Galkin and A.~P.~Martynenko,
  Phys.\ Rev.\ D {\bf 70}, 014018 (2004)
  [Phys.\ Rev.\ D {\bf 77}, 079903 (2008)]
  [hep-ph/0404280].

\bibitem{Guo:doubly}
  X.~H.~Guo, H.~Y.~Jin and X.~Q.~Li,
  Phys.\ Rev.\ D {\bf 58}, 114007 (1998)
  [hep-ph/9805301].

\bi{Cheng93} H.~Y.~Cheng, C.~Y.~Cheung, G.~L.~Lin, Y.~C.~Lin, T.~M.~Yan and H.~L.~Yu,
  Phys.\ Rev.\ D {\bf 47}, 1030 (1993)
  [hep-ph/9209262].


\bi{ChengSU3}  H.~Y.~Cheng, C.~Y.~Cheung, G.~L.~Lin, Y.~C.~Lin, T.~M.~Yan and H.~L.~Yu,
  Phys.\ Rev.\ D {\bf 49}, 5857 (1994)
  [Phys.\ Rev.\ D {\bf 55}, 5851 (1997)]
  [hep-ph/9312304].

\bibitem{Jiang}
  N.~Jiang, X.~L.~Chen and S.~L.~Zhu,
  arXiv:1505.02999 [hep-ph].

\bi{Cheng97}
  H.~Y.~Cheng,
  Phys.\ Lett.\ B {\bf 399}, 281 (1997)
  [hep-ph/9701234].

\bibitem{Dey}
  J.~Dey, M. Dey, V.~Shevchenko and P.~Volkovitsky,
  Phys.\ Lett.\ B {\bf 337}, 185 (1994).

\bi{Tawfiq01} S. Tawfiq, J.G. K\"orner and  P.J. O'Donnell, Phys.
Rev. D {\bf 63}, 034005 (2001).

\bi{Savage95} M.J. Savage, Phys. Lett. B {\bf 345}, 61 (1995).

\bi{Pich} M.C. Ba\~uls, A. Pich, and I. Scimemi, Phys. Rev. D {\bf
61}, 094009 (2000).

\bibitem{Aliev}
  T.~M.~Aliev, K.~Azizi and A.~Ozpineci,
  Phys.\ Rev.\ D {\bf 79}, 056005 (2009)
  [arXiv:0901.0076 [hep-ph]].

\bibitem{ZGWang}
  Z.~G.~Wang,
  Phys.\ Rev.\ D {\bf 81}, 036002 (2010)
  [arXiv:0909.4144 [hep-ph]];
  Eur.\ Phys.\ J.\ A {\bf 44}, 105 (2010)
  [arXiv:0910.2112 [hep-ph]].

\bibitem{Bernotas}
  A.~Bernotas and V.~\v Simonis,
  Phys.\ Rev.\ D {\bf 87}, no. 7, 074016 (2013)
  [arXiv:1302.5918 [hep-ph]].

\bibitem{Majethiya}
  A.~Majethiya, B.~Patel and P.~C.~Vinodkumar,
  Eur. Phys. J. A {\bf 42}, 213 (2009).

\bibitem{Bahtiyar}
 H.~Bahtiyar, K.~U.~Can, G.~Erkol and M.~Oka,
  Phys.\ Lett.\ B {\bf 747}, 281 (2015)
  [arXiv:1503.07361 [hep-lat]].

\bi{Lu} M. Lu, M.J. Savage, and J. Walden, Phys. Lett. B {\bf
369}, 337 (1996).

\bi{Cohen}
  Z.~Aziza Baccouche, C.~K.~Chow, T.~D.~Cohen and B.~A.~Gelman,
  Nucl.\ Phys.\ A {\bf 696}, 638 (2001)
  [hep-ph/0105148]

\bibitem{Zhu:2000}
  S.~L.~Zhu,
  Phys.\ Rev.\ D {\bf 61}, 114019 (2000)
  [hep-ph/0002023].

\bibitem{Chow}
  C.~K.~Chow,
  Phys.\ Rev.\ D {\bf 54}, 3374 (1996)
  [hep-ph/9510421].

\bibitem{Gamermann}
  D.~Gamermann, C.~E.~Jimenez-Tejero and A.~Ramos,
  Phys.\ Rev.\ D {\bf 83}, 074018 (2011)
  [arXiv:1011.5381 [hep-ph]].

\bibitem{Dong:rad}
  Y.~B. Dong, A.~Faessler, T.~Gutsche, S.~Kumano and V.~E.~Lyubovitskij,
  Phys.\ Rev.\ D {\bf 82}, 034035 (2010)
  [arXiv:1006.4018 [hep-ph]].

\bibitem{Branz}
  T.~Branz, A.~Faessler, T.~Gutsche, M.~A.~Ivanov, J.~G.~Korner, V.~E.~Lyubovitskij and B.~Oexl,
  Phys.\ Rev.\ D {\bf 81}, 114036 (2010)
  [arXiv:1005.1850 [hep-ph]].

\bi{Kamal83} A.N. Kamal and Riazuddin, Phys. Rev. D {\bf 28}, 2317
(1983).

\bi{Kamal82} A.N. Kamal and R.C. Verma, Phys. Rev. D {\bf 26}, 190
(1982).

\bi{Hua} L.C. Hua, Phys. Rev. D {\bf 26}, 199 (1982).

\bi{Chengweakrad}
  H.~Y.~Cheng, C.~Y.~Cheung, G.~L.~Lin, Y.~C.~Lin, T.~M.~Yan and H.~L.~Yu,
  Phys.\ Rev.\ D {\bf 51}, 1199 (1995)
  [hep-ph/9407303].

\bi{Uppalrad} T. Uppal and R.C. Verma, Phys. Rev. D {\bf 47}, 2858
(1993).

\end{thebibliography}
\end{document}